\documentclass[prd,twocolumn,nofootinbib,superscriptaddress]{revtex4-2}
\usepackage{preamble}
\graphicspath{{images/}} 

\begin{document}

\title{Inspiral tests of general relativity and waveform geometry}

\author{Brian C. Seymour\,\orcidlink{0000-0002-7865-1052}} 
\affiliation{TAPIR, Walter Burke Institute for Theoretical Physics, California Institute of Technology, Pasadena, CA 91125, USA}
\affiliation{Center of Gravity, Niels Bohr Institute, Blegdamsvej 17, 2100 Copenhagen, Denmark}
\author{Jacob Golomb\,\orcidlink{0000-0002-6977-670X}} 
\affiliation{TAPIR, Walter Burke Institute for Theoretical Physics, California Institute of Technology, Pasadena, CA 91125, USA}
\author{Yanbei Chen\,\orcidlink{0000-0002-9730-9463}} 
\affiliation{TAPIR, Walter Burke Institute for Theoretical Physics, California Institute of Technology, Pasadena, CA 91125, USA}

\date{\today}

\begin{abstract} 
    The phase evolution of gravitational waves encodes critical information about the orbital dynamics of binary systems. In this work, we test the robustness of parameterized tests against unmodeled deviations from general relativity. We demonstrate that these parameterized tests are flexible and sensitive in detecting generic deviations in the waveform using the Cutler-Vallisneri bias formalism. This universality arises from examining the inherent geometry of the waveform signal and understanding how biases manifest. We show how Bayes factors are governed by the intrinsic geometry of the waveform signal manifold when parameterized tests are used to approximate generic violations of GR. We use the singular value decomposition to propose templates that are orthogonal to parameterized tests, identifying degeneracies and enhancing the detection of potential deviations. More broadly, the geometric framework developed here clarifies---at a fundamental level---how subtle waveform effects (including orbital eccentricity, spin precession, waveform systematics, and instrumental glitches) can mimic one another in data, and when they are intrinsically distinguishable.
    \href{https://github.com/BrianCSeymour/waveform-geometry-testing-gr}{\faGithub}
\end{abstract}
\maketitle

\section{Introduction}
The detections of gravitational waves (GW) from compact binary coalescences \cite{LIGOScientific:2016aoc,LIGOScientific:2017vwq,LIGOScientific:2018mvr,LIGOScientific:2020ibl,LIGOScientific:2021djp} have provided new ways to directly test the behavior of strong gravity in general relativity (GR). Despite many searches in LIGO-Virgo-KAGRA (LVK) data, no evidence for deviations has been found \cite{LIGOScientific:2016lio,LIGOScientific:2018dkp,LIGOScientific:2019fpa,LIGOScientific:2020tif,LIGOScientific:2021sio,Yunes:2016jcc,Nair:2019iur,Silva:2020acr}. The network currently consists of two LIGO interferometers \cite{LIGOScientific:2014pky}, Virgo \cite{VIRGO:2014yos}, KAGRA \cite{KAGRA:2020tym}, and eventually LIGO India \cite{iyer2013ligo,Saleem:2021iwi}. To maximize the scientific reach of these current detectors, there are proposals for A\# \cite{lsc2022report-asharp} and LIGO Voyager \cite{ligo-voyager-dcc,LIGO:2020xsf}. Third-generation ground-based facilities---Cosmic Explorer \cite{Reitze:2019iox,Evans:2021gyd} and the Einstein Telescope \cite{Punturo:2010zz,ET:2019dnz,Branchesi:2023mws}---are in the planning stages, which are expected to improve the sensitivity compared to LIGO A+ by an order of magnitude. There are a number of space-based detectors planned of which LISA is expected to launch in 2035 \cite{LISA:2017pwj,LISACosmologyWorkingGroup:2022jok}. There are other concepts including TianQin \cite{TianQin:2015yph}, Taiji \cite{Hu:2017mde,Ruan:2018tsw}, B-DECIGO/DECIGO \cite{Seto:2001qf,Sato:2017dkf,Kawamura:2020pcg}, and TianGO \cite{Kuns:2019upi,kunsthesis}.

One of the most prominent techniques for searching for deviations to GR is the parameterized post-Einsteinian (ppE) framework \cite{Yunes:2009ke,Cornish:2011ys,Li:2011cg,Agathos:2013upa,Meidam:2017dgf,Yunes:2025xwp}. This formalism searches for deviations to the phase of the waveform that appear at particular orders in velocity of the waveform. This framework was also extended in a number of ways throughout the years, including additional polarizations \cite{Chatziioannou:2012rf}, precessing waveforms \cite{Loutrel:2022xok}, higher-order modes \cite{Mezzasoma:2022pjb,Mehta:2022pcn}, eccentric signals \cite{Bhat:2024hyb}, approaches for parameterized searches in the plunge-merger phase \cite{Bonilla:2022dyt,Maggio:2022hre,Watarai:2023yky,Watarai:2025hsb}, and a neural post-Einstein framework \cite{Xie:2024ubm}. It is noted that such an expansion in $v/c$ does not work for all types of perturbations, namely logarithmic or screened terms \cite{Sampson:2013jpa,Yunes:2025xwp}. It also needs the perturbations to the binding energy and GW fluxes to be able to be expressed as a factor of $(M/r)^\alpha$ which we have found is not possible for, e.g., nonviolent nonlocality that has an essential singularity \cite{Seymour:2024kcd}.

These tests of GR are intimately connected to how well these beyond GR deviations are observable in data. It has been noted for a long time that waveform distinguishability \cite{Flanagan:1997kp} and thus correspondingly accuracy requirements \cite{Lindblom:2008cm,Kumar:2015tha,Chatziioannou:2017tdw,Purrer:2019jcp,Hu:2022rjq} are tied to the geometry of the signal manifold \cite{McWilliams:2010eq,Toubiana:2024car,Knapp:2025ecw,Thompson:2025hhc}. Systematic bias to the waveform model directly manifests as biases to the measured values of the parameters \cite{Flanagan:1997kp,Cutler:2007mi,Hu:2022bji,Lindblom:2008cm,Kumar:2015tha,Chatziioannou:2017tdw,Purrer:2019jcp,Hu:2022rjq,Owen:2023mid,Kapil:2024zdn,Dhani:2024jja,Capuano:2025kkl,Volkel:2025jdx}. Building upon approaches to understand how waveforms are biased due to unmodeled signals \cite{Cutler:2007mi,Hu:2022bji}, Vallisneri found that the GR parameters aim to mask the beyond GR signal deviation \cite{Vallisneri:2012qq,Vallisneri:2013rc} which is called \textit{stealth bias} \cite{Vallisneri:2013rc,Vitale:2013bma}. It is shown that it is only the perpendicular portion of the waveform that contributes to the detectability of GR violations. As one would expect, both systematics of the GR template waveform \cite{Moore:2021eok,Perkins:2022fhr,Favata:2013rwa,Chandramouli:2024vhw,Gupta:2024gun,Saini:2022igm,Bhat:2022amc,Saini:2023rto,Narayan:2023vhm,Garg:2024qxq}, external astrophysical environment \cite{Kejriwal:2023djc,Yunes:2010sm,Chamberlain:2018snj,Robson:2018svj,Yu:2020dlm,Yu:2021dqx,Laeuger:2023qyz,Deme:2020ewx,Chandramouli:2021kts,Gupta:2019unn}, and even instrumental glitches \cite{LIGOScientific:2016gtq,Ghonge:2023ksb,Kwok:2021zny,Ashton:2022ztk,Spadaro:2023muy} can show up as false positives deviations to GR.

In addition to this statistical hypothesis testing, eventually the community wants to find ways to better test GR than measuring a single parameter. If multiple ppE parameters are attempted to be measured for a particular waveform, the covariance between each of them means that constraints are markedly reduced. The singular value decomposition (SVD) approach to searching for deviations to GR was originally devised by Pai and Arun \cite{Pai:2012mv,Arun:2013bp} which is an approach similar in spirit to surrogate modeling \cite{Cannon:2010qh,Cannon:2011xk, Cannon:2011rj,Keppel:2012nb,Tiglio:2021ysj}. This approach used features of ppE deviations from GR to identify the features that are most common and most precisely measured \cite{Pai:2012mv,Arun:2013bp}. Additionally, if one attempts to measure multiple ppE parameters in data, a principal component analysis (PCA) can be performed to identify which directions that the covariance matrix is best measured \cite{Datta:2022izc,Saleem:2021nsb,Datta:2023muk,Ma:2024kkz,Mahapatra:2025cwk,Shoom:2021mdj,Arun:2006yw,Arun:2006hn,Volkel:2022aca}. While the SVD and PCA differ due in origin from modeling with least parameters versus the statistical relationship of measuring multiple parameters, they are related to one another.

In this paper, we will build upon the existing literature and relate the tests of GR to the geometry of the signal manifold from the noise-weighted inner product. We will begin by identifying how GR parameters are biased if a beyond GR signal \cite{Vallisneri:2012qq} is introduced and show how the ppE formalism can capture generic deviations remarkably well due to the behavior of the geometrical picture. We will show how multiparameter tests of GR are difficult because the ppE deviations to GR have similar features when orthogonalized away from GR. Finally, we will build upon the work of Pai \textit{et al.}~\cite{Pai:2012mv,Arun:2013bp} and introduce a new form of the SVD that identifies the common features of the ppE tests. Throughout, we will use geometric units where $c= G = 1$. The Python notebooks and environment used to generate the main figures of this paper is given in the Github repository \textsc{waveform-geometry-testing-gr} \href{https://github.com/BrianCSeymour/waveform-geometry-testing-gr}{\faGithub}.

\begin{figure}[htbp]
    \centering
    \includegraphics[width=\linewidth]{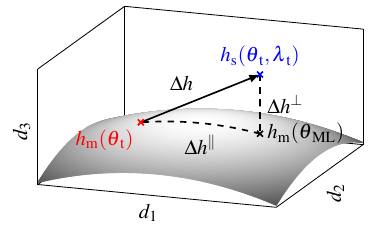}
    \caption{
        Illustration of degeneracy when testing GR. We show the injected signal (blue) which depends on the true GR parameters $\bf\theta_\t$ and the beyond GR parameters $\bf\lambda_\t$. The model signal at the true GR parameters $\bf\theta_\t$ (red) is shown and the best fit signal is at the maximum likelihood point  $\bf\theta_\ml$ (black). The GR waveform is modified by $\Delta h$ which causes biases to the GR waveform, thus residual signal to measure beyond GR deviations is given by the perpendicular signal $\Delta h^\perp$. Note that this is a high dimensional manifold where $(d_1, d_2, d_3)$ are the values of the signal at particular frequency bins.
    }
    \label{fig:intro-geo-man}
\end{figure}

\section{Background and methods}

This analysis is motivated by work done on calibration errors in a waveform using the Cutler-Vallisneri bias formalism \cite{Cutler:2007mi,Hu:2022bji}. In Fig.~\ref{fig:intro-geo-man}, we show a visualization of how the true GR parameters $\bf\theta_\t$ are biased due to a beyond GR signal $h_\mathrm{s}(\bf\theta_\t ,\bf\lambda_\t)$ (blue) where $\bf\lambda_\t$ are the beyond GR parameters. The possible values of the model waveform $h_\m(\bf\theta_\t)$ are shown in the gray manifold while the best fitting parameter is $\bf\theta_\ml$. In this section, we will derive how the GR parameters are biased replicating \cite{Cutler:2007mi,Hu:2022bji} and show how the Bayes factor depends strongly on the residual deviation from GR $\Delta h^\perp$ as found in \cite{Vallisneri:2012qq,Vallisneri:2013rc}.

\subsection{Review of biased parameter estimation}
Suppose we have a true waveform that is $s_t \equiv h_\s(\bf{\theta}_t,\bf{\lambda}_t)$ and we are attempting to recover it with a model $\hm(\bf\theta)$. The likelihood in GR is just the true waveform plus Gaussian noise
\begin{equation}
    d = s_t + n \, , 
\end{equation}
while our template is $\hm(\bf{\theta})$. For stationary and Gaussian noise the likelihood for a waveform is equal to
\begin{equation}
    \log L(\bf\theta) \propto - \frac{1}{2}\left( d-\hm(\bf{\theta})|d-\hm(\bf{\theta}) \right) \, ,
\end{equation}
where the noise-weighted inner product is defined as 
\begin{equation}
    \left( a| b \right)\equiv 4 \,\text{Re} \int_0^\infty \frac{a^*(f) b(f)}{S_n(f)}df \, ,
\end{equation}
where $S_n(f)$ is the power spectral density (PSD) of the detector.

We wish to find the Fisher information about the point to characterize detectability \cite{Cutler:1994ys,Flanagan:1997kp,Cutler:2007mi,Vallisneri:2007ev,Rodriguez:2013mla}. The maximum likelihood estimator is found where the derivative of the likelihood is zero
\begin{equation}\label{eq:MLE-eq}
    \left.\partial_{i} \log L\right|_{\bf\theta_\text{ML}} = \left.\left( \partial_{i} \hm(\bf\theta)| d-\hm(\bf\theta) \right)\right|_{\bf\theta_\text{ML}} = 0 \, .
\end{equation}
In general, this is a nonlinear equation to solve for $\bf\theta_\ml$, but we make the standard assumption that it is sharply peaked in $\bf\theta$ so that $\bf \theta_\ml - \bf\theta_\t \ll 1$. Let us define $\Delta h$ as 
\begin{equation}\label{eq:lambdadh}
    \Delta h = h_\s(\bf \theta_\t, \bf \lambda_\t) - \hm(\bf\theta_\t) \,.
\end{equation}
We also assume that $\Delta h$ is small\footnote{Note that stability of this equation is poor for actual numerical implementation, especially when $\Delta h_\parallel \gg \Delta h_\perp$. Appendix \ref{app:bias-equation-numerical-stability} discusses how this is rectified by modifying the form of $\Delta h$ used in \eqref{eq:statbias}.}. Therefore, we can use $d-\hm(\bf\theta) = n + \Delta h + \Delta \theta^i \partial_i \hm $. The maximum likelihood estimate is thus equal to 
\begin{equation}\label{eq:gr-ml-value}
    \theta_\ml^i = \theta_\t^i + \Delta \theta_\bias^i + \Delta \theta_\stat^i \, ,
\end{equation}
where the bias and statistical errors are equal to 
\begin{align}\label{eq:statbias}
    \Delta \theta^i_\stat &= \Sigma^{ij}\left( \partial_j \hm | n \right) \, , \\
    \Delta \theta_\bias^i &= \Sigma^{ij}\left( \partial_j \hm |\Delta h \right)\, ,
\end{align}
where $\Sigma^{ij}$ is the matrix inverse of the Fisher matrix $\Gamma_{ij} \equiv \left( \partial_i \hm |\partial_j\hm \right)$. One can show that the covariance of the $\Delta \theta^i_\stat$ random variable is equal to the inverse of the Fisher matrix $E\left[ \Delta\theta_\stat^i\Delta\theta_\stat^j \right] \equiv \Sigma^{ij} = \left( \Gamma^{-1} \right)^{ij}$ \cite{Finn:1992wt,Cutler:1994ys}. We emphasize that the $ \bf{\Delta\theta}_\bias$ comes from the mismodeling of the waveform and is independent of SNR. 

Suppose that we are not in the nested case as described above, where our modeled waveform actually depends on some other parameters which are not in the signal. In this case, we model it as $\hm(\bf\theta,\bf\mu)$ where $\bf\mu$ are some small parameters which we want to measure in the waveform. In this case, we can perform the same analysis as above. If we define $\Theta^I = (\bf\theta,\bf\mu)$ as the full parameter set, the maximum likelihood estimate is 
\begin{equation}\label{eq:thetaI-equation-ml-stat}
    \Theta^I_\ml = \Theta^I_\t + \Delta\Theta_\stat^I + \Delta\Theta_\bias^I \, ,
\end{equation}
where $\Theta^I_\t = \left( \bf\theta_\t, \bf0 \right)$ so we are assuming the true value of $\bf\mu_\t \sim 0$. The statistical error is the same $\Delta\Theta_\stat^I = \Sigma^{IJ}\left( \partial_J \hm | n \right)$ while the bias is 
\begin{equation}\label{eq:param-bias}
    \Delta \Theta_\bias^I = \Sigma^{IJ}\left( \partial_J \hm |\Delta h \right)\, .
\end{equation}
We emphasize that at high SNR, our small model parameters will scale like
\begin{equation}\label{eq:param-bias-mu}
    \bf\mu_\ml = \Delta\bf\mu_\bias(\bf\theta_\t,\bf\lambda_\t) + \Delta\bf\mu_\stat \, ,
\end{equation} 
where the bias term is linearly proportional to the $\bf\lambda_\t$ parameter $\Delta\bf\mu_\bias \propto \bf\lambda_\t$.
This key fact will be used to explain the \textit{unreasonable effectiveness of ppE templates for testing GR}. If you imagine that there is a generic phase deviation in the true waveform $\Delta\Psi_\t$, this will be accessible if we measure any $\delta\varphi_k$ since it is responsive to any modification to the waveform.

\subsection{Geometric interpretation of bias equation }

In the previous section, we reviewed how an injection can bias the parameter estimation in a mismodeled waveform. Now, we will look at Eq.~\eqref{eq:param-bias} and discover how it can be understood more deeply by thinking about the geometry of the waveform. We define the parallel and perpendicular components as  
\begin{align} \label{eq:biasgeo}
    \Delta h_\parallel &= \Delta \Theta_\bias^J \partial_J \hm \, , \nn\\
    \Delta h_\perp &=\Delta h - \Delta \Theta_\bias^J \partial_J \hm \, .
\end{align}
Using these identities, one can see that Eq.~\eqref{eq:param-bias} becomes 
\begin{equation}\label{eq:bias-parallel-equation}
    \Delta \Theta_\bias^I = \Sigma^{IJ}\left( \partial_J \hm | \Delta h_\parallel \right) \,.
\end{equation}
This demonstrates that \textit{only deviations parallel to waveform manifold} bias the parameter estimation. 

Let us now investigate the case that we have a waveform with a bias that is due to a small parameter $\bf\lambda$ and our model has a small parameter $\bf\mu$. For simplicity assume that $\bf\lambda$ and $\bf\mu$ are both one dimensional parameters. Then, one can show that the estimate for $\mu$ is 
\begin{equation}\label{eq:mubias-single}
    \Delta\mu_\bias = \frac{\left( \left( \partial_{\mu} \hm \right)^{\perp \bf\theta}|\left( \Delta h \right)^{\perp \bf\theta} \right)}{\Vert \left( \partial_{\mu} \hm \right)^{\perp \bf\theta}\Vert^2} \, ,
\end{equation}
where the $\perp \bf\theta$ denotes removing the part of the signal which is parallel to the main $\bf\theta$ parameters. Thus, we see evidence for $\bf\mu$ when both the waveform $\hm(\bf\theta,\bf\mu)$ and the residual $\Delta h(\bf\theta_t,\bf\lambda)$ cannot eliminate the effect by a redefinition of $\bf\theta$. 

In Eq.~\eqref{eq:mubias-single}, we performed this calculation for the simplest case that you are trying to measure a single parameter $\mu$. If $\bf\mu = \mu^a$ is an array of beyond GR parameters, the equation above can be generalized using Schur decomposition which yields 
\begin{equation}\label{eq:mubias-multi}
    \Delta \mu_\bias^a = \left( \Gamma_\mathrm{red}^{-1} \right)_{ab} \left( \left( \partial_{\mu^b} \hm\right)^{\perp \bf\theta} | \left( \Delta h \right)^{\perp\bf\theta} \right) \, ,
\end{equation}
where the reduced Fisher matrix is equal to 
\begin{equation}
    \Gamma^\mathrm{red}_{ab} = \left(\left( \partial_{\mu^a} \hm \right)^{\perp \bf\theta}|\left( \partial_{\mu^b} \hm \right)^{\perp \bf\theta} \right) \,.
\end{equation}
One can see in the single $\mu$ case the equation reduces to Eq.~\eqref{eq:mubias-single} since $\Gamma^\mathrm{red}_{\mu\mu} = \Vert \left( \partial_{\mu} \hm \right)^{\perp \bf\theta}\Vert^2$. In Appendix~\ref{app:multibias}, we derive Eq.~\eqref{eq:mubias-multi} using the Schur decomposition identities for a matrix consisting of blocks.

\subsection{Multidetector geometry}

The manifold language introduced in this chapter straightforwardly generalizes to the case of multiple detectors observing one GW event. The likelihood for parameters is 
\begin{equation}
    p(d|\bf\theta,\bf\mu) \propto \exp\left[ -\frac{1}{2}\sum_A\Vert d_A - \hm^A(\bf\theta,\bf\mu) \Vert^2 \right]\, .
\end{equation}
It is useful to introduce the summed network inner product and use square brackets to represent this 
\begin{equation}
    \left[ a | b \right] = \sum_{A}\left(a_A | b_A  \right)_{A} \, ,
\end{equation}
then the maximum likelihood point is the solution to 
\begin{equation}
    \Gamma_{IJ} \Delta\Theta^J = \left[\partial_I \hm| \Delta h\right] + \left[\partial_I \hm|n\right] \,,
\end{equation}
where the network Fisher information matrix is $\Gamma_{IJ} \equiv \left[\partial_I\hm|\partial_J\hm\right]$. The equations for the statistical error and bias straightforwardly generalize from Eq.~\eqref{eq:statbias} and are equal to 
\begin{align}
    \Delta \Theta^I_\stat &= \left( \Gamma^{-1} \right)^{IJ}\left[ \partial_I \hm | n \right] \, , \\
    \Delta \Theta_\bias^I &= \left( \Gamma^{-1} \right)^{IJ}\left[ \partial_I \hm |\Delta h \right]\, .
\end{align}
Henceforth, we will continue to do calculations as if they were a single detector and use the normal inner product $\left( \cdot | \cdot\right)$ notation, but the geometric description of testing GR in multiple detectors follows the same paradigm.

\subsection{Bayes factors for tests of GR}\label{sec:bayes-tgr}

We wish to compare whether there is support for beyond GR (bGR) signals in the data. Vallisneri \cite{Vallisneri:2012qq,Vallisneri:2013rc} showed how the Bayes factor provides support for the bGR signal and identified the geometric meaning. If we have two hypotheses (a) GR $\left( \bf\theta \right)$ and (b) bGR $\left( \bf\theta,\bf\lambda \right)$, the Bayes factor is equal to 
\begin{equation}\label{eq:vallisneri-bf}
    \mathcal{B}^\bGR_\GR \equiv  \frac{p(d|\bGR)}{p(d|\GR)}\,,
\end{equation}
where the evidence of data $d$ in model $\mathcal{M}$ with parameters $\bf\Theta$ is 
\begin{equation}
    p(d|\mathcal{M}) = \int d\bf\Theta p(d|\bf\Theta,\mathcal{M}) p(\bf\Theta) \, ,
\end{equation}
where $p(\bf\Theta)$ is the prior and $p(d|\bf\Theta,\mathcal{M})$ is the likelihood. Vallisneri computed that the Bayes factor is 
\begin{equation}\label{eq:optimalBayes}
    \left.\mathcal{B}^\bGR_\GR\right|_{s_\bGR } =\frac{(2 \pi)^{1 / 2} \Delta \lambda_\stat}{\Delta \lambda_\prior} e^{\rho_\perp^2 / 2+x\rho_\perp+x^2 / 2}\,,
\end{equation}
where $x\equiv \left( \Delta h_\perp | n \right)/\Vert \Delta h _\perp\Vert \sim \mathcal{N}(0,1)$ is a standard normal random variable, $\Delta \lambda_\stat$ is the statistical error on $\bf\lambda$, and we assumed a flat prior with width $\Delta \lambda_\prior$. The residual SNR is 
\begin{equation}
    \rho_\perp \equiv \Vert \Delta h_{\perp}\Vert
\end{equation}
and $\Delta h_{\perp}$ is the \textit{true} perpendicular signal to the GR waveform as defined in Eq.~\eqref{eq:lambdadh} and Eq.~\eqref{eq:biasgeo} due to $\bf\lambda_t$. One can see that the residual SNR is what provides evidence for a GR violation. If the true signal is a GR signal, the Bayes factor is just 
\begin{equation}
    \left.\mathcal{B}^\bGR_\GR\right|_{s_\GR }= \frac{(2 \pi)^{1 / 2} \Delta \lambda_\stat}{\Delta \lambda_\prior} e^{x^2 / 2}\,,
\end{equation}
which has a randomly distributed exponent $x^2 / 2$ and an \textit{Occam factor} prefactor $\propto \Delta \lambda_\stat/\Delta \lambda_\prior$ that will cause it to favor GR on average. 

Suppose we searched for deviations from GR with post-Newtonian (PN) deviations from GR in the formalism of the parameterized post-Einsteinian/TIGER/FTI tests of GR \cite{Yunes:2009ke,Cornish:2011ys,Li:2011cg,Agathos:2013upa,Meidam:2017dgf,Mehta:2022pcn} which mismodeled the bGR waveform. We will now generalize the Bayes factor formula in this case. We assume that the injected signal is $s_\bGR^t = h(\bf\theta_\t, \bf\lambda_\t)$. We will compare two hypotheses (a) GR $\left( \bf\theta \right)$ and (b) ppE $\left( \bf\theta,\bf\mu \right)$. Careful computation reveals that the Bayes factor is 
\begin{equation}\label{eq:bayes-pn-gr}
    \left.\mathcal{B}^\ppe_\GR\right|_{s_\bGR } = \frac{(2 \pi)^{1 / 2} \Delta \mu_\stat}{\Delta \mu_\prior} e^{\left( \rho_\perp^\ppe \right)^2 / 2+y\rho_\perp^\ppe+y^2 / 2}\,,
\end{equation}
where $y\equiv \left( \Delta h_\ppe^\perp | n \right)/\Vert\Delta h_\ppe^\perp\Vert$ is a unit normal Gaussian random variable and the captured residual SNR $\rho_\perp^\ppe$ is defined via
\begin{equation}\label{eq:captured-residual}
    \rho_\perp^\ppe = \mathcal{O} \rho_\perp \, ,
\end{equation}
which depends on the overlap of the bGR and ppE waveform deviations. The overlap is equal to 
\begin{equation}\label{eq:overlap-bgr-pn}
    \mathcal{O}(\Delta h_\ppe^\perp,\Delta h_\bGR^\perp) = \frac{\left( \Delta h_\ppe^\perp|\Delta h_\bGR^\perp \right)}{\Vert \Delta h_\ppe^\perp \Vert \Vert \Delta h_\bGR^\perp \Vert} \, .
\end{equation}
Note that this overlap is between 0 and 1, and quantifies how similar two signals are. We use notation $\Delta h_\ppe^\perp \equiv \left( \Delta h_\ppe^\ml \right)^\perp = \Delta \mu_\mathrm{bias} (\partial_\mu h_\m)^{\perp}$ where $\Delta \mu_\mathrm{bias}$ is the best fit for the ppE test as a function of $\Delta h_\bGR^\perp$. The lost residual SNR is then equal to
\begin{align}\label{eq:rho-lost}
    \rho^\mathrm{diff}_\perp &\equiv \Vert \Delta h^\perp_\bGR - \Delta h_\ppe^\perp\Vert \,, \nn\\
    &= \rho_\perp \sqrt{1-\mathcal{O}^2}\,.
\end{align}
In Appendix \ref{app:bayesfactor}, we rederive the optimal Bayes factor in Eq.~\eqref{eq:optimalBayes} using methods from \cite{Vallisneri:2012qq} and generalize it to the parameterized test Bayes factor in Eq.~\eqref{eq:bayes-pn-gr}. One can see that the geometric structure comes from features related to how Fisher matrices invert when you have blocks using the Schur decomposition. 

Finally, if we were to detect deviations from GR using e.g.~one of the parameterized tests, it is important to see how well we could distinguish a bGR model from a ppE model. This would be important in the hypothetical case that we need to identify which particular bGR models are consistent with the data and favored. We show that the Bayes factor between bGR $\left( \bf\theta,\bf\lambda \right)$ and ppE $\left( \bf\theta,\bf\mu \right)$ is equal to 
\begin{equation}\label{eq:bayes-factor-bgr-pn}
    \left.\mathcal{B}^\bGR_\ppe\right|_{s_\bGR } = \frac{\Delta \lambda_\mathrm{stat} \Delta \mu_\mathrm{prior}}{\Delta \lambda_\mathrm{prior}\Delta \mu_\mathrm{stat}} e^{(x+\rho_\perp)^2/2- (y+\mathcal{O}\rho_\perp)^2/2}  \,,
\end{equation}
where the $x,y$ are unit normal Gaussian random variables as defined below Eq.~\eqref{eq:optimalBayes} and Eq.~\eqref{eq:bayes-pn-gr}. Note that $E[xy] = \sqrt{1-\mathcal{O}^2}$ so these variables are correlated with one another. 
Additionally, we note that the fidelity at which we can distinguish between bGR and ppE is determined by the quantity $\rho^\mathrm{diff}_\perp$ that we derived in Eq.~\eqref{eq:rho-lost}. This expression is fully derived in Appendix \ref{app:bayesfactor}, but it involves using the fact that Bayes factor between bGR and ppE in Eq.~\eqref{eq:bayes-factor-bgr-pn} can be expressed as a ratio of the Bayes factors in Eq.~\eqref{eq:optimalBayes} and Eq.~\eqref{eq:bayes-pn-gr}. We also note that there is no simple expression for this Bayes factor, but it can be written as a sum of two uncorrelated random variables as we detail in Eq.~\eqref{eq:bgr-pn-bayes-appendix-white-decorrelate} of Appendix \ref{sec:bayes-factor-twotheories}. 

In the case of high residual SNR, we can express Eq.~\eqref{eq:bayes-factor-bgr-pn} more simply and provide a z-score for how much the theory deviates. If we take the limit that $\rho_\perp \gg 1$, the difference between the chi-squared random variables can be expressed in terms of unit normal Gaussian random variable $z\sim\mathcal{N}(0,1)$
\begin{equation}\label{eq:bayes-factor-bgr-pn-limit}
    \log \left.\mathcal{B}^\bGR_\ppe\right|_{s_\bGR } \approx \mathcal{C} + \frac{\left( \rho^\mathrm{diff}_\perp \right)^2}{2} + \sqrt{1+\mathcal{O}^2 - 2 \mathcal{O} \sqrt{1-\mathcal{O}^2}}\rho_\perp z \,,
\end{equation}
where the normal approximation was used for the chi-squared variables and $\mathcal{C}$ is the constant coming from prior terms of Eq.~\eqref{eq:bayes-factor-bgr-pn}. One can see that the z-score for the terms in the exponential is 
\begin{equation}
    Z = \frac{1-\mathcal{O}^2}{2 \sqrt{1+\mathcal{O}^2-2 \mathcal{O}\sqrt{1-\mathcal{O}^2} }}\rho_\perp \,,
\end{equation}
which characterizes the degree of preference for bGR vs ppE in the presence of noise. This example demonstrates that the smaller the overlap is, the easier it is to discern two signals while the significance scales linearly with the residual SNR.

\begin{figure}[htbp]
    \centering
    \includegraphics[width=1.\linewidth]{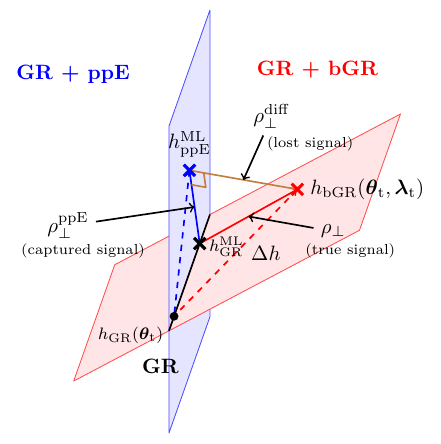}
    \caption{In this plot, we visually show how the ppE tests of GR can capture generic bGR deviations. The GR manifold (black) is a line that is at the intersection between the true bGR manifold (red) and the ppE manifold (blue). One can see that the perpendicular part of the signal from GR is $\rho_\perp = \Vert \Delta h_\perp\Vert$ which is the residual SNR after allowing the GR parameters to be biased $h_\GR^\ml = h_\GR(\theta_\t + \Delta\theta_\bias)$. One can see that the best fit ppE parameter is located at the blue mark and has residual SNR $\rho_\perp^\ppe = \mathcal{O} \rho_\perp$ as given in Eq.~\eqref{eq:captured-residual}. Finally, the brown line is the missed signal from our ppE model. One can see that the best fit for the ppE model corresponds to a value such that the residual (brown line) is perpendicular to the ppE manifold. With this picture in mind, we can explore how well tests of GR capture generic deviations.}
    \label{fig:intersection-tgr}
\end{figure}

In Fig.~\ref{fig:intersection-tgr}, we illustrate the geometric properties of the Bayes factor derivations that we have done. The GR waveform manifold is represented by the black line between the GR + ppE and GR + bGR manifolds (blue and red, respectively). The bGR signal (red x) is injected at the true parameter values $\left( \bf\theta_\t, \bf\lambda_\t \right)$. We also show the value of the GR waveform at $h_\GR(\bf\theta_\t)$ at true parameters and also its value which happens at the maximum likelihood point $h_\GR(\bf\theta_\ml)$ as computed in Eq.~\eqref{eq:gr-ml-value} where we ignore the statistical noise. We can see that the maximum likelihood point in the GR manifold minimizes the value of $\rho_\perp$.

We also show the best fit value of the waveform if our recovery model is GR + ppE. This will peak at the location $h_\ppe^\ml = h_\GR^\ml + \Delta h_\ppe^\perp$ which is the point which minimizes the distance from injected bGR signal $h_\bGR(\bf\theta_\t,\bf\lambda_\t)$. The distance between the best fit value of $h_\ppe(\bf\theta_\ml,\bf\mu_\ml)$ and $h_\bGR(\bf\theta_\t,\bf\lambda_\t)$ is equal to the signal which was `lost' by the ppE model. The size of this is given by the length of the brown line and size $\rho^\mathrm{diff}_\perp$ in Fig.~\ref{fig:intersection-tgr}. In fact, as we showed in Eqs.~\eqref{eq:bayes-factor-bgr-pn} and \eqref{eq:bayes-factor-bgr-pn-limit}, the size of $\rho^\mathrm{diff}_\perp$ characterizes how well one can distinguish between our true bGR model and the ppE model. Also note that the brown line is perpendicular to the ppE manifold\footnote{Note that $\left( \Delta h^\perp_\bGR - \Delta h^\perp_\ppe | \Delta h^\perp_\ppe \right) = 0$ since it is a ML point. This can also be shown explicitly via $\Vert\Delta h^\perp_\ppe\Vert = \mathcal{O} \rho_\perp$ and $(\Delta h^\perp_\bGR | \Delta h^\perp_\ppe) = \mathcal{O}^2 \rho_\perp^2$.} as shown in Fig.~\ref{fig:intersection-tgr} which physically makes sense because the best fit point optimizes the location where the blue manifold is closest to the red signal injection. This \textit{geometrical picture shows that the systematic error of using parameterized tests is primarily characterized by the overlap} $\mathcal{O}$ between the ppE and bGR manifolds.

The fitting factor is a number that characterizes how good of a fit the waveform is, and is related to the residual SNR. The fitting factor is defined as 
\begin{equation}
    \mathrm{FF}_\mathcal{M}=\max_{\Delta \theta^i} \frac{\left(h_\mathcal{M}+\Delta h, h_\mathcal{M}+\Delta \theta^i \partial_i h\right)}{\left|h_\mathcal{M}+\Delta h\right| \cdot\left|h_\mathcal{M}+\Delta \theta^i \partial_i h\right|}\,,
\end{equation}
where $\mathcal{M}$ is the model. If $\mathcal{M} = \GR$ and we want to see how similar a bGR signal is, the fitting factor is given by \cite{Vallisneri:2012qq}
\begin{equation}
    1 - \mathrm{FF}_\GR = \frac{1}{2} \frac{\rho_\perp^2}{\Vert h_\GR \Vert^2} \, .
\end{equation}
For the case of a mismodeled bGR search with $\mathcal{M} = \ppe$, the fitting factor is now 
\begin{equation}
    1 - \mathrm{FF}_\ppe = \frac{1}{2} \frac{ \left( \rho^\mathrm{diff}_\perp \right)^2}{\Vert h_\GR \Vert^2} \, ,
\end{equation}
so the mismatch of mismodeling a bGR signal is small and depends directly proportionally to the overlap in Eq.~\eqref{eq:overlap-bgr-pn}.

\section{PPE parameters capture generic phase deviations}

In the previous section, we showed that the perpendicular signal is what dominates the test of GR. So far, we have laid the groundwork where we stressed the relationship between the statistical nature of the problem and how it relates to the intrinsic behavior of the frequency domain waveforms. We will now show what the residual of the ppE tests of GR look like in the frequency domain.

\begin{figure*}
    \centering
    \includegraphics{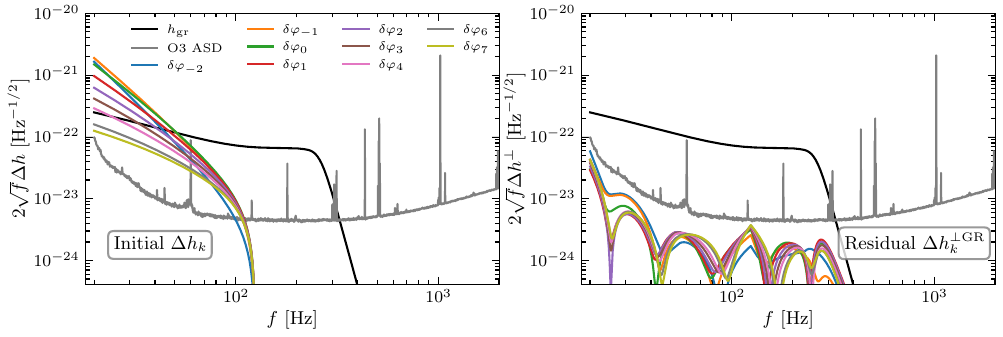}
    \caption{Residual amplitude for ppE injected deviations from GR for a GW150914-like detection. On the left and right, we include the GR waveform (black) and O3 Livingston ASD (gray). On the left, we show the waveform residuals that would be caused by injection what we show in Fig.~\ref{fig:dephasing-PN-150914} from parameterized tests. On the right, we show what the residual deviation is after the stealth biases in the GR parameters are accounted for. This plot is made by tuning the ppE coefficients so that their residuals $\Vert\Delta h_k^{\perp \GR}\Vert=1$ are of unit size. This figure illustrates that the primary observable in these parameterized tests is the magnitude of the residual deviation (right), which is markedly reduced compared to the nominal deviations (left). 
    \href{https://github.com/BrianCSeymour/waveform-geometry-testing-gr/blob/main/ppe-perpendicular-plots.ipynb}{\faFileCodeO}
    }
    \label{fig:amplitude-PN-150914}
\end{figure*}

We will use the ppE framework to add PN deviations to the GR waveform phase. We will use the normalizations that the LVK Collaboration uses \cite{LIGOScientific:2021sio}. During the inspiral, the GR phase is
\begin{align}
    \Psi_{\mathrm{PN}}(f) &=  2 \pi f t_{\mathrm{c}}-\varphi_{\mathrm{c}}-\frac{\pi}{4}  \\
    &+\frac{3}{128 \eta}(\pi \tilde{f})^{-5 / 3} \sum_{k=0}^7\left[\varphi_k+\varphi_{k l} \log (\pi \tilde{f})\right](\pi \tilde{f})^{k / 3} \,,
\end{align}
where $t_c,\varphi_c$ are the time and phase of coalescence at the geocenter, $\eta$ is the symmetric mass ratio, and $\tilde{f}= M(1+z) f$ is the dimensionless GW frequency. $M$ is the source frame mass while the detector frame total mass is $(1+z) M$ which accounts for how the detector frame mass changes with redshift $z$. The GR phasing coefficients are $\varphi_k$ and $\varphi_{k l}$ which depend on the GR parameters $\bf\theta$ (see \cite{Mishra:2010tp} for explicit expressions). In this work, we will focus on the nonlog terms and look for dephasing that occurs as 
\begin{equation}
    \Delta \Psi_k(f) = \frac{3}{128 \eta} \delta\varphi_k \varphi_k  (\pi \tilde{f})^{(k-5) / 3} \, ,
\end{equation}
where $\delta\varphi_k$ is the $k$th order fractional deviation to the GR phase. The goal is to see how well this parameterized template can capture the effects of a generic bGR deviation that is parameterized like 
\begin{equation}
    \Delta \Psi_\bGR(f) = \lambda \psi_\lambda(f) \, .
\end{equation}
Both the parameterized test and true bGR waveform are assumed to be phase deviations in the frequency domain so the waveforms are 
\begin{equation}\label{eq:dh-tgr-exponential-phase}
    h(f) = h_\GR(f) e^{i \Delta\Psi(f)} \, ,
\end{equation}
so the nominal waveform discrepancy is 
\begin{equation}\label{eq:dh-tgr-nominal}
    \Delta h(f) \approx i \Delta \Psi(f) h_\GR(f;\bf\theta_\t) \, ,
\end{equation}
because the phase deviation to the waveform is small. In fact, using Eq.~\eqref{eq:dh-tgr-nominal} as the definition for $\Delta h$ is more numerically stable than Eq.~\eqref{eq:dh-tgr-exponential-phase} as is discussed in Sec.~II.D of Ref.~\cite{Cutler:2007mi} and in Appendix \ref{app:bias-equation-numerical-stability} of this paper.

\begin{figure*}
    \centering
    \includegraphics{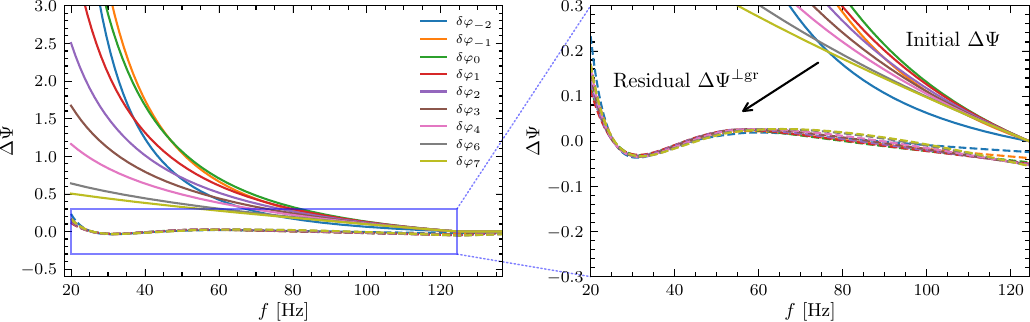}
    \caption{This figure shows how the ppE behave after GR parameter projection.
    On the left, we show the ppE dephasing for a GW150914-like detection and with $\delta\varphi_k$ normalized equivalently to Fig.~\ref{fig:amplitude-PN-150914}. On the right, we show the residual (perpendicular) phase deviation in the injection after the GR biases are taken into account. 
    The total dephasing of the perpendicular waveforms is noticeably smaller with the residual phase deviations $\Delta \Psi^{\perp \GR}$ displaying multiple zero crossings, akin to the behavior of an oscillatory polynomial. The near-identical appearance of the ppE dephasing residuals on the right visually demonstrates why leading-order deviations from GR appear similar to one another.
    \href{https://github.com/BrianCSeymour/waveform-geometry-testing-gr/blob/main/ppe-perpendicular-plots.ipynb}{\faFileCodeO}
    }
    \label{fig:dephasing-PN-150914}
\end{figure*}

We can visualize how the parameterized test coefficients affect the waveform by plotting the nominal and residual deviations. On the left side of Fig.~\ref{fig:amplitude-PN-150914}, we plot nominal $\Delta h$ as shown from Eq.~\eqref{eq:dh-tgr-nominal} using GR parameters given by GW150914. We normalize the value of $\delta \varphi_k$ so that the residual SNR has unit size $\Vert \Delta h_k^\perp\Vert=1$. As we have shown throughout this paper, it is the perpendicular component of the waveform that really influences observation. On the right side of the plot we show the residual component of $\Delta h_\perp^k$. One can see that the amount of residual signal in the right panel is significantly reduced compared to the original signal in the left panel---i.e.~$\Delta h_k^{\perp \GR} \ll \Delta h_k$. Additionally, the signals $\Delta h_k^{\perp\GR}$ are overlapping and look very similar to one another as can be seen on the right. This illustrates that \textit{there is a similarity between each perpendicular ppE test}. This occurs because the GR parameters are biased---notably the masses $(M,\eta)$ and spins $(\chi_\mathrm{eff},\chi_\mathrm{p})$ compensate for this deviation. Eliminating the biases in the GR parameters leads to a marked suppression of the residuals compared to the nominal deviation $\Delta h_k$ for a GW150914-like injection.

Figure \ref{fig:amplitude-PN-150914} is difficult to interpret because we are used to seeing the phase plotted instead of the amplitude. If we instead look for the perpendicular phase, we can see how biased GR parameters change the effective power law behavior of the ppE tests. We can read off the phase in the following fashion
\begin{equation}\label{eq:deltapsi-perp}
    \Delta h_\perp = i \Delta \Psi_\perp(f) h_\GR(f) \, ,
\end{equation}
where $\Delta \Psi_\perp(f)$ is the phase deviation after removing the GR biases. Note that in general, there could be a second term $\mathcal{O}(\Delta A\cdot h_\GR(f))$ if we were adding additional amplitude corrections. In practice since our perturbations from GR are purely phase deviations, this has a negligible effect. In Fig.~\ref{fig:dephasing-PN-150914}, we show the nominal phase deviations $\Delta \Psi_k$ (left) and the perpendicular phase deviations $\Delta \Psi_k^\perp$ (right). One can see that the power law behavior of the $\Delta \Psi \propto f^\alpha$  does not exist in the residuals because GR parameters are biased so it has an alternating pattern on the right.

\subsection{Overlap between different ppE orders}\label{sec:overlap-ppe}

In Sec.~\ref{sec:bayes-tgr}, we studied how good a parameterized template is at capturing the bGR waveform deviation and its relation to the Bayes factor. One can compute the overlap between two different ppE tests as given in Eq.~\eqref{eq:overlap-bgr-pn}.

\begin{figure}[htbp]
    \centering
    \includegraphics{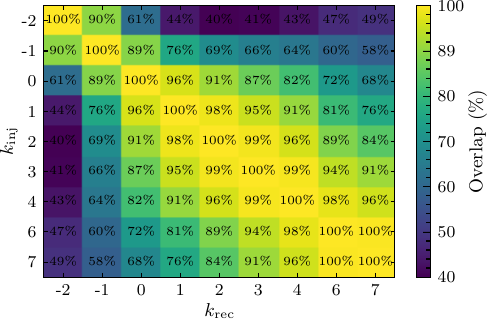}
    \caption{The overlap between deviations injected (y-axis) and the recovery model (x-axis). We use GW150914 event parameters with three detectors at O3 Livingston sensitivity, all extrinsic parameters measured and intrinsic parameters $(\mathcal{M}_c,q,\chi_\mathrm{eff},\chi_\mathrm{p})$. 
    One can see that injecting a deviation at $k/2$-th PN order fractionally from GR is perfectly captured on the diagonal but has less overlap as the injection and recovery order grow. This means that if there is a true deviation, the significance drops slowly as you search for the wrong PN order $k_\mathrm{rec}\ne k_\mathrm{inj}$ because the intrinsic parameters will capture the difference. 
    Note that we do not include $k=5$ in either of these plots because it is nearly completely degenerate with phase of coalescence.
    \href{https://github.com/BrianCSeymour/waveform-geometry-testing-gr/blob/main/overlap-plots.ipynb}{\faFileCodeO}}
    \label{fig:overlap-ppe-tests-spin}
\end{figure}

We compute the overlap between different ppE tests of GR as a simple way to see how important mismodeling is for constraints on bGR theories. In Fig.~\ref{fig:overlap-ppe-tests-spin}, we compute the overlap between $\Delta h_{k_\mathrm{inj}}^\perp$ injected signal residual and the $\Delta h_{k_\mathrm{rec}}^\perp$ which is the perpendicular template. We compute this plot using $\bf\theta = \left(\bf\theta_\mathrm{ext}, \bf\theta_\mathrm{int}, \delta\varphi_k \right)$. $\bf\theta_\mathrm{ext}$ is the extrinsic parameters
\begin{equation}
    \bf\theta_\mathrm{ext}= \left(d_L, \mathrm{ra},\mathrm{dec}, \iota, \psi, t_c, \phi_c \right)\, ,
\end{equation}
which are the luminosity distance, right ascension, declination, inclination, polarization, time of coalescence, and phase of coalescence, respectively. The intrinsic parameters are 
\begin{equation}\label{eq:intrinsic-par-list}
    \bf\theta_\mathrm{int} = \left(\mathcal{M}_c, \eta, \chi_\mathrm{eff},\chi_\mathrm{p} \right)\, ,
\end{equation}
which are chirp mass, symmetric mass ratio, effective inspiral spin parameter, and effective precession spin parameter, respectively. If the binary has total mass $M=m_1 + m_2$, the symmetric mass ratio is $\eta = m_1 m_2 / M^2$ and the chirp mass is $\mathcal{M}_c = \eta^{3/5}M$. The effective spin is $\chi_\mathrm{eff}\equiv(m_1 \chi_{1z} + m_2 \chi_{2z})/M$ where $ \chi_{i z}$ is the z-component of the spin of body $i$ \cite{Ajith:2009bn}, and effective precession spin $\chi_\mathrm{p}$ captures the in-plane spin dynamics as defined in Ref.~\cite{Schmidt:2014iyl}. We used GW150914 event parameters but use Hanford, Livingston, and Virgo network operating at O3 Livingston sensitivity. We compute the residual waveform via Eq.~\eqref{eq:biasgeo} for both the injected and recovery models and then plot the overlap in Fig.~\ref{fig:overlap-ppe-tests-spin}.  One can see that if $k_\mathrm{rec} = k_\mathrm{inj}$, the overlap is 1 so there is no bias since we are searching with the same model as we are recovering with. In contrast, if we had a true deviation at $k_\mathrm{inj} = -2$ and tried to recover with $k_\mathrm{rec} = 7$, we would miss out in about $50\%$ of the residual SNR. Note that the values of this plot strongly depend on the values of $\bf\theta_\mathrm{int}$. If we had picked a chirp mass which was in the binary neutron star (BNS) range $\mathcal{M}_c \sim 1.2 M_\odot$, this would change significantly. This case is further discussed in Appendix \ref{app:overlap-bns} where we show the overlap plots.

\begin{figure}[htbp]
    \centering
    \includegraphics{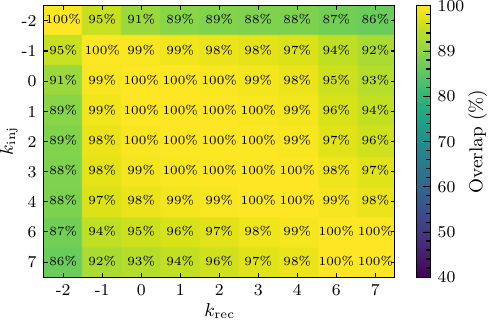}
    \caption{The overlap between deviations injected (y-axis) and the recovery model (x-axis). We use GW150914 event parameters with three detectors at O3 Livingston sensitivity, all extrinsic parameters measured and intrinsic parameters $(\mathcal{M}_c,q)$. This can be directly compared to Fig.~\ref{fig:overlap-ppe-tests-spin} which measures the parameters $(\chi_\mathrm{eff},\chi_\mathrm{p})$ additionally. One can see that ignoring the parameter uncertainty of the spin parameters means that the overlap is higher for these waveforms since the spins cannot bias the waveforms.
    \href{https://github.com/BrianCSeymour/waveform-geometry-testing-gr/blob/main/overlap-plots.ipynb}{\faFileCodeO}}
    \label{fig:overlap-ppe-tests-nospin}
\end{figure}

We also stress that the result of this overlap computation depends on which GR parameters are measured. If we take our intrinsic parameters to be just the masses
\begin{equation}
    \bf\theta_\mathrm{int}= \left(\mathcal{M}_c, \eta \right)\, ,
\end{equation} 
then the size of the overlaps computed would be modified from what is shown Fig.~\ref{fig:overlap-ppe-tests-spin}. In Fig.~\ref{fig:overlap-ppe-tests-nospin}, we show the result of this computation where we only allow the masses to be biased. You can see that including fewer GR parameters means that the overlaps between mismodeled ppE waveforms are much higher. The parameterized tests thus appear \textit{fractionally} more similar when excluding spin biases---in the sense that the overlap is larger for different ppE tests. However note that the residual SNR $\rho^\perp$ is higher if you have fewer total GR parameters since the spins cannot compensate to reduce the residual GW strain.

The results of what we see in Fig.~\ref{fig:overlap-ppe-tests-spin} and Fig.~\ref{fig:overlap-ppe-tests-nospin} highlights why it is difficult to do multiparameter tests of GR with the ppE parameters. Since their overlaps are so high, this means that the parameter combinations are highly degenerate with one another. This means that the errors are highly correlated and inflated when the Fisher matrix is inverted. In a later part of this manuscript Sec.~\ref{sec:svd}, we will address how to distinguish the independent information in a data-driven way.

\subsection{Capturing nonviolent nonlocality with ppE}

In a previous paper \cite{Seymour:2024kcd}, we studied the effects of nonviolent nonlocality on the waveform for stellar mass BBH systems. We found that the ppE parameters were pretty successful at capturing nonviolent nonlocality deviations from GR when stacking across many events. This result is not very surprising in light of what we have presented about the overlap between different ppE tests in Sec.~\ref{sec:overlap-ppe} which for a GW150914-like signal the ppE tests have overlaps of at least $40\%$. In Ref.~\cite{Seymour:2024kcd}, we measured the stochastic effects of nonviolent nonlocality hierarchically over a large number of events and the constraint is about a factor of $30\%$ better when we use the optimal waveform compared to just using the parameterized tests. In this section, we will explain why this is successful.

If we examine only the principal component of the stochastic deviation, one can show that nonviolent nonlocality has phase deviation that occurs at order 
\begin{equation}\label{eq:nvnl-phase-deviation}
    \Delta \Psi_\mathrm{NVNL}(f) \sim \left( \pi M f \right)^\alpha e^{-\frac{1}{8} \left( \pi M f \right)^{-4/3}}
\end{equation}
where $\alpha$ is some generic power parameter\footnote{This expression is implicitly in $\Delta \Psi(f)$ in Eq.~(35) and (36) of \cite{Seymour:2024kcd}. Solving Eq.~(35) would result in an equation which is some power in $f$ times $f_{\ell m}(r(f))$ where $f_{\ell m}$ given in Eq.~(2) of \cite{Seymour:2024kcd}.}. Notice however that Eq.~\eqref{eq:nvnl-phase-deviation} power series has an essential singularity as $Mf \rightarrow 0 $, so we cannot identify the ``leading-order'' ppE term. The exponential in Eq.~\eqref{eq:nvnl-phase-deviation} turns on too rapidly to identify such a leading order expression.

\begin{figure}[htbp]
    \centering
    \includegraphics{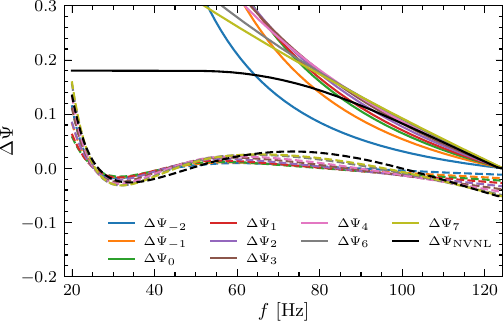}
    \caption{The ppE phase deviations (colored) and the nonviolent nonlocality deviation (black) for both nominal $\Delta \Psi(f)$ (solid) and the perpendicular signal $\Delta \Psi^{\perp\GR}(f)$ (dashed).
    While the nominal dephasing differs significantly at low frequencies between the ppE templates and the injected nonviolent nonlocality deviation, this difference can be compensated by adjusting intrinsic parameters such as masses and spins.
    As a result, the perpendicular dephasing becomes very similar across cases. This plot demonstrates the flexibility of the ppE templates to capture dephasing patterns that are otherwise quite distinct, owing to the bias framework developed in this paper.
    \href{https://github.com/BrianCSeymour/waveform-geometry-testing-gr/blob/main/ppe-nvnl-overlap.ipynb}{\faFileCodeO}}
    \label{fig:fig7-nvnl-fitting-illustration-150914}
\end{figure}

While Eq.~\eqref{eq:nvnl-phase-deviation} is manifestly not of the form of a power law in $f$ as prescribed by the parameterized tests, we can use the bias framework developed in this paper to see how well nonviolent nonlocality deviations can be captured by parameterized tests. In Fig.~\ref{fig:fig7-nvnl-fitting-illustration-150914}, we show the nominal deviations for a GW150914-like event normalizing the waveform residual to one. One can see that while the nominal phase deviations between ppE (colored solid) are not very similar in shape to the nonviolent nonlocality phase deviation (black solid), the residual phase deviations (dashed) overlap and agree to a much larger extent. While nonviolent nonlocality exhibits nonpower-law behavior, the perpendicular phase deviations (dashed black) still shows a remarkable similarity to the ppE perpendicular deviations [$\Delta\Psi_\perp(f)$ as defined in Eq.~\eqref{eq:deltapsi-perp}]. For the ppE dephasing, we plot the best fit values of $\delta \varphi_k$ for an injected nonviolent nonlocality deviation.

\section{SVD approach}\label{sec:svd}

As we have seen in the previous sections, the ppE waveform deviations look relatively similar after removing the effects of GR biases. One can see that the right side of Fig.~\ref{fig:dephasing-PN-150914} has some properties that look similar across each PN order. The consequence of this is that attempting to measure multiple $\delta\varphi_k$ at once would make the covariance blow up and thus it is difficult to perform multiparameter tests of GR. Due to the allure of measuring multiple ppE parameters simultaneously, there has been a large amount of work in this area with the singular value decomposition (SVD) approach \cite{Pai:2012mv, Arun:2013bp} and with the similar but distinct principal component analysis approach \cite{Ohme:2013nsa, Saleem:2021nsb,Datta:2022izc,Datta:2023muk,Ma:2024kkz,Gupta:2020lxa,Seymour:2024kcd}. In this section, we will propose an improved SVD test of GR that builds upon the groundwork of the original Pai results \cite{Pai:2012mv}.

In this work, we want to find the common features of $\Delta h_k^\perp$ where each $k$ is a different PN order deviation from GR. The SVD finds the best projector \cite{Tiglio:2021ysj} such that 
\begin{equation}\label{eq:svd-mineq}
    C(v_\alpha) =\sum_{k}\left\|\Delta h_k^\perp-P_n \Delta h_k^\perp\right\|^2\,,
\end{equation}
where we assume that $P_n$ is an orthogonal projector onto the SVD basis and $\Vert \cdot \Vert$ is a norm. The orthogonal projector is equal to
\begin{equation}
    P_n h_a = \sum_{\alpha=1}^n\left( v_\alpha |h_a\right) v_\alpha \,,
\end{equation}
and $v_\alpha$ are the SVD basis elements. These elements are orthonormal for the inner product 
\begin{equation}\label{eq:svd-orthogonality}
    \left( v_\alpha|v_{\beta} \right) = \delta_{\alpha\beta}\,,
\end{equation}
We will demonstrate that the SVD described by Pai \textit{et al.}~\cite{Pai:2012mv} is equivalent to minimizing Eq.~\eqref{eq:svd-mineq}. Thus, \textit{the SVD corresponds to finding an orthogonal basis that best projects all input waveforms}.

Let us lay the groundwork for the SVD by noting properties of the noise-weighted inner product. In this section, we describe the general procedure to describe how to \textit{pack} a frequency domain signal which has the symmetry $h(-f) = - h^*(f)$ corresponding to a real time domain signal. Note that the noise-weighted inner product is equal to 
\begin{equation}\label{eq:continuous-inner-prod}
    \left( h_1|h_2 \right) = 4 \Re \int_{0}^\infty df \frac{h_1^*(f) h_2(f)}{S(f)}\,,
\end{equation}
If we break it up into finite differences, the noise-weighted inner product is 
\begin{equation}\label{eq:discrete-inner-prod}
    \left( h_1|h_2 \right) = \sum_{i=0}^{N_f-1} \frac{h_1^*(f_i) h_2(f_i) + h_1(f_i) h_2^*(f_i)}{2} w^2(f_i) \, ,
\end{equation}
where the weight is defined via 
\begin{equation}
    w(f_i) = \frac{2 \sqrt{\Delta f_i}}{\sqrt{S(f_i)}} \, ,
\end{equation}
where the frequency spacing is $\Delta f_i = f_{i+1}-f_i$ and $S(f_i)$ is the value of the PSD at frequency bin $i$. Note that the discrete inner product in Eq.~\eqref{eq:discrete-inner-prod} does not have the form of the Hermitian inner product that we know and expect. This is just due to the fact that we simplified the original inner product integral in Eq.~\eqref{eq:continuous-inner-prod} using the fact that the time domain signals are real. Therefore, we need to construct an inner product where it is properly packed so that it is of the form that the SVD theorem applies\footnote{Note that we could have done this by including negative frequencies and doing the SVD on this case which does have a Hermitian inner product $(a|b) \sim a^\dagger b$, but this would need extra transformations to read out the real form of the singular vectors.}. Let us introduce the packing operator $\mathcal{P}$ as
\begin{equation}\label{eq:packing-simple-def}
    \mathcal{P}\left[ \Delta h \right] =
    \begin{bmatrix}
        \Re\,\Delta h(f_0)\, w(f_0) \\
        \vdots \\
        \Re\,\Delta h(f_{N_f-1})\, w(f_{N_f-1}) \\[6pt]
        \Im\,\Delta h(f_0)\, w(f_0) \\
        \vdots \\
        \Im\,\Delta h(f_{N_f-1})\, w(f_{N_f-1})
    \end{bmatrix} \in \mathbb{R}^{2N_f}
\end{equation}
which is just a vector of length $2 N_f$ which consists of the real and imaginary parts of $\Delta h_k$ times the weight. This works because if we choose $h_{1,2} = a_{1,2} + i b_{1,2}$, then 
\begin{align}\label{eq:packing-requirement-real}
    \left( h_1|h_2 \right) &=\sum_{i=0}^{N_f-1} \left( a_1^i a_2^i + b_1^i b_2^i \right) w^2(f_i)\,,\nn\\
    &\equiv \mathcal{P}[h_1]^T \mathcal{P}[h_2] \,,
\end{align}
where the first equality is via Eq.~\eqref{eq:discrete-inner-prod} and second is via the definition of the packing operator in Eq.~\eqref{eq:packing-simple-def}. With these definitions, the packing operator now explicitly has the Euclidean inner product form. Thus, we define the data matrix $\bf H$ via 
\begin{equation}\label{eq:H-pn-svd}
    H_{kI} \equiv \mathcal{P}[\Delta h_k^\perp ]\,,
\end{equation}
where $0\leq I<2 N_f$ indexes packed frequency and $k$ indexes PN order. Note that since we are working in the perpendicular basis, the covariance matrix is decoupled from the GR parameters.

\begin{figure*}
    \centering
    \includegraphics[width=\linewidth]{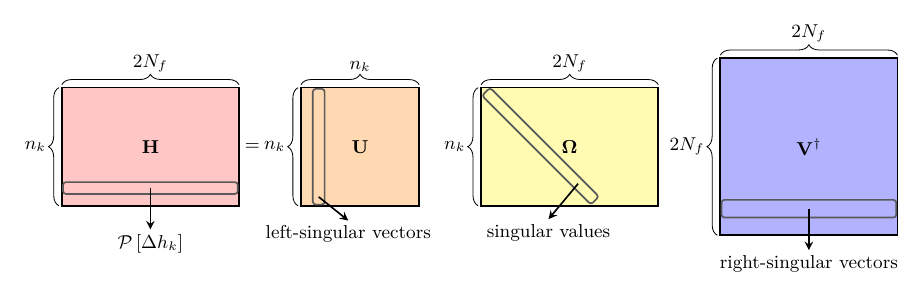}
    \caption{We show the visualization of the SVD operation. For our choice of the SVD, we use $H_{ik} = \mathcal{P}\left[ \Delta h_k^\perp \right]$ as given in Eqs.~\eqref{eq:H-pn-svd} and \eqref{eq:svd-modified-gravity}. The matrix $\bf U$ is a square matrix that represents linear redefinitions of $\delta\varphi_k$ while the matrix $\bf V$ is linear combinations of $\Delta h_k^\perp(f_i) w(f_i)$ which diagonalize $\bf H$. Finally, the matrix $\bf\Sigma$ is ordered by singular value that ranks the importance of each principal mode. This figure is written when we assume that we are using a single detector, but in the multiple detector case, the number of $\bf H$ columns would be increased as described in Eq.~\eqref{eq:multidetector-packing}. We use the standard SVD algorithm on the matrix $\bf H$ to get the singular vectors $v_\alpha(f)$ from $\bf V^\dagger$ rows.
    }
    \label{fig:svd-matrix-tikz-visualization}
\end{figure*}

Since we use the ``snapshots'' of the data matrix (appropriately packing them together), the data matrix is
\begin{equation}
    \bf{H}=\left[\dots, \Delta h_{k}^\perp(f) w(f), \dots\right] \in \mathbb{C}^{n_k \times 2N_f}\,,
\end{equation}
where $n_k$ is the number of ppE waveform deviations and $N_f$ is the number of frequency bins. This snapshot matrix $\bf H$ can be best captured by projectors from the SVD formalism \cite{Pai:2012mv,Tiglio:2021ysj}
\begin{equation}\label{eq:cpx-svd}
    \bf H = \bf U \bf \Omega \bf V^\dagger \,,
\end{equation}
where $\bf U$ and $\bf V^\dagger$ are left and right unitary square matrices of $n_k$ and $2 N_f$ dimensions, respectively. $\bf \Omega$ is a diagonal matrix consisting of the singular values of $\bf H$ in descending order. The SVD can be equivalently represented as 
\begin{equation}
    \bf H = \sum_\alpha^{n_k} \omega_\alpha \bf u_\alpha \bf v^*_\alpha \,,
\end{equation}
since the rank of the matrix is the number of ppE waveform deviations. The $\bf u_\alpha$ are the left-singular vectors and the $\bf v^*_\alpha$ are the right-singular vectors. This is shown in Fig.~\ref{fig:svd-matrix-tikz-visualization}, which has the $\bf H$ matrix in red, and the left-singular and right-singular vectors denoted.

Let us also give some comments about how to best use the SVD with modified gravity waveforms. In our work, we have been generically assuming that there are only phase deviations. If there are amplitude and phase deviations (e.g.~in a true bGR waveform), the deviations are of the form
\begin{equation}
    \Delta h^\perp(f_i) = \left( \Delta A^\perp + i \Delta\Psi^\perp(f_i) A(f_i) \right)e^{i \Psi(f_i)}\,.
\end{equation}
It is actually more useful to note that the inner product has invariance corresponding to $(h_1|h_2) = (h_1(f) e^{- i \Psi(f)}|h_2(f)e^{- i \Psi(f)})$. Therefore, one can define the packing operator as 
\begin{equation}\label{eq:svd-modified-gravity}
    \mathcal{P}\left[ \Delta h^\perp \right] = 
    \begin{bmatrix}
        \vdots\\
        \Delta A^\perp(f_i) w(f_i) \\
        \vdots \\
        \vdots \\
        \Delta\Psi^\perp(f_i) A(f_i) w(f_i)\\
        \vdots 
    \end{bmatrix} \,,
\end{equation}
where we explicitly removed the GR phase. Notice that our definition for Eq.~\eqref{eq:svd-modified-gravity} is a good representation of the GW inner product because it is real and well-behaved as stipulated in Eq.~\eqref{eq:packing-requirement-real}. Since we have been focusing on ppE phase deviations corrections to GR, the first column actually reduces to zero for this paper, but we give the generic formalism here. Additionally, the case of multiple detectors proceeds straightforwardly. 
\begin{equation}
    \sum_{\mathrm{det}\, A } \left( a|b \right)_A = \sum_{\mathrm{det}\, A } \sum_{i=0}^{N_f-1}   a_A^*(f_i) b_A(f_i) w_A^2(f_i) \, ,
\end{equation}
where the weights $w_A(f_i)$ depend on the individual detector sensitivity. Therefore, the packing operator for a collection of gravitational wave detectors is just
\begin{equation}\label{eq:multidetector-packing}
    \mathcal{P}\left[ \Delta h^\perp \right] \equiv 
    \begin{bmatrix}
        \vdots \\
        \mathcal{P}_{\mathrm{det} \,A}\left[ \Delta h^\perp_{\mathrm{det}\, A} \right] \\
        \vdots 
    \end{bmatrix} \in \mathbb{R}^{2 N_f \, N_\mathrm{det}}\,,
\end{equation}
where this formalism now directly generalizes to this case. This also generalizes trivially to detectors of different bandwidths (e.g. LISA + LIGO) where the frequency bins are not the same in each detector.

\begin{figure*}
    \centering
    \includegraphics[width=0.49\linewidth]{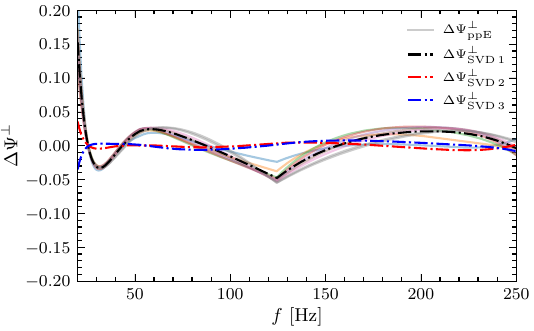}
    \includegraphics[width=0.49\linewidth]{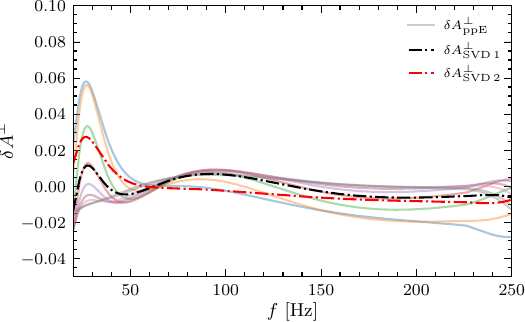}
    \caption{Demonstration of the SVD on different ppE waveform deviations. In transparent solid, different $\Delta \Psi_k^\perp$ are shown and the dashed lines are the SVD modes for GW150914-like event. Note that we normalize each $\Delta \Psi_k^\perp$ so that $\Vert \Delta h_k^\perp\Vert = 1$ so each signal is equally weighted in the SVD. One can see that the dashed black line is the dominant SVD term that best fits each waveform. Additionally the red and blue curves represent subdominant SVD dephasing that are orthogonal to each other. One can see that each of these ppE waveform deviations is captured by only a few SVD modes, which explains how degeneracies would form when using the original parameter space.
    \href{https://github.com/BrianCSeymour/waveform-geometry-testing-gr/blob/main/ppe-perpendicular-plots.ipynb}{\faFileCodeO}}
    \label{fig:svd-plot-example}
\end{figure*}

We have introduced the SVD to identify the common features in the ppE templates. We are specializing this analysis to the case of using all perpendicular ppE waveform templates. $\bf U$ is a unitary matrix that converts between the ppE template parameters $\delta\varphi_k$ to the SVD basis $\delta\varphi_\alpha$. If one looks at the rows of the matrix $\bf V$ and adds the real part of the first half plus imaginary part of the second half, we would have 
\begin{equation}\label{eq:vector-svd-amplitude-phase}
    v_\alpha(f_i) =\left( \Delta A^\perp(f_i)  + i \Delta \Psi_\alpha^\perp(f_i) A(f_i) \right)e^{i \Psi(f_i)} w(f_i)\,,
\end{equation}
which allows unique identification of the dominant parts of the SVD as $\Delta \Psi_\alpha^\perp(f_i)$. Note that $\Delta A^\perp(f_i) = 0$ since we originally started with only phase deviations. Therefore we can easily read off the SVD phase deviation $\Delta \Psi_\alpha^\perp(f_i)$ by looking at the singular vector. Note that when amplitude deviations are set to zero, Eq.~\eqref{eq:vector-svd-amplitude-phase} is the same from of what Pai \textit{et al.}~\cite{Pai:2012mv} found. This proves that the SVD operation in \cite{Pai:2012mv} is equivalent to minimizing Eq.~\eqref{eq:svd-mineq}.

Let us now show what the SVD looks like for a particular signal. In Fig.~\ref{fig:svd-plot-example}, we took the example of GW150914 again and computed the SVD of $\Delta h_k^\perp$ between $k=-2,-1,\dots,4,6,7$. We normalized each of the perpendicular strains such that they are equally important ($\Vert \Delta h_k^\perp \Vert = 1$). On the left panel of Fig.~\ref{fig:svd-plot-example}, we plot the phase modulation $\Delta \Psi_k^\perp(f)$ in solid, semitransparent lines. We plot the three dominant SVD lines with dot dashed (black, red, blue). You can see that black SVD tracks each line very well while the blue and red ones are orthogonal to it. This demonstrates that there is a main principal feature of the ppE tests that exists in all of them. We plot the three dominant SVD modes because they account for 99.9\% of the total variance; consequently, truncating the expansion at these three eigenvectors results in an information loss of only 0.1\%. Additionally, on the right side of Fig.~\ref{fig:svd-plot-example}, the fractional amplitude modulations $\delta A_k^\perp \equiv \Delta A_k / A_\GR$ are shown in solid, semitransparent lines while the dominant SVD modes are dot dashed for the SVD vectors. Note that $\delta A_k = 0$ originally---\textit{since we only considered phase modulations to the parameterized tests}. However $\delta A_k^\perp\ne 0$ because the GR parameter biases\footnote{Notably precession parameters such as $\chi_p$ are mixing phase and amplitude in the perpendicular waveform.} shift some the phase modulation in $\Delta h_k$ into \textit{a mixture of phase and amplitude modulation in} $\Delta h_k^\perp$. This figure directly demonstrates why we saw such high overlaps between each of the ppE tests in Fig.~\ref{fig:overlap-ppe-tests-spin}. 

\begin{figure*}
    \centering
    \includegraphics{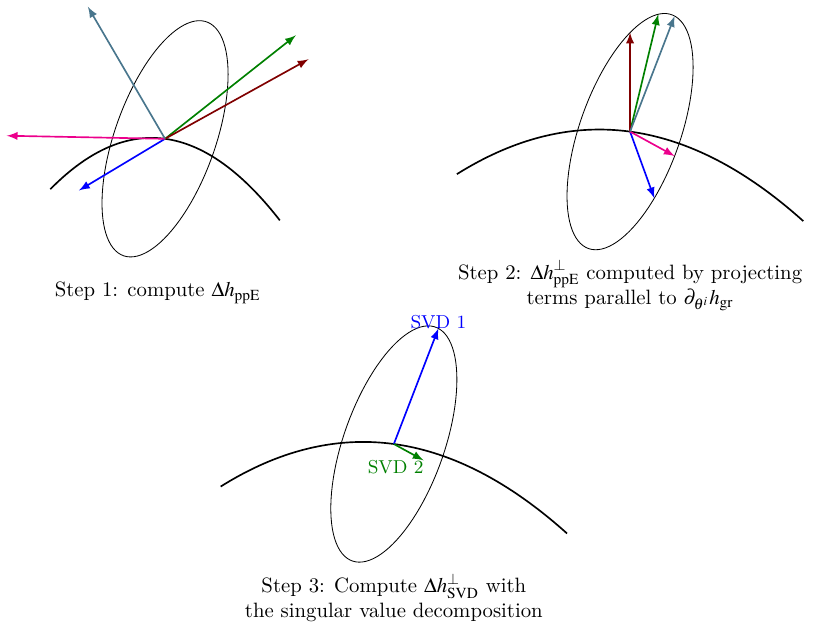}
    \caption{Construction of SVD multiparameter tests. We start by orthogonalizing the ppE deviations from GR. Next, we construct the SVD basis by choosing the directions that are best fit by looking at the perpendicular ppE components. One can see that the SVD 1 in the third step is perpendicular to SVD 2, but \textit{the length is shorter} because most of the SVD vectors point upwards. This figure highlights our intuition that the natural basis is to do the SVD on $\Delta h_k^\perp$ and not on the original $\Delta h_k$. By construction, the SVD vectors lie in the subspace that is perpendicular to GR. }
    \label{fig:svd-pn-demo}
\end{figure*}

In Fig.~\ref{fig:svd-pn-demo}, we visually demonstrate how our proposal for using the SVD to identify nondegenerate directions for the parameterized tests of GR. One can see that originally, each $\Delta h_k$ (colored arrow) lies off the GR manifold (black curved line). In step 2, the parameterized tests are projected to be perpendicular to GR manifold while being unit length. In step 3, the singular value decomposition examines these collection of parameterized test templates and identifies the orthogonal basis which is weighted by importance.

\section{Conclusion} \label{sec:conclusion}
In this work, we have shown that there is a relationship between searches for beyond GR morphologies and the geometric description of noise in parameter estimation. We showed that the ppE parameters are able to capture generic beyond GR deviations due to the bias mechanism described above. We also showed that these parameterized tests should do a good job in real data analysis and there is not too big of a loss by having systematic issues in the beyond GR template. We demonstrated that the techniques for multiparameter tests of GR suffer from degeneracies, so this means measuring multiple ones simultaneously is difficult. Finally, we proposed a new template for searching for deviations from GR using the SVD. The SVD introduced in this paper generalizes previous approaches by allowing for amplitude + phase modulations and multiple detectors.

In this work, we have provided a geometric explanation for why parameterized tests are unexpectedly effective at capturing smooth, monotonic dephasing away from GR. It remains to be seen whether there are some dephasing terms that could not be captured by the parameterized tests. The approaches that are developed in this paper for geometrically understanding how similar small effects such as GR violation are to one another can be applied to other situations that GW astronomy is in. For example, this could help understand when it is possible to disentangle the observational signatures of spin precession, orbital eccentricity, waveform-model systematics, non-Gaussian instrumental glitches, and genuine deviations from GR in a more categorical manner. Many of the most informative measurements in GW astronomy hinge on resolving such subtle effects, and the geometric framework developed here provides a principled way to assess their detectability.

\section*{ACKNOWLEDGMENTS}
We thank Michele Vallisneri, Ethan Payne, Taylor Knapp, Max Isi, Daiki Watarai, Bangalore Sathyaprakash, and Katerina Chatziioannou for their insightful discussions. B.S. acknowledges support by the National Science Foundation Graduate Research Fellowship under Grant No. DGE-1745301. Y.C. and B.S. acknowledge support from the Brinson Foundation, the Simons Foundation (Award No.~568762), and by NSF Grants No.~PHY-2309211 and No.~PHY-2309231. J.G. would like to gratefully acknowledge the support from the National Science Foundation through the Grant NSF No.~PHY-2207758. We also thank the anonymous referee for their feedback on this work.

\section*{DATA AVAILABILITY}
The data that support the findings of this article are openly 
available for Zenodo \cite{seymour_2026_waveform_geometry} and Github \href{https://github.com/BrianCSeymour/waveform-geometry-testing-gr}{\faGithub}.

\appendix

\section{Comment on numerical stability}\label{app:bias-equation-numerical-stability}
We stated in Eq.~\eqref{eq:lambdadh} that the best way to parameterize the bias equation is in terms of 
\begin{equation}
    \Delta h_\mathrm{raw} \equiv h_\s(\bf \theta_\t, \bf \lambda_\t) - \hm(\bf\theta_\t) \,,
\end{equation}
and then ``raw'' bias to the GR parameters is equal to 
\begin{equation}\label{eq:rawdeltah}
    \left( \Delta \theta_\bias^i \right)_\mathrm{raw} = \Sigma^{ij}\left( \partial_j \hm |\Delta h_\mathrm{raw} \right) \,,
\end{equation}
In fact, as first derived by Cutler-Vallisneri, the expression for Eq.~\eqref{eq:rawdeltah} has a limited range of validity. In fact, it is more numerically stable to use the following expression for $\Delta h$ as 
\begin{equation}
    \Delta h_\mathrm{stab} \equiv \left[ \Delta A + i \Delta \Psi A_\s \right] e^{i \Psi_\s}
\end{equation}
then the ``stabilized'' bias equation is
\begin{equation}\label{eq:stabdeltah}
    \left( \Delta \theta_\bias^i \right)_\mathrm{stab} = \Sigma^{ij}\left( \partial_j \hm |\Delta h_\mathrm{stab} \right) \,,
\end{equation}
where $h_s(\bf\theta_\t,\bf\lambda_\t) \equiv A_\s e^{i\Psi_\s}$, 
\begin{align}
    \Delta A &\equiv A_\s(\bf\theta_\t,\bf\lambda_\t) - A_\m(\bf\theta_\t) \,,\\
    \Delta \Psi &\equiv \Psi_\s(\bf\theta_\t,\bf\lambda_\t) - \Psi_\m(\bf\theta_\t) \,.
\end{align}
This is derived in their paper, but one can show that the error for these expressions is 
\begin{equation}
    \text{error of }\left( \Delta \theta_\bias^\mathrm{raw} \right)_\mathrm{error} \sim \Sigma^{ij} \left( \partial_j \hm | \Delta \Psi^2 h_\s \right) 
\end{equation}
Let us define alternative expressions for amplitude and phase deviation which are the difference between true and ML point
\begin{align}
    \delta A &\equiv A_\s(\bf\theta_\t,\bf\lambda_\t) - A_\m(\bf\theta_\ml) \,, \\
    \delta \Psi &\equiv \Psi_\s(\bf\theta_\t,\bf\lambda_\t) - \Psi_\m(\bf\theta_\ml) \,.
\end{align}
In fact, note that $\delta \Psi$ is small compared to $\Delta \Psi$ so expanding \textit{in these has a much greater regime of validity}. One can show that 
\begin{equation}
    \text{error of }\left( \Delta \theta_\bias^i \right)_\mathrm{stab} \sim \Sigma^{ij} \left( \partial_j \hm | \delta \Psi^2 h_\s \right) \,,
\end{equation}
which is valid over a much larger part of the parameter space.

Note that the discussion above admits a simple geometric interpretation. Using the ``raw'' expression for the bias equation in Eq.~\eqref{eq:rawdeltah} effectively assumes that the model waveform never departs from the true waveform by more than $\sim 1$ radian in phase, $\Delta\Psi \lesssim 1$, which corresponds schematically to requiring $\|\Delta h\| \lesssim 1$.  However, what actually controls the validity of the expansion is not the total model error, but the \emph{residual} error after refitting the GR parameters. In other words, it is sufficient that the ML phase only deviates from the true phase by less than a radian, $\delta\Psi \lesssim 1$, which corresponds to $\|\Delta h_\perp\| \lesssim 1$, where $\Delta h_\perp$ is the component of the model error orthogonal to the GR tangent space $\{\partial_i h_{\mathrm m}\}$.

Therefore, the regime of validity of the expressions in this paper is set by the smallness of the residual mismatch, $\|\Delta h_\perp\|$, rather than the total mismatch, $\|\Delta h\|$.  By working with the ``stabilized'' expression in Eq.~\eqref{eq:stabdeltah}, the bias formula remains accurate even in situations with $\|\Delta h\| \gtrsim 1$, i.e.\ in the presence of substantial stealth bias \cite{Vallisneri:2013rc}, as long as the residual component $\Delta h_\perp$ remains small.

\section{Bias formula for multiple \texorpdfstring{$\mu^a$}{mu\^a} parameters}\label{app:multibias}
This can generically be proven in the following way. Assume $\bf \mu = \mu^a$ and $\bf \lambda = \lambda^A$. Then the estimate for $\mu^a$ is 
\begin{equation}
    \Gamma_{I J}\mu_\bias^J = \left( \partial_I \hm | \Delta h \right) \, .
\end{equation}
Let us now write this equation as a matrix equation in block form 
\begin{equation}
    \begin{pmatrix}
        \Gamma_{ij} & \Gamma_{ib} \\
        \Gamma_{ai} & \Gamma_{ab}
    \end{pmatrix} 
    \begin{pmatrix}
        \Delta\theta^i_\bias \\
        \Delta\mu^b_\bias
    \end{pmatrix}= 
    \begin{pmatrix}
        \left( \partial_i \hm | \Delta h \right) \\
        \left( \partial_a \hm | \Delta h \right)
    \end{pmatrix}\, .
\end{equation}
Using properties of the Schur complement, we can solve for $\Delta\mu_\bias^a$ where the degeneracy with $\bf\theta$ is removed
\begin{equation}
    \left[ \Gamma_{ab}-\Gamma_{ai}\left( \Gamma_{ij} \right)^{-1}\Gamma_{jb} \right]\Delta \mu_\bias^a = \left( \partial_a \hm | \Delta h \right) - \Gamma_{ai}\left( \Gamma_{ij} \right)^{-1}\left( \partial_j \hm | \Delta h \right) \, .
\end{equation}
The part which is perpendicular to $\bf\theta$ is $\left( \partial_a \hm \right)^{\perp\bf\theta} = \partial_a \hm - \Gamma_{aj} \left( \Gamma_{ij} \right)^{-1} \partial_i \hm$ (this can be proven by a simple calculation). We also note that the reduced Fisher matrix is defined to be 
\begin{equation}
    \Gamma_{ab}^\text{red} = \left[ \Gamma_{ab}-\Gamma_{ai}\left( \Gamma_{ij} \right)^{-1}\Gamma_{jb} \right] = \left( \left( \partial_a \hm \right)^{\perp\bf\theta}|\left( \partial_b \hm \right)^{\perp\bf\theta} \right) \, .
\end{equation}
Thus the reduced Fisher matrix is the inner product of the derivatives of $\mu^a$ which have been made orthogonal to $\bf\theta$.
Therefore, the geometric value for the bias to $\bf\mu$ is 
\begin{equation}
    \Delta \mu_\bias^a = \Sigma^{ab}_{\text{red}} \left( \left( \partial_b \hm\right)^{\perp \bf\theta} | \left( \Delta h \right)^{\perp\bf\theta} \right)\,.
\end{equation}
This equation describes how biases can affect our measurement of small parameters. For example, a true deviation from GR $\Delta h$ at a particular PN order can find evidence of deviation at a different PN order. Additionally, it explains how eccentricity and precession can be confused with one another.

\section{Derivation of Bayes factors}\label{app:bayesfactor}

In this Appendix, we will compute the Bayes factor in two different cases. The first is if you have a bGR theory, and you know it \textit{a priori} and search with the correct waveform which was originally done in \cite{Vallisneri:2012qq}. The second case, we will generalize this equation to the case where we are searching for a violation with an incorrect model (for example a parameterized test of GR). Generally, the Bayes factor is a ratio of the evidences 
\begin{equation}
    \mathcal{B}_{\mathcal{M}_1}^{\mathcal{M}_2} \equiv \frac{p(d|\mathcal{M}_2)}{p(d|\mathcal{M}_1)} \,,
\end{equation}
where $d$ is the data and $\mathcal{M}_i$ are the models. The evidence is equal to the integral over all the parameters
\begin{equation}
    p(d|\mathcal{M}) = \int d^{n} \theta\, p(\theta|\mathcal{M}) p(d|\theta,\mathcal{M})\,,
\end{equation}
and $n$ is the number of parameters in the model.

\subsection{Bayes factor for a correctly modeled theory}

We will first derive the Bayes factor for the case of a beyond-GR (bGR) injection where we have perfectly modeled the bGR morphology. If we inject $s_\mathrm{bGR} = h_\mathrm{bGR}(\bf\theta_\t,\lambda_\t)$, in the Fisher matrix limit the likelihood is
\begin{equation}
    p\left(s_\mathrm{bGR} \mid \delta \theta^\mu\right)=\mathcal{N} e^{-|n|^2 / 2+\left(G^{-1}\right)^{\mu \nu}\left(n| h_\mu\right)\left(n| h_\nu\right) / 2-G_{\mu \nu} \delta \theta^\mu \delta \theta^\nu / 2} \,,
\end{equation}
where $G_{\mu\nu} = \left( h_\mu | h_\nu \right)$ is the $(m+1)$ dimensional bGR Fisher matrix and Greek letters $(\mu,\nu)$ range over $(\theta^i,\lambda)$ parameters. The value of $\delta \theta^\mu$ is defined as
\begin{equation}
    \delta \theta^\mu = \left(G^{-1}\right)^{\mu\nu}\left(n| h_\nu\right) \,.
\end{equation}
If we assume flat priors, the evidence is 
\begin{equation}
    \begin{aligned}
        &p\left(s_\mathrm{bGR} \mid \mathrm{bGR}\right)=   \int p\left(\theta^\mu\mid \mathrm{bGR}\right) p\left(\delta \theta^\mu\mid s_\mathrm{bGR}  \right) \,,\\
        &=  \frac{(2 \pi)^{(m+1) / 2} \sqrt{\left|G^{-1}\right|}}{\prod_\mu \Delta \theta_{\mathrm{prior}}^\mu}  \times \mathcal{N} e^{-|n|^2 / 2+\left(G^{-1}\right)^{\mu \nu}\left(n| h_\mu\right)\left(n| h_\nu\right) / 2}\,.
    \end{aligned}
\end{equation}
Next, we will compute the evidence for GR when we have beyond GR morphology 
\begin{align}\label{eq:gr-evidence}
    p\left(s_\mathrm{bGR} \mid \GR\right)= & \frac{(2 \pi)^{m / 2} \sqrt{\left|F^{-1}\right|}}{\prod_i \Delta \theta_{\text {prior }}^i} \nn\\
    & \times \mathcal{N} e^{-|n+\Delta h|^2 / 2+\left(F^{-1}\right)^{i j}\left(n+\Delta h| h_i\right)\left(n+\Delta h| h_j\right) / 2}\,,
\end{align}
where $F_{ij}$ is the $(m)$ dimensional GR Fisher matrix and $(i,j)$ range of $\theta^i$. After further simplification, the GR evidence can be compactly written as
\begin{align}
    p\left(s_\mathrm{bGR} \mid \GR\right)= & \frac{(2 \pi)^{m / 2} \sqrt{\left|F^{-1}\right|}}{\prod_i \Delta \theta_{\text {prior }}^i} \nn\\
    & \times \mathcal{N} e^{-|n|^2/2-|\Delta h_{\perp \GR}|^2/2- x |\Delta h_{\perp\GR}|  +(F^{-1})^{ij}n_i n_j} \,,
\end{align}
where $\Delta h = h_\mathrm{bGR}(\bf\theta_\t,\lambda_\t) - h_\GR(\bf\theta_\t)$. Therefore the Bayes factor is equal to 
\begin{equation}\label{eq:or-bgr-gr}
    \mathcal{B}^\mathrm{bGR}_\GR= \frac{(2 \pi)^{1 / 2} \Delta \lambda_\mathrm{est}}{\Delta \lambda_{\mathrm{prior}}^{\mathrm{bGR}}} e^{|\Delta h_{\perp \GR}|^2 /2+x |\Delta h_{\perp\GR}|+x^2/2 }\,,
\end{equation}
where $x=\left(\Delta h_{\perp}, n\right) /\left|\Delta h_{\perp}\right|$ is a normal random variable and the standard deviation on $\lambda$ from the observation is equal to
\begin{align}
    \Delta \lambda_\mathrm{est} &\equiv \sqrt{\left|F\right|/\left|G\right|}\,, \nn \\
    &= \frac{1}{\Vert \left( \partial_\lambda h_\mathrm{bGR} \right)^{\perp\GR}\Vert}\,,
\end{align}
where $\left( \partial_\lambda h_\mathrm{bGR} \right)^{\perp\GR}$ is the component of the derivative that is perpendicular to GR. This equation matches the results of \cite{Vallisneri:2012qq} in Eq.~(11). One can see that the prefactor $\Delta \lambda_\mathrm{est}/\Delta \lambda_{\mathrm{prior}}^{\mathrm{bGR}}$ outside Eq.~\eqref{eq:or-bgr-gr} is an Occam factor that disfavors more complicated models while the evidence for beyond-GR morphology comes from the term $\Delta h_{\perp \GR}$ in the exponential.

\subsection{Bayes factor for a mismodeled theory}

Let us now turn to the case that we are using a ppE model to search for beyond GR morphology which has mismodeled it. The GR evidence is the same as in Eq.~\eqref{eq:gr-evidence} while the evidence for ppE is equal to 
\begin{align}
    p\left(s_\GR \mid \ppe\right)&=  \frac{(2 \pi)^{(m+1) / 2} \sqrt{\left|H^{-1}\right|}}{\prod_\alpha \Delta \theta_{\text {prior }}^\alpha} \nn\\
    & \times \mathcal{N} e^{-|n|^2/2-|\Delta h_{\perp \ppe}|^2/2- x |\Delta h_{\perp\ppe}|  +(H^{-1})^{\alpha\beta}n_\alpha n_\beta} \,,
\end{align}
where $H^{\alpha\beta}$  is the $(m+1)$ dimensional ppE Fisher matrix and $\left( \alpha,\beta \right)$ range over $(\theta^i,\mu)$. We are using notation $\Delta h_{\perp\ppe}$ to mean perpendicular to both GR and $\partial_\mu h$. The odds ratio between ppE and GR for a bGR injection is thus
\begin{widetext}\label{eq:pn-gr-bayesfactor-appendix}
    \begin{align}
        \left.\mathcal{B}^{\ppe}_\mathrm{GR}\right|_{s_\bGR}&=\frac{ (2 \pi)^{1 / 2} \Delta \mu_\mathrm{est}}{\Delta \mu_{\mathrm{prior}}^{\ppe}} \exp \Big[ -|\Delta h_{\perp \ppe}|^2/2- y |\Delta h_{\perp\ppe}|  +y^2 /2+ |\Delta h_{\perp \GR}|^2/2+x |\Delta h_{\perp\GR}| \Big]\,, \nn\\
        &=  \frac{(2 \pi)^{1 / 2}  \Delta \mu_\mathrm{est}}{\Delta \mu_{\mathrm{prior}}^{\ppe}} \exp \Big[ \frac{1}{2} \left( \rho_\ppe^\perp \right)^2  - \rho_\ppe^\perp y + \frac{1}{2} y^2 \Big]\,,
    \end{align}
\end{widetext}
where covariance on the ppE parameter is equal to
\begin{align}
    \Delta \mu_\mathrm{est} &= \sqrt{\frac{\vert F\vert}{\vert H\vert}}\,, \nn\\
    &= \frac{1}{\Vert \left( \partial_\mu h_\ppe \right)^{\perp \GR}\Vert} \, .
\end{align}
The quantities $(x,y)$ are random unit variables
\begin{align}
    x&\equiv\left(\Delta h_{\perp \GR}| n\right) /\left|\Delta h_{\perp \GR}\right|\,,\nn \\
    y&\equiv\left(\Delta h_{\perp \ppe}| n\right) /\left|\Delta h_{\perp \ppe}\right|\,, \nn
\end{align}
note that $(x,y)$ have some nonvanishing correlation. The \textit{captured residual SNR} is 
\begin{equation}
    \rho_\perp^\ppe = \mathcal{O}(\Delta h_\ppe^{\perp\GR},\Delta h_\bGR^{\perp\GR}) \rho_\perp  \,,
\end{equation}
and the \textit{overlap} between the bGR and ppE waveforms is defined via 
\begin{equation}
    \mathcal{O} \equiv \frac{\left( \Delta h^{\perp \GR} | \partial_\mu h^{\perp \GR}\right)}{\Vert \Delta h^{\perp \GR}\Vert \Vert \partial_\mu h^{\perp \GR}\Vert} \,.
\end{equation}
In this derivation we used the block matrices to show that the random unit variables are related via 
\begin{equation}
    x \Vert\Delta h_{\perp\GR}\Vert = y \Vert\Delta h_{\perp\ppe}\Vert + \mathcal{O}(\Delta h_\ppe^{\perp\GR},\Delta h_\bGR^{\perp\GR}) (\Delta h_\ppe^{\perp \GR} | n) \,,
\end{equation}
which allowed simplification of the equation above. This can be written in a simpler form in the following way
\begin{equation}\label{eq:xyz-correlation}
    x = \sqrt{1-\mathcal{O}^2} y  + \mathcal{O} z
\end{equation}
this follows from $\Vert\Delta h_{\perp\ppe}\Vert = \sqrt{1-\mathcal{O}^2} \rho_\perp$ and defining $z \equiv (\Delta h_\ppe^{\perp \GR} | n) / \Vert \Delta h_\ppe^{\perp \GR}\Vert$. Looking at Fig.~\ref{fig:intersection-tgr}, we see that $\Delta h_\ppe^{\perp \GR}$ is the blue line, $\Delta h_\bGR^{\perp\GR}$ is the red line, and $\Delta h_{\perp\ppe}$ is the brown line. Note that $x,y,z$ are correlated with one another, one can show that $E\left[ xz \right] = \mathcal{O}$ directly by using $E[(a|n)(b|n)] = (a|b)$ identify for Gaussian noise. Using this property, $E[yz]=0$ and $E[xy]=\sqrt{1-\mathcal{O}^2}$. 

\subsection{Bayes factor between two beyond GR theories}\label{sec:bayes-factor-twotheories}

Finally, let us compute the Bayes factor that compares a beyond GR signal to that of a parameterized model as we did in Eq.~\eqref{eq:bayes-factor-bgr-pn}. This can be done using 
\begin{equation}
    \left.\mathcal{B}^\bGR_\ppe\right|_{s_\bGR } =  \left.\frac{\mathcal{B}^\GR_\bGR}{\mathcal{B}^\GR_\ppe}\right|_{s_\bGR }\,,
\end{equation}
since the Bayes factors are the ratio of the evidences. This results in the expression
\begin{equation}\label{eq:bgr-pn-bayes-appendix}
    \left.\mathcal{B}^\bGR_\ppe\right|_{s_\bGR } = \frac{\Delta \lambda_\mathrm{stat} \Delta \mu_\mathrm{prior}}{\Delta \lambda_\mathrm{prior}\Delta \mu_\mathrm{stat}} e^{(x+\rho_\perp)^2/2- (y+\mathcal{O}\rho_\perp)^2/2}  \,,
\end{equation}
where $x,y$ are correlated as discussed in Eq.~\eqref{eq:xyz-correlation} and the Bayes factors on the right side were derived in previous Eq.~\eqref{eq:or-bgr-gr} and Eq.~\eqref{eq:pn-gr-bayesfactor-appendix}. While we have written down an expression for the Bayes factor, in this section we will (a) perform a rotation so it can be written as sum of decorrelated unit normals and (b) perform a high SNR expansion which approximates the z-score for how big the bias is.

Let us start with the log Bayes factor and decorrelate it (excluding constant Occam factor terms of Eq.~\eqref{eq:bgr-pn-bayes-appendix}), which we write as
\begin{equation}
    \ell = \frac{1}{2} \rho_\perp^2 + \rho_\perp x + \frac{1}{2} x^2 - \left( \frac{1}{2} \mathcal{O}^2\rho_\perp^2 + \rho_\perp \mathcal{O} y + \frac{1}{2} y^2 \right)\,,
\end{equation}
where $\ell \equiv \left.\mathcal{B}^\bGR_\ppe\right|_{s_\bGR } + \text{priors}$. Since $x,y$ are correlated it is convenient to convert to $y,z$ as given in Eq.~\eqref{eq:xyz-correlation}. The quadratic form of the log Bayes factor can be thus written as 
\begin{equation}
    \ell = \frac{1}{2}\left( 1 - \mathcal{O}^2 \right)\rho_\perp^2 + b^T v + \frac{1}{2}v^T H v
\end{equation}
where $v = (y,z)$ and 
\begin{align}
    b &= \rho_\perp \begin{bmatrix}
        \sqrt{1-\mathcal{O}^2} - \mathcal{O} \\
        \mathcal{O}
    \end{bmatrix} \,,\\
    H &=\begin{bmatrix}
        - \mathcal{O}^2 & \mathcal{O} \sqrt{1-\mathcal{O}^2} \\
        \mathcal{O} \sqrt{1-\mathcal{O}^2} & \mathcal{O}^2
    \end{bmatrix} \,.
\end{align}
Since $H$ is a symmetric matrix, there exists a rotation matrix which diagonalizes $H$. One can show that the likelihood equals 
\begin{equation}\label{eq:chi-square-decorrelated}
    \ell = \frac{1}{2}\mathcal{O} \left[ \left( w_1 + \mu_1 \right)^2 - \left( w_2 + \mu_2 \right)^2 \right]\,,
\end{equation}
where $w_1,w_2$ are unit Gaussian random variables that are uncorrelated which are a rotated basis from $(y,z)$\footnote{Explicitly $(w_1,w_2) \equiv R \cdot (y,z)$.}. The rotation matrix which accomplishes this decorrelation and whitening operation is
\begin{equation}
    R = \begin{bmatrix}
        \sqrt{\frac{1-\mathcal{O}}{2}} & \sqrt{\frac{1+\mathcal{O}}{2}} \\
        -\sqrt{\frac{1+\mathcal{O}}{2}} & \sqrt{\frac{1-\mathcal{O}}{2}}
    \end{bmatrix}\,,
\end{equation}
and the parameters $\mu_1,\mu_2$ are defined via $\left( \mu_1, - \mu_2 \right) \equiv \frac{1}{\mathcal{O}} R\cdot b$ and are equal to
\begin{align}
    \mu_1 &\equiv  \frac{\sqrt{1+\mathcal{O}} - \mathcal{O} \sqrt{1-\mathcal{O}}}{\sqrt{2} \mathcal{O}} \,,\\
    \mu_2 &\equiv \frac{\sqrt{1-\mathcal{O}} - \mathcal{O} \sqrt{1+\mathcal{O}}}{\sqrt{2} \mathcal{O}} \,.
\end{align}
By doing this rotation and whitening operation, we have written this in a simple form. The results of Eq.~\eqref{eq:chi-square-decorrelated} demonstrate that \textit{the log Bayes factor can be written as a difference between two noncentral chi-square distributions} where the noncentrality parameters are $\lambda_1 = \mu_1^2$ and $\lambda_2 = \mu_2^2$. Therefore, we have derived the expression for the Bayes factor between two theories
\begin{equation}\label{eq:bgr-pn-bayes-appendix-white-decorrelate}
    \left.\mathcal{B}^\bGR_\ppe\right|_{s_\bGR } = \frac{\Delta \lambda_\mathrm{stat} \Delta \mu_\mathrm{prior}}{\Delta \lambda_\mathrm{prior}\Delta \mu_\mathrm{stat}} e^{(w_1+\mu_1)^2/2- (w_2+\mu_2)^2/2}  \,.
\end{equation}
Since the difference between two noncentral chi-square random variables is not a well-known distribution, this is the most simplified form for the Bayes factor here.

While Eq.~\eqref{eq:bgr-pn-bayes-appendix-white-decorrelate} depends on two random variables, we can write this as a simplified distribution if we assume that the residual SNR is large ($\rho_\perp\gg1$). If the residual SNR is large (and correspondingly $|\mu_1|,|\mu_2| \gg1$\footnote{Note that for $\mathcal{O}=1$, $\mu_1=-\mu_2=\rho_\perp$ while for smaller $\mathcal{O}$ this can be less valid as $\mu_2$ can get small and cross zero. Thus the validity of $\rho_\perp\gg1 \mathrel{{\ooalign{\hidewidth$\not\phantom{=}$\hidewidth\cr$\implies$}}} |\mu_2|\gg1$, but for most signals the overlap is large enough that this is a good heuristic.}) then the log Bayes factor can be written as 
\begin{equation}
    \ell \approx \frac{1}{2} \left( 1-\mathcal{O}^2 \right)\rho_\perp^2 + \mathcal{O}\left( \mu_1 w_1 - \mu_2 w_2 \right) \,.
\end{equation}
Note that we can define a new unit normal random variable 
\begin{equation}
    W \equiv \frac{\mu_1 w_1 - \mu_2 w_2}{\sqrt{\mu_1^2 + \mu_2^2}}\sim \mathcal{N}(0,1)\,,
\end{equation}
thus the log Bayes factor is distributed as
\begin{equation}
    \ell \approx \frac{1}{2} \left( 1-\mathcal{O}^2 \right)\rho_\perp^2 + \mathcal{O}\sqrt{\mu_1^2 + \mu_2^2} W
\end{equation}
This means the z-score is equal to 
\begin{align}\label{eq:z-score-approximation-app}
    Z &= \frac{\left( \rho_\perp^\mathrm{diff} \right)^2/2}{|\mathcal{O}|\sqrt{\mu_1^2 + \mu_2^2}}\,, \nn\\
    &= \frac{1-\mathcal{O}^2}{2 \sqrt{1+\mathcal{O}^2-2 \mathcal{O}\sqrt{1-\mathcal{O}^2} }}\rho_\perp\,.
\end{align}
One can see that the z-score is linearly proportional to the residual SNR, but also depends on the overlap between the ppE and beyond GR waveforms. Explicit evaluation of the prefactor yields an answer about $Z/\rho_\perp \sim 0.3-0.6$ when $\mathcal{O}\lesssim 0.7$ while for $\mathcal{O} \sim 1$ then $Z/\rho_\perp = 0$. This makes sense that the z-score is zero and there is no way to discern between these two theories for perfectly aligned theories (i.e.~the angle between the manifolds is zero in Fig.~\ref{fig:intersection-tgr}). Note that the expression in Eq.~\eqref{eq:z-score-approximation-app} is probably only exactly valid in the case of an exceedingly loud beyond GR signal ($\rho_\perp\gg1$), so in realistic scenarios one must use Eq.~\eqref{eq:bgr-pn-bayes-appendix-white-decorrelate} which contains two random variables.

\begin{figure*}[htbp]
    \centering
    \includegraphics{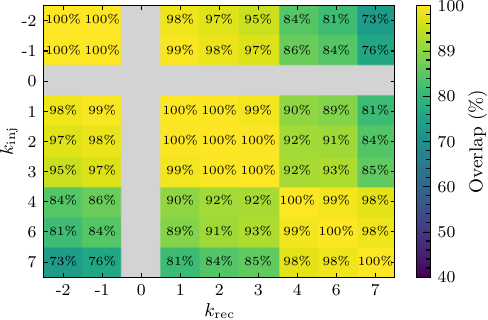}
    \includegraphics{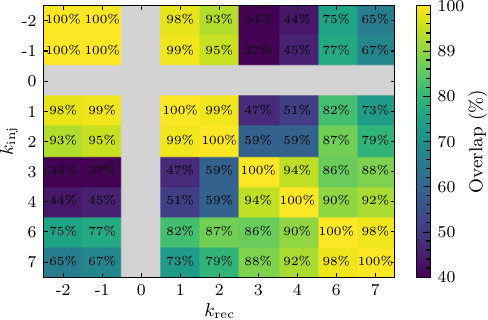}
    \caption{Residual amplitude for ppE injected deviations from GR for a GW170817-like detection. On the left is case of nonspinning inference while on the right is the spinning case. Compared to the BBH case given in Sec.~\ref{sec:overlap-ppe}, this figure highlights how the overlap depends on the choice of the GR parameters. One can see that the fact that $\chi_\mathrm{eff}$ enters the BNS waveform primarily at 1.5PN while $\chi_\mathrm{p}$ enters at 2PN causes the overlaps to reduce on the right side. Due to the fact that BNS systems do not have much merger phenomenology in current ground-based detectors, the $k=0$ term is highly degenerate with the chirp mass which is thus excluded from this plot.
    \href{https://github.com/BrianCSeymour/waveform-geometry-testing-gr/blob/main/overlap-plots.ipynb}{\faFileCodeO}}
    \label{fig:overlap-ppe-tests-bns}
\end{figure*}

\section{Multidetector geometry}

Let us use terminology that is from the polarization part of the LVK GWTC-3 paper \cite{LIGOScientific:2021sio}. A multidetector signal looks like 
\begin{equation}
    d^A(f) = F^{A \alpha} h_\alpha(f,\;\bf\theta_\t) + n^A(f)\,,
\end{equation}
where $A$ is the detector index and $\alpha=(+,\times)$ is the polarization index. The antenna responses are $F^{A\alpha} = D_A^{ij} \epsilon_{ij}^{\alpha}$ where $D_A^{ij}$ is the $A$th detector tensor and $\epsilon_{ij}^{\alpha}$ is the GW polarization tensor of the $\alpha$ polarization. If we introduce the square brackets to represent the sum over detectors
\begin{equation}
    \left[ a | b \right] = \sum_{A}\left(a_A | b_A  \right)_{A} \, ,
\end{equation}
then the maximum likelihood point is the solution to 
\begin{equation}
    \Gamma_{IJ} \Delta\Theta^J = \left[\partial_I \hm| \Delta h\right] + \left[\partial_I \hm|n\right]\,,
\end{equation}
where the network Fisher information matrix is $\Gamma_{IJ} \equiv \left[\partial_I\hm|\partial_J\hm\right]$. The equations for the statistical error and bias straightforwardly generalize from Eq.~\eqref{eq:statbias} and are equal to 
\begin{align}
    \Delta \Theta^I_\stat &= \Sigma^{IJ}\left[ \partial_I \hm | n \right] \, , \\
    \Delta \Theta_\bias^I &= \Sigma^{IJ}\left[ \partial_I \hm |\Delta h \right]\, .
\end{align}
Thus, the effect of searching for deviations $\Delta\mu^a$ in a waveform described by true parameters $\left(\theta^i_\t,\lambda^A_\t\right)$ is 
\begin{equation}
    \Delta \mu_\bias^a = \Sigma^{ab}_{\text{red}} \left[ \left( \partial_b \hm\right)^{\perp \bf\theta} | \left( \Delta h \right)^{\perp\bf\theta} \right]\,,
\end{equation}
where $\Sigma^{ab}_{\text{red}}$ is the reduced network covariance matrix and the sum is understood.

\section{Overlaps for GW170817}\label{app:overlap-bns}

As we discussed in the main part of the paper in Sec.~\ref{sec:overlap-ppe}, the results of how strong the overlaps are depends on the values of the intrinsic GR parameters for the event. In this Appendix, we will provide context for this by giving an example for a GW170817-like event. 

The left panel of Fig.~\ref{fig:overlap-ppe-tests-bns} shows the overlap between the ppE tests computed using only the masses as measured GR parameters, analogous to Fig.~\ref{fig:overlap-ppe-tests-nospin}. In contrast, the right panel allows $\chi_\mathrm{eff}$ and $\chi_\mathrm{p}$ to vary, making it analogous to Fig.~\ref{fig:overlap-ppe-tests-spin} in the main text. The long inspiral time of a BNS event changes how well the ppE tests can compensate for one another. One can see that generally, the overlaps between the theories are smaller than what we saw in Sec.~\ref{sec:overlap-ppe}. For both of these plots, we have not included the $k = 0$ case for the overlap computation, which is because this 0PN test is nearly perfectly correlated with the chirp mass in the Fisher matrix. This degeneracy arises because current ground-based GW detectors don't measure much of the merger phase for BNS systems, and has been highlighted in the literature before in App.~B of \cite{Payne:2023kwj}. One can also see that in the right panel of Fig.~\ref{fig:overlap-ppe-tests-bns}, the overlaps for the $k=3,4$---corresponding to 1.5PN and 2PN terms---are significantly reduced compared to the nonspinning case. This reduction occurs because $\chi_\mathrm{eff}$ and $\chi_\mathrm{p}$ first enter the waveform at these PN orders respectively.

\FloatBarrier 
\bibliography{bibliography.bib}

\begin{thebibliography}{126}%
\makeatletter
\providecommand \@ifxundefined [1]{%
 \@ifx{#1\undefined}
}%
\providecommand \@ifnum [1]{%
 \ifnum #1\expandafter \@firstoftwo
 \else \expandafter \@secondoftwo
 \fi
}%
\providecommand \@ifx [1]{%
 \ifx #1\expandafter \@firstoftwo
 \else \expandafter \@secondoftwo
 \fi
}%
\providecommand \natexlab [1]{#1}%
\providecommand \enquote  [1]{``#1''}%
\providecommand \bibnamefont  [1]{#1}%
\providecommand \bibfnamefont [1]{#1}%
\providecommand \citenamefont [1]{#1}%
\providecommand \href@noop [0]{\@secondoftwo}%
\providecommand \href [0]{\begingroup \@sanitize@url \@href}%
\providecommand \@href[1]{\@@startlink{#1}\@@href}%
\providecommand \@@href[1]{\endgroup#1\@@endlink}%
\providecommand \@sanitize@url [0]{\catcode `\\12\catcode `\$12\catcode `\&12\catcode `\#12\catcode `\^12\catcode `\_12\catcode `\%12\relax}%
\providecommand \@@startlink[1]{}%
\providecommand \@@endlink[0]{}%
\providecommand \url  [0]{\begingroup\@sanitize@url \@url }%
\providecommand \@url [1]{\endgroup\@href {#1}{\urlprefix }}%
\providecommand \urlprefix  [0]{URL }%
\providecommand \Eprint [0]{\href }%
\providecommand \doibase [0]{https://doi.org/}%
\providecommand \selectlanguage [0]{\@gobble}%
\providecommand \bibinfo  [0]{\@secondoftwo}%
\providecommand \bibfield  [0]{\@secondoftwo}%
\providecommand \translation [1]{[#1]}%
\providecommand \BibitemOpen [0]{}%
\providecommand \bibitemStop [0]{}%
\providecommand \bibitemNoStop [0]{.\EOS\space}%
\providecommand \EOS [0]{\spacefactor3000\relax}%
\providecommand \BibitemShut  [1]{\csname bibitem#1\endcsname}%
\let\auto@bib@innerbib\@empty
\bibitem [{\citenamefont {Abbott}\ \emph {et~al.}(2016{\natexlab{a}})\citenamefont {Abbott} \emph {et~al.}}]{LIGOScientific:2016aoc}%
  \BibitemOpen
  \bibfield  {author} {\bibinfo {author} {\bibfnamefont {B.~P.}\ \bibnamefont {Abbott}} \emph {et~al.} (\bibinfo {collaboration} {LIGO Scientific, Virgo}),\ }\bibfield  {title} {\bibinfo {title} {{Observation of Gravitational Waves from a Binary Black Hole Merger}},\ }\href {https://doi.org/10.1103/PhysRevLett.116.061102} {\bibfield  {journal} {\bibinfo  {journal} {Phys. Rev. Lett.}\ }\textbf {\bibinfo {volume} {116}},\ \bibinfo {pages} {061102} (\bibinfo {year} {2016}{\natexlab{a}})},\ \Eprint {https://arxiv.org/abs/1602.03837} {arXiv:1602.03837 [gr-qc]} \BibitemShut {NoStop}%
\bibitem [{\citenamefont {Abbott}\ \emph {et~al.}(2017)\citenamefont {Abbott} \emph {et~al.}}]{LIGOScientific:2017vwq}%
  \BibitemOpen
  \bibfield  {author} {\bibinfo {author} {\bibfnamefont {B.~P.}\ \bibnamefont {Abbott}} \emph {et~al.} (\bibinfo {collaboration} {LIGO Scientific, Virgo}),\ }\bibfield  {title} {\bibinfo {title} {{GW170817: Observation of Gravitational Waves from a Binary Neutron Star Inspiral}},\ }\href {https://doi.org/10.1103/PhysRevLett.119.161101} {\bibfield  {journal} {\bibinfo  {journal} {Phys. Rev. Lett.}\ }\textbf {\bibinfo {volume} {119}},\ \bibinfo {pages} {161101} (\bibinfo {year} {2017})},\ \Eprint {https://arxiv.org/abs/1710.05832} {arXiv:1710.05832 [gr-qc]} \BibitemShut {NoStop}%
\bibitem [{\citenamefont {Abbott}\ \emph {et~al.}(2019{\natexlab{a}})\citenamefont {Abbott} \emph {et~al.}}]{LIGOScientific:2018mvr}%
  \BibitemOpen
  \bibfield  {author} {\bibinfo {author} {\bibfnamefont {B.~P.}\ \bibnamefont {Abbott}} \emph {et~al.} (\bibinfo {collaboration} {LIGO Scientific, Virgo}),\ }\bibfield  {title} {\bibinfo {title} {{GWTC-1: A Gravitational-Wave Transient Catalog of Compact Binary Mergers Observed by LIGO and Virgo during the First and Second Observing Runs}},\ }\href {https://doi.org/10.1103/PhysRevX.9.031040} {\bibfield  {journal} {\bibinfo  {journal} {Phys. Rev. X}\ }\textbf {\bibinfo {volume} {9}},\ \bibinfo {pages} {031040} (\bibinfo {year} {2019}{\natexlab{a}})},\ \Eprint {https://arxiv.org/abs/1811.12907} {arXiv:1811.12907 [astro-ph.HE]} \BibitemShut {NoStop}%
\bibitem [{\citenamefont {Abbott}\ \emph {et~al.}(2021{\natexlab{a}})\citenamefont {Abbott} \emph {et~al.}}]{LIGOScientific:2020ibl}%
  \BibitemOpen
  \bibfield  {author} {\bibinfo {author} {\bibfnamefont {R.}~\bibnamefont {Abbott}} \emph {et~al.} (\bibinfo {collaboration} {LIGO Scientific, Virgo}),\ }\bibfield  {title} {\bibinfo {title} {{GWTC-2: Compact Binary Coalescences Observed by LIGO and Virgo During the First Half of the Third Observing Run}},\ }\href {https://doi.org/10.1103/PhysRevX.11.021053} {\bibfield  {journal} {\bibinfo  {journal} {Phys. Rev. X}\ }\textbf {\bibinfo {volume} {11}},\ \bibinfo {pages} {021053} (\bibinfo {year} {2021}{\natexlab{a}})},\ \Eprint {https://arxiv.org/abs/2010.14527} {arXiv:2010.14527 [gr-qc]} \BibitemShut {NoStop}%
\bibitem [{\citenamefont {Abbott}\ \emph {et~al.}(2023)\citenamefont {Abbott} \emph {et~al.}}]{LIGOScientific:2021djp}%
  \BibitemOpen
  \bibfield  {author} {\bibinfo {author} {\bibfnamefont {R.}~\bibnamefont {Abbott}} \emph {et~al.} (\bibinfo {collaboration} {KAGRA, VIRGO, LIGO Scientific}),\ }\bibfield  {title} {\bibinfo {title} {{GWTC-3: Compact Binary Coalescences Observed by LIGO and Virgo during the Second Part of the Third Observing Run}},\ }\href {https://doi.org/10.1103/PhysRevX.13.041039} {\bibfield  {journal} {\bibinfo  {journal} {Phys. Rev. X}\ }\textbf {\bibinfo {volume} {13}},\ \bibinfo {pages} {041039} (\bibinfo {year} {2023})},\ \Eprint {https://arxiv.org/abs/2111.03606} {arXiv:2111.03606 [gr-qc]} \BibitemShut {NoStop}%
\bibitem [{\citenamefont {Abbott}\ \emph {et~al.}(2016{\natexlab{b}})\citenamefont {Abbott} \emph {et~al.}}]{LIGOScientific:2016lio}%
  \BibitemOpen
  \bibfield  {author} {\bibinfo {author} {\bibfnamefont {B.~P.}\ \bibnamefont {Abbott}} \emph {et~al.} (\bibinfo {collaboration} {LIGO Scientific, Virgo}),\ }\bibfield  {title} {\bibinfo {title} {{Tests of general relativity with GW150914}},\ }\href {https://doi.org/10.1103/PhysRevLett.116.221101} {\bibfield  {journal} {\bibinfo  {journal} {Phys. Rev. Lett.}\ }\textbf {\bibinfo {volume} {116}},\ \bibinfo {pages} {221101} (\bibinfo {year} {2016}{\natexlab{b}})},\ \bibinfo {note} {[Erratum: Phys.Rev.Lett. 121, 129902 (2018)]},\ \Eprint {https://arxiv.org/abs/1602.03841} {arXiv:1602.03841 [gr-qc]} \BibitemShut {NoStop}%
\bibitem [{\citenamefont {Abbott}\ \emph {et~al.}(2019{\natexlab{b}})\citenamefont {Abbott} \emph {et~al.}}]{LIGOScientific:2018dkp}%
  \BibitemOpen
  \bibfield  {author} {\bibinfo {author} {\bibfnamefont {B.~P.}\ \bibnamefont {Abbott}} \emph {et~al.} (\bibinfo {collaboration} {LIGO Scientific, Virgo}),\ }\bibfield  {title} {\bibinfo {title} {{Tests of General Relativity with GW170817}},\ }\href {https://doi.org/10.1103/PhysRevLett.123.011102} {\bibfield  {journal} {\bibinfo  {journal} {Phys. Rev. Lett.}\ }\textbf {\bibinfo {volume} {123}},\ \bibinfo {pages} {011102} (\bibinfo {year} {2019}{\natexlab{b}})},\ \Eprint {https://arxiv.org/abs/1811.00364} {arXiv:1811.00364 [gr-qc]} \BibitemShut {NoStop}%
\bibitem [{\citenamefont {Abbott}\ \emph {et~al.}(2019{\natexlab{c}})\citenamefont {Abbott} \emph {et~al.}}]{LIGOScientific:2019fpa}%
  \BibitemOpen
  \bibfield  {author} {\bibinfo {author} {\bibfnamefont {B.~P.}\ \bibnamefont {Abbott}} \emph {et~al.} (\bibinfo {collaboration} {LIGO Scientific, Virgo}),\ }\bibfield  {title} {\bibinfo {title} {{Tests of General Relativity with the Binary Black Hole Signals from the LIGO-Virgo Catalog GWTC-1}},\ }\href {https://doi.org/10.1103/PhysRevD.100.104036} {\bibfield  {journal} {\bibinfo  {journal} {Phys. Rev. D}\ }\textbf {\bibinfo {volume} {100}},\ \bibinfo {pages} {104036} (\bibinfo {year} {2019}{\natexlab{c}})},\ \Eprint {https://arxiv.org/abs/1903.04467} {arXiv:1903.04467 [gr-qc]} \BibitemShut {NoStop}%
\bibitem [{\citenamefont {Abbott}\ \emph {et~al.}(2021{\natexlab{b}})\citenamefont {Abbott} \emph {et~al.}}]{LIGOScientific:2020tif}%
  \BibitemOpen
  \bibfield  {author} {\bibinfo {author} {\bibfnamefont {R.}~\bibnamefont {Abbott}} \emph {et~al.} (\bibinfo {collaboration} {LIGO Scientific, Virgo}),\ }\bibfield  {title} {\bibinfo {title} {{Tests of general relativity with binary black holes from the second LIGO-Virgo gravitational-wave transient catalog}},\ }\href {https://doi.org/10.1103/PhysRevD.103.122002} {\bibfield  {journal} {\bibinfo  {journal} {Phys. Rev. D}\ }\textbf {\bibinfo {volume} {103}},\ \bibinfo {pages} {122002} (\bibinfo {year} {2021}{\natexlab{b}})},\ \Eprint {https://arxiv.org/abs/2010.14529} {arXiv:2010.14529 [gr-qc]} \BibitemShut {NoStop}%
\bibitem [{\citenamefont {Abbott}\ \emph {et~al.}(2025)\citenamefont {Abbott} \emph {et~al.}}]{LIGOScientific:2021sio}%
  \BibitemOpen
  \bibfield  {author} {\bibinfo {author} {\bibfnamefont {R.}~\bibnamefont {Abbott}} \emph {et~al.} (\bibinfo {collaboration} {LIGO Scientific, VIRGO, KAGRA}),\ }\bibfield  {title} {\bibinfo {title} {{Tests of General Relativity with GWTC-3}},\ }\href {https://doi.org/10.1103/PhysRevD.112.084080} {\bibfield  {journal} {\bibinfo  {journal} {Phys. Rev. D}\ }\textbf {\bibinfo {volume} {112}},\ \bibinfo {pages} {084080} (\bibinfo {year} {2025})},\ \Eprint {https://arxiv.org/abs/2112.06861} {arXiv:2112.06861 [gr-qc]} \BibitemShut {NoStop}%
\bibitem [{\citenamefont {Yunes}\ \emph {et~al.}(2016)\citenamefont {Yunes}, \citenamefont {Yagi},\ and\ \citenamefont {Pretorius}}]{Yunes:2016jcc}%
  \BibitemOpen
  \bibfield  {author} {\bibinfo {author} {\bibfnamefont {N.}~\bibnamefont {Yunes}}, \bibinfo {author} {\bibfnamefont {K.}~\bibnamefont {Yagi}},\ and\ \bibinfo {author} {\bibfnamefont {F.}~\bibnamefont {Pretorius}},\ }\bibfield  {title} {\bibinfo {title} {{Theoretical Physics Implications of the Binary Black-Hole Mergers GW150914 and GW151226}},\ }\href {https://doi.org/10.1103/PhysRevD.94.084002} {\bibfield  {journal} {\bibinfo  {journal} {Phys. Rev. D}\ }\textbf {\bibinfo {volume} {94}},\ \bibinfo {pages} {084002} (\bibinfo {year} {2016})},\ \Eprint {https://arxiv.org/abs/1603.08955} {arXiv:1603.08955 [gr-qc]} \BibitemShut {NoStop}%
\bibitem [{\citenamefont {Nair}\ \emph {et~al.}(2019)\citenamefont {Nair}, \citenamefont {Perkins}, \citenamefont {Silva},\ and\ \citenamefont {Yunes}}]{Nair:2019iur}%
  \BibitemOpen
  \bibfield  {author} {\bibinfo {author} {\bibfnamefont {R.}~\bibnamefont {Nair}}, \bibinfo {author} {\bibfnamefont {S.}~\bibnamefont {Perkins}}, \bibinfo {author} {\bibfnamefont {H.~O.}\ \bibnamefont {Silva}},\ and\ \bibinfo {author} {\bibfnamefont {N.}~\bibnamefont {Yunes}},\ }\bibfield  {title} {\bibinfo {title} {{Fundamental Physics Implications for Higher-Curvature Theories from Binary Black Hole Signals in the LIGO-Virgo Catalog GWTC-1}},\ }\href {https://doi.org/10.1103/PhysRevLett.123.191101} {\bibfield  {journal} {\bibinfo  {journal} {Phys. Rev. Lett.}\ }\textbf {\bibinfo {volume} {123}},\ \bibinfo {pages} {191101} (\bibinfo {year} {2019})},\ \Eprint {https://arxiv.org/abs/1905.00870} {arXiv:1905.00870 [gr-qc]} \BibitemShut {NoStop}%
\bibitem [{\citenamefont {Silva}\ \emph {et~al.}(2021)\citenamefont {Silva}, \citenamefont {Holgado}, \citenamefont {C\'ardenas-Avenda\~no},\ and\ \citenamefont {Yunes}}]{Silva:2020acr}%
  \BibitemOpen
  \bibfield  {author} {\bibinfo {author} {\bibfnamefont {H.~O.}\ \bibnamefont {Silva}}, \bibinfo {author} {\bibfnamefont {A.~M.}\ \bibnamefont {Holgado}}, \bibinfo {author} {\bibfnamefont {A.}~\bibnamefont {C\'ardenas-Avenda\~no}},\ and\ \bibinfo {author} {\bibfnamefont {N.}~\bibnamefont {Yunes}},\ }\bibfield  {title} {\bibinfo {title} {{Astrophysical and theoretical physics implications from multimessenger neutron star observations}},\ }\href {https://doi.org/10.1103/PhysRevLett.126.181101} {\bibfield  {journal} {\bibinfo  {journal} {Phys. Rev. Lett.}\ }\textbf {\bibinfo {volume} {126}},\ \bibinfo {pages} {181101} (\bibinfo {year} {2021})},\ \Eprint {https://arxiv.org/abs/2004.01253} {arXiv:2004.01253 [gr-qc]} \BibitemShut {NoStop}%
\bibitem [{\citenamefont {Aasi}\ \emph {et~al.}(2015)\citenamefont {Aasi} \emph {et~al.}}]{LIGOScientific:2014pky}%
  \BibitemOpen
  \bibfield  {author} {\bibinfo {author} {\bibfnamefont {J.}~\bibnamefont {Aasi}} \emph {et~al.} (\bibinfo {collaboration} {LIGO Scientific}),\ }\bibfield  {title} {\bibinfo {title} {{Advanced LIGO}},\ }\href {https://doi.org/10.1088/0264-9381/32/7/074001} {\bibfield  {journal} {\bibinfo  {journal} {Class. Quant. Grav.}\ }\textbf {\bibinfo {volume} {32}},\ \bibinfo {pages} {074001} (\bibinfo {year} {2015})},\ \Eprint {https://arxiv.org/abs/1411.4547} {arXiv:1411.4547 [gr-qc]} \BibitemShut {NoStop}%
\bibitem [{\citenamefont {Acernese}\ \emph {et~al.}(2015)\citenamefont {Acernese} \emph {et~al.}}]{VIRGO:2014yos}%
  \BibitemOpen
  \bibfield  {author} {\bibinfo {author} {\bibfnamefont {F.}~\bibnamefont {Acernese}} \emph {et~al.} (\bibinfo {collaboration} {VIRGO}),\ }\bibfield  {title} {\bibinfo {title} {{Advanced Virgo: a second-generation interferometric gravitational wave detector}},\ }\href {https://doi.org/10.1088/0264-9381/32/2/024001} {\bibfield  {journal} {\bibinfo  {journal} {Class. Quant. Grav.}\ }\textbf {\bibinfo {volume} {32}},\ \bibinfo {pages} {024001} (\bibinfo {year} {2015})},\ \Eprint {https://arxiv.org/abs/1408.3978} {arXiv:1408.3978 [gr-qc]} \BibitemShut {NoStop}%
\bibitem [{\citenamefont {Akutsu}\ \emph {et~al.}(2021)\citenamefont {Akutsu} \emph {et~al.}}]{KAGRA:2020tym}%
  \BibitemOpen
  \bibfield  {author} {\bibinfo {author} {\bibfnamefont {T.}~\bibnamefont {Akutsu}} \emph {et~al.} (\bibinfo {collaboration} {KAGRA}),\ }\bibfield  {title} {\bibinfo {title} {{Overview of KAGRA: Detector design and construction history}},\ }\href {https://doi.org/10.1093/ptep/ptaa125} {\bibfield  {journal} {\bibinfo  {journal} {PTEP}\ }\textbf {\bibinfo {volume} {2021}},\ \bibinfo {pages} {05A101} (\bibinfo {year} {2021})},\ \Eprint {https://arxiv.org/abs/2005.05574} {arXiv:2005.05574 [physics.ins-det]} \BibitemShut {NoStop}%
\bibitem [{\citenamefont {Iyer}\ \emph {et~al.}(2013)\citenamefont {Iyer}, \citenamefont {Souradeep}, \citenamefont {Unnikrishnan}, \citenamefont {Dhurandhar}, \citenamefont {Raja},\ and\ \citenamefont {Sengupta}}]{iyer2013ligo}%
  \BibitemOpen
  \bibfield  {author} {\bibinfo {author} {\bibfnamefont {B.}~\bibnamefont {Iyer}}, \bibinfo {author} {\bibfnamefont {T.}~\bibnamefont {Souradeep}}, \bibinfo {author} {\bibfnamefont {C.}~\bibnamefont {Unnikrishnan}}, \bibinfo {author} {\bibfnamefont {S.}~\bibnamefont {Dhurandhar}}, \bibinfo {author} {\bibfnamefont {S.}~\bibnamefont {Raja}},\ and\ \bibinfo {author} {\bibfnamefont {A.}~\bibnamefont {Sengupta}},\ }\href@noop {} {\emph {\bibinfo {title} {LIGO-India: proposal of the consortium for Indian initiative in gravitational-wave observations (IndIGO)}}},\ \bibinfo {type} {Tech. Rep.}\ (\bibinfo  {institution} {LIGO Tech. Rep. LIGO-M1100296-v2 https://dcc. ligo. org/LIGO-M1100296-v2/public},\ \bibinfo {year} {2013})\BibitemShut {NoStop}%
\bibitem [{\citenamefont {Saleem}\ \emph {et~al.}(2022{\natexlab{a}})\citenamefont {Saleem} \emph {et~al.}}]{Saleem:2021iwi}%
  \BibitemOpen
  \bibfield  {author} {\bibinfo {author} {\bibfnamefont {M.}~\bibnamefont {Saleem}} \emph {et~al.},\ }\bibfield  {title} {\bibinfo {title} {{The science case for LIGO-India}},\ }\href {https://doi.org/10.1088/1361-6382/ac3b99} {\bibfield  {journal} {\bibinfo  {journal} {Class. Quant. Grav.}\ }\textbf {\bibinfo {volume} {39}},\ \bibinfo {pages} {025004} (\bibinfo {year} {2022}{\natexlab{a}})},\ \Eprint {https://arxiv.org/abs/2105.01716} {arXiv:2105.01716 [gr-qc]} \BibitemShut {NoStop}%
\bibitem [{\citenamefont {Fritschel}\ \emph {et~al.}(2022)\citenamefont {Fritschel} \emph {et~al.}}]{lsc2022report-asharp}%
  \BibitemOpen
  \bibfield  {author} {\bibinfo {author} {\bibfnamefont {P.}~\bibnamefont {Fritschel}} \emph {et~al.},\ }\bibfield  {title} {\bibinfo {title} {Report of the lsc post-o5 study group},\ }\href@noop {} {\bibfield  {journal} {\bibinfo  {journal} {LIGO Document: LIGO-T2200287}\ } (\bibinfo {year} {2022})}\BibitemShut {NoStop}%
\bibitem [{\citenamefont {Adhikari}\ \emph {et~al.}(2024)\citenamefont {Adhikari} \emph {et~al.}}]{ligo-voyager-dcc}%
  \BibitemOpen
  \bibfield  {author} {\bibinfo {author} {\bibfnamefont {R.~X.}\ \bibnamefont {Adhikari}} \emph {et~al.},\ }\bibfield  {title} {\bibinfo {title} {Ligo voyager upgrade concept},\ }\href {https://dcc.ligo.org/LIGO-T1400226} {\bibfield  {journal} {\bibinfo  {journal} {LIGO Document: LIGO-T1400226}\ } (\bibinfo {year} {2024})}\BibitemShut {NoStop}%
\bibitem [{\citenamefont {Adhikari}\ \emph {et~al.}(2020)\citenamefont {Adhikari} \emph {et~al.}}]{LIGO:2020xsf}%
  \BibitemOpen
  \bibfield  {author} {\bibinfo {author} {\bibfnamefont {R.~X.}\ \bibnamefont {Adhikari}} \emph {et~al.} (\bibinfo {collaboration} {LIGO}),\ }\bibfield  {title} {\bibinfo {title} {{A cryogenic silicon interferometer for gravitational-wave detection}},\ }\href {https://doi.org/10.1088/1361-6382/ab9143} {\bibfield  {journal} {\bibinfo  {journal} {Class. Quant. Grav.}\ }\textbf {\bibinfo {volume} {37}},\ \bibinfo {pages} {165003} (\bibinfo {year} {2020})},\ \Eprint {https://arxiv.org/abs/2001.11173} {arXiv:2001.11173 [astro-ph.IM]} \BibitemShut {NoStop}%
\bibitem [{\citenamefont {Reitze}\ \emph {et~al.}(2019)\citenamefont {Reitze} \emph {et~al.}}]{Reitze:2019iox}%
  \BibitemOpen
  \bibfield  {author} {\bibinfo {author} {\bibfnamefont {D.}~\bibnamefont {Reitze}} \emph {et~al.},\ }\bibfield  {title} {\bibinfo {title} {{Cosmic Explorer: The U.S. Contribution to Gravitational-Wave Astronomy beyond LIGO}},\ }\href@noop {} {\bibfield  {journal} {\bibinfo  {journal} {Bull. Am. Astron. Soc.}\ }\textbf {\bibinfo {volume} {51}},\ \bibinfo {pages} {035} (\bibinfo {year} {2019})},\ \Eprint {https://arxiv.org/abs/1907.04833} {arXiv:1907.04833 [astro-ph.IM]} \BibitemShut {NoStop}%
\bibitem [{\citenamefont {Evans}\ \emph {et~al.}(2021)\citenamefont {Evans} \emph {et~al.}}]{Evans:2021gyd}%
  \BibitemOpen
  \bibfield  {author} {\bibinfo {author} {\bibfnamefont {M.}~\bibnamefont {Evans}} \emph {et~al.},\ }\href@noop {} {\bibinfo {title} {{A Horizon Study for Cosmic Explorer: Science, Observatories, and Community}}} (\bibinfo {year} {2021}),\ \Eprint {https://arxiv.org/abs/2109.09882} {arXiv:2109.09882 [astro-ph.IM]} \BibitemShut {NoStop}%
\bibitem [{\citenamefont {Punturo}\ \emph {et~al.}(2010)\citenamefont {Punturo} \emph {et~al.}}]{Punturo:2010zz}%
  \BibitemOpen
  \bibfield  {author} {\bibinfo {author} {\bibfnamefont {M.}~\bibnamefont {Punturo}} \emph {et~al.},\ }\bibfield  {title} {\bibinfo {title} {{The Einstein Telescope: A third-generation gravitational wave observatory}},\ }\href {https://doi.org/10.1088/0264-9381/27/19/194002} {\bibfield  {journal} {\bibinfo  {journal} {Class. Quant. Grav.}\ }\textbf {\bibinfo {volume} {27}},\ \bibinfo {pages} {194002} (\bibinfo {year} {2010})}\BibitemShut {NoStop}%
\bibitem [{\citenamefont {Maggiore}\ \emph {et~al.}(2020)\citenamefont {Maggiore} \emph {et~al.}}]{ET:2019dnz}%
  \BibitemOpen
  \bibfield  {author} {\bibinfo {author} {\bibfnamefont {M.}~\bibnamefont {Maggiore}} \emph {et~al.} (\bibinfo {collaboration} {ET}),\ }\href {https://doi.org/10.1088/1475-7516/2020/03/050} {\bibinfo {title} {{Science Case for the Einstein Telescope}}} (\bibinfo {year} {2020}),\ \Eprint {https://arxiv.org/abs/1912.02622} {arXiv:1912.02622 [astro-ph.CO]} \BibitemShut {NoStop}%
\bibitem [{\citenamefont {Branchesi}\ \emph {et~al.}(2023)\citenamefont {Branchesi} \emph {et~al.}}]{Branchesi:2023mws}%
  \BibitemOpen
  \bibfield  {author} {\bibinfo {author} {\bibfnamefont {M.}~\bibnamefont {Branchesi}} \emph {et~al.},\ }\href {https://doi.org/10.1088/1475-7516/2023/07/068} {\bibinfo {title} {{Science with the Einstein Telescope: a comparison of different designs}}} (\bibinfo {year} {2023}),\ \Eprint {https://arxiv.org/abs/2303.15923} {arXiv:2303.15923 [gr-qc]} \BibitemShut {NoStop}%
\bibitem [{\citenamefont {Amaro-Seoane}\ \emph {et~al.}(2017)\citenamefont {Amaro-Seoane} \emph {et~al.}}]{LISA:2017pwj}%
  \BibitemOpen
  \bibfield  {author} {\bibinfo {author} {\bibfnamefont {P.}~\bibnamefont {Amaro-Seoane}} \emph {et~al.} (\bibinfo {collaboration} {LISA}),\ }\href@noop {} {\bibinfo {title} {{Laser Interferometer Space Antenna}}} (\bibinfo {year} {2017}),\ \Eprint {https://arxiv.org/abs/1702.00786} {arXiv:1702.00786 [astro-ph.IM]} \BibitemShut {NoStop}%
\bibitem [{\citenamefont {Auclair}\ \emph {et~al.}(2023)\citenamefont {Auclair} \emph {et~al.}}]{LISACosmologyWorkingGroup:2022jok}%
  \BibitemOpen
  \bibfield  {author} {\bibinfo {author} {\bibfnamefont {P.}~\bibnamefont {Auclair}} \emph {et~al.} (\bibinfo {collaboration} {LISA Cosmology Working Group}),\ }\bibfield  {title} {\bibinfo {title} {{Cosmology with the Laser Interferometer Space Antenna}},\ }\href {https://doi.org/10.1007/s41114-023-00045-2} {\bibfield  {journal} {\bibinfo  {journal} {Living Rev. Rel.}\ }\textbf {\bibinfo {volume} {26}},\ \bibinfo {pages} {5} (\bibinfo {year} {2023})},\ \Eprint {https://arxiv.org/abs/2204.05434} {arXiv:2204.05434 [astro-ph.CO]} \BibitemShut {NoStop}%
\bibitem [{\citenamefont {Luo}\ \emph {et~al.}(2016)\citenamefont {Luo} \emph {et~al.}}]{TianQin:2015yph}%
  \BibitemOpen
  \bibfield  {author} {\bibinfo {author} {\bibfnamefont {J.}~\bibnamefont {Luo}} \emph {et~al.} (\bibinfo {collaboration} {TianQin}),\ }\bibfield  {title} {\bibinfo {title} {{TianQin: a space-borne gravitational wave detector}},\ }\href {https://doi.org/10.1088/0264-9381/33/3/035010} {\bibfield  {journal} {\bibinfo  {journal} {Class. Quant. Grav.}\ }\textbf {\bibinfo {volume} {33}},\ \bibinfo {pages} {035010} (\bibinfo {year} {2016})},\ \Eprint {https://arxiv.org/abs/1512.02076} {arXiv:1512.02076 [astro-ph.IM]} \BibitemShut {NoStop}%
\bibitem [{\citenamefont {Hu}\ and\ \citenamefont {Wu}(2017)}]{Hu:2017mde}%
  \BibitemOpen
  \bibfield  {author} {\bibinfo {author} {\bibfnamefont {W.-R.}\ \bibnamefont {Hu}}\ and\ \bibinfo {author} {\bibfnamefont {Y.-L.}\ \bibnamefont {Wu}},\ }\bibfield  {title} {\bibinfo {title} {{The Taiji Program in Space for gravitational wave physics and the nature of gravity}},\ }\href {https://doi.org/10.1093/nsr/nwx116} {\bibfield  {journal} {\bibinfo  {journal} {Natl. Sci. Rev.}\ }\textbf {\bibinfo {volume} {4}},\ \bibinfo {pages} {685} (\bibinfo {year} {2017})}\BibitemShut {NoStop}%
\bibitem [{\citenamefont {Ruan}\ \emph {et~al.}(2020)\citenamefont {Ruan}, \citenamefont {Guo}, \citenamefont {Cai},\ and\ \citenamefont {Zhang}}]{Ruan:2018tsw}%
  \BibitemOpen
  \bibfield  {author} {\bibinfo {author} {\bibfnamefont {W.-H.}\ \bibnamefont {Ruan}}, \bibinfo {author} {\bibfnamefont {Z.-K.}\ \bibnamefont {Guo}}, \bibinfo {author} {\bibfnamefont {R.-G.}\ \bibnamefont {Cai}},\ and\ \bibinfo {author} {\bibfnamefont {Y.-Z.}\ \bibnamefont {Zhang}},\ }\bibfield  {title} {\bibinfo {title} {{Taiji program: Gravitational-wave sources}},\ }\href {https://doi.org/10.1142/S0217751X2050075X} {\bibfield  {journal} {\bibinfo  {journal} {Int. J. Mod. Phys. A}\ }\textbf {\bibinfo {volume} {35}},\ \bibinfo {pages} {2050075} (\bibinfo {year} {2020})},\ \Eprint {https://arxiv.org/abs/1807.09495} {arXiv:1807.09495 [gr-qc]} \BibitemShut {NoStop}%
\bibitem [{\citenamefont {Seto}\ \emph {et~al.}(2001)\citenamefont {Seto}, \citenamefont {Kawamura},\ and\ \citenamefont {Nakamura}}]{Seto:2001qf}%
  \BibitemOpen
  \bibfield  {author} {\bibinfo {author} {\bibfnamefont {N.}~\bibnamefont {Seto}}, \bibinfo {author} {\bibfnamefont {S.}~\bibnamefont {Kawamura}},\ and\ \bibinfo {author} {\bibfnamefont {T.}~\bibnamefont {Nakamura}},\ }\bibfield  {title} {\bibinfo {title} {{Possibility of direct measurement of the acceleration of the universe using 0.1-Hz band laser interferometer gravitational wave antenna in space}},\ }\href {https://doi.org/10.1103/PhysRevLett.87.221103} {\bibfield  {journal} {\bibinfo  {journal} {Phys. Rev. Lett.}\ }\textbf {\bibinfo {volume} {87}},\ \bibinfo {pages} {221103} (\bibinfo {year} {2001})},\ \Eprint {https://arxiv.org/abs/astro-ph/0108011} {arXiv:astro-ph/0108011} \BibitemShut {NoStop}%
\bibitem [{\citenamefont {Sato}\ \emph {et~al.}(2017)\citenamefont {Sato} \emph {et~al.}}]{Sato:2017dkf}%
  \BibitemOpen
  \bibfield  {author} {\bibinfo {author} {\bibfnamefont {S.}~\bibnamefont {Sato}} \emph {et~al.},\ }\bibfield  {title} {\bibinfo {title} {{The status of DECIGO}},\ }\href {https://doi.org/10.1088/1742-6596/840/1/012010} {\bibfield  {journal} {\bibinfo  {journal} {J. Phys. Conf. Ser.}\ }\textbf {\bibinfo {volume} {840}},\ \bibinfo {pages} {012010} (\bibinfo {year} {2017})}\BibitemShut {NoStop}%
\bibitem [{\citenamefont {Kawamura}\ \emph {et~al.}(2021)\citenamefont {Kawamura} \emph {et~al.}}]{Kawamura:2020pcg}%
  \BibitemOpen
  \bibfield  {author} {\bibinfo {author} {\bibfnamefont {S.}~\bibnamefont {Kawamura}} \emph {et~al.},\ }\bibfield  {title} {\bibinfo {title} {{Current status of space gravitational wave antenna DECIGO and B-DECIGO}},\ }\href {https://doi.org/10.1093/ptep/ptab019} {\bibfield  {journal} {\bibinfo  {journal} {PTEP}\ }\textbf {\bibinfo {volume} {2021}},\ \bibinfo {pages} {05A105} (\bibinfo {year} {2021})},\ \Eprint {https://arxiv.org/abs/2006.13545} {arXiv:2006.13545 [gr-qc]} \BibitemShut {NoStop}%
\bibitem [{\citenamefont {Kuns}\ \emph {et~al.}(2020)\citenamefont {Kuns}, \citenamefont {Yu}, \citenamefont {Chen},\ and\ \citenamefont {Adhikari}}]{Kuns:2019upi}%
  \BibitemOpen
  \bibfield  {author} {\bibinfo {author} {\bibfnamefont {K.~A.}\ \bibnamefont {Kuns}}, \bibinfo {author} {\bibfnamefont {H.}~\bibnamefont {Yu}}, \bibinfo {author} {\bibfnamefont {Y.}~\bibnamefont {Chen}},\ and\ \bibinfo {author} {\bibfnamefont {R.~X.}\ \bibnamefont {Adhikari}},\ }\bibfield  {title} {\bibinfo {title} {{Astrophysics and cosmology with a decihertz gravitational-wave detector: TianGO}},\ }\href {https://doi.org/10.1103/PhysRevD.102.043001} {\bibfield  {journal} {\bibinfo  {journal} {Phys. Rev. D}\ }\textbf {\bibinfo {volume} {102}},\ \bibinfo {pages} {043001} (\bibinfo {year} {2020})},\ \Eprint {https://arxiv.org/abs/1908.06004} {arXiv:1908.06004 [gr-qc]} \BibitemShut {NoStop}%
\bibitem [{\citenamefont {Kuns}(2019)}]{kunsthesis}%
  \BibitemOpen
  \bibfield  {author} {\bibinfo {author} {\bibfnamefont {K.~A.}\ \bibnamefont {Kuns}},\ }\emph {\bibinfo {title} {Future Networks of Gravitational Wave Detectors: Quantum Noise and Space Detectors}},\ \href@noop {} {Ph.D. thesis},\ \bibinfo  {school} {University of California, Santa Barbara} (\bibinfo {year} {2019})\BibitemShut {NoStop}%
\bibitem [{\citenamefont {Yunes}\ and\ \citenamefont {Pretorius}(2009)}]{Yunes:2009ke}%
  \BibitemOpen
  \bibfield  {author} {\bibinfo {author} {\bibfnamefont {N.}~\bibnamefont {Yunes}}\ and\ \bibinfo {author} {\bibfnamefont {F.}~\bibnamefont {Pretorius}},\ }\bibfield  {title} {\bibinfo {title} {{Fundamental Theoretical Bias in Gravitational Wave Astrophysics and the Parameterized Post-Einsteinian Framework}},\ }\href {https://doi.org/10.1103/PhysRevD.80.122003} {\bibfield  {journal} {\bibinfo  {journal} {Phys. Rev. D}\ }\textbf {\bibinfo {volume} {80}},\ \bibinfo {pages} {122003} (\bibinfo {year} {2009})},\ \Eprint {https://arxiv.org/abs/0909.3328} {arXiv:0909.3328 [gr-qc]} \BibitemShut {NoStop}%
\bibitem [{\citenamefont {Cornish}\ \emph {et~al.}(2011)\citenamefont {Cornish}, \citenamefont {Sampson}, \citenamefont {Yunes},\ and\ \citenamefont {Pretorius}}]{Cornish:2011ys}%
  \BibitemOpen
  \bibfield  {author} {\bibinfo {author} {\bibfnamefont {N.}~\bibnamefont {Cornish}}, \bibinfo {author} {\bibfnamefont {L.}~\bibnamefont {Sampson}}, \bibinfo {author} {\bibfnamefont {N.}~\bibnamefont {Yunes}},\ and\ \bibinfo {author} {\bibfnamefont {F.}~\bibnamefont {Pretorius}},\ }\bibfield  {title} {\bibinfo {title} {{Gravitational Wave Tests of General Relativity with the Parameterized Post-Einsteinian Framework}},\ }\href {https://doi.org/10.1103/PhysRevD.84.062003} {\bibfield  {journal} {\bibinfo  {journal} {Phys. Rev. D}\ }\textbf {\bibinfo {volume} {84}},\ \bibinfo {pages} {062003} (\bibinfo {year} {2011})},\ \Eprint {https://arxiv.org/abs/1105.2088} {arXiv:1105.2088 [gr-qc]} \BibitemShut {NoStop}%
\bibitem [{\citenamefont {Li}\ \emph {et~al.}(2012)\citenamefont {Li}, \citenamefont {Del~Pozzo}, \citenamefont {Vitale}, \citenamefont {Van Den~Broeck}, \citenamefont {Agathos}, \citenamefont {Veitch}, \citenamefont {Grover}, \citenamefont {Sidery}, \citenamefont {Sturani},\ and\ \citenamefont {Vecchio}}]{Li:2011cg}%
  \BibitemOpen
  \bibfield  {author} {\bibinfo {author} {\bibfnamefont {T.~G.~F.}\ \bibnamefont {Li}}, \bibinfo {author} {\bibfnamefont {W.}~\bibnamefont {Del~Pozzo}}, \bibinfo {author} {\bibfnamefont {S.}~\bibnamefont {Vitale}}, \bibinfo {author} {\bibfnamefont {C.}~\bibnamefont {Van Den~Broeck}}, \bibinfo {author} {\bibfnamefont {M.}~\bibnamefont {Agathos}}, \bibinfo {author} {\bibfnamefont {J.}~\bibnamefont {Veitch}}, \bibinfo {author} {\bibfnamefont {K.}~\bibnamefont {Grover}}, \bibinfo {author} {\bibfnamefont {T.}~\bibnamefont {Sidery}}, \bibinfo {author} {\bibfnamefont {R.}~\bibnamefont {Sturani}},\ and\ \bibinfo {author} {\bibfnamefont {A.}~\bibnamefont {Vecchio}},\ }\bibfield  {title} {\bibinfo {title} {{Towards a generic test of the strong field dynamics of general relativity using compact binary coalescence}},\ }\href {https://doi.org/10.1103/PhysRevD.85.082003} {\bibfield  {journal} {\bibinfo  {journal} {Phys. Rev. D}\ }\textbf {\bibinfo {volume} {85}},\ \bibinfo {pages} {082003} (\bibinfo {year} {2012})},\ \Eprint {https://arxiv.org/abs/1110.0530} {arXiv:1110.0530 [gr-qc]} \BibitemShut {NoStop}%
\bibitem [{\citenamefont {Agathos}\ \emph {et~al.}(2014)\citenamefont {Agathos}, \citenamefont {Del~Pozzo}, \citenamefont {Li}, \citenamefont {Van Den~Broeck}, \citenamefont {Veitch},\ and\ \citenamefont {Vitale}}]{Agathos:2013upa}%
  \BibitemOpen
  \bibfield  {author} {\bibinfo {author} {\bibfnamefont {M.}~\bibnamefont {Agathos}}, \bibinfo {author} {\bibfnamefont {W.}~\bibnamefont {Del~Pozzo}}, \bibinfo {author} {\bibfnamefont {T.~G.~F.}\ \bibnamefont {Li}}, \bibinfo {author} {\bibfnamefont {C.}~\bibnamefont {Van Den~Broeck}}, \bibinfo {author} {\bibfnamefont {J.}~\bibnamefont {Veitch}},\ and\ \bibinfo {author} {\bibfnamefont {S.}~\bibnamefont {Vitale}},\ }\bibfield  {title} {\bibinfo {title} {{TIGER: A data analysis pipeline for testing the strong-field dynamics of general relativity with gravitational wave signals from coalescing compact binaries}},\ }\href {https://doi.org/10.1103/PhysRevD.89.082001} {\bibfield  {journal} {\bibinfo  {journal} {Phys. Rev. D}\ }\textbf {\bibinfo {volume} {89}},\ \bibinfo {pages} {082001} (\bibinfo {year} {2014})},\ \Eprint {https://arxiv.org/abs/1311.0420} {arXiv:1311.0420 [gr-qc]} \BibitemShut {NoStop}%
\bibitem [{\citenamefont {Meidam}\ \emph {et~al.}(2018)\citenamefont {Meidam} \emph {et~al.}}]{Meidam:2017dgf}%
  \BibitemOpen
  \bibfield  {author} {\bibinfo {author} {\bibfnamefont {J.}~\bibnamefont {Meidam}} \emph {et~al.},\ }\bibfield  {title} {\bibinfo {title} {{Parametrized tests of the strong-field dynamics of general relativity using gravitational wave signals from coalescing binary black holes: Fast likelihood calculations and sensitivity of the method}},\ }\href {https://doi.org/10.1103/PhysRevD.97.044033} {\bibfield  {journal} {\bibinfo  {journal} {Phys. Rev. D}\ }\textbf {\bibinfo {volume} {97}},\ \bibinfo {pages} {044033} (\bibinfo {year} {2018})},\ \Eprint {https://arxiv.org/abs/1712.08772} {arXiv:1712.08772 [gr-qc]} \BibitemShut {NoStop}%
\bibitem [{\citenamefont {Yunes}\ \emph {et~al.}(2025)\citenamefont {Yunes}, \citenamefont {Siemens},\ and\ \citenamefont {Yagi}}]{Yunes:2025xwp}%
  \BibitemOpen
  \bibfield  {author} {\bibinfo {author} {\bibfnamefont {N.}~\bibnamefont {Yunes}}, \bibinfo {author} {\bibfnamefont {X.}~\bibnamefont {Siemens}},\ and\ \bibinfo {author} {\bibfnamefont {K.}~\bibnamefont {Yagi}},\ }\bibfield  {title} {\bibinfo {title} {{Gravitational-wave tests of general relativity with ground-based detectors and pulsar-timing arrays}},\ }\href {https://doi.org/10.1007/s41114-024-00054-9} {\bibfield  {journal} {\bibinfo  {journal} {Living Rev. Rel.}\ }\textbf {\bibinfo {volume} {28}},\ \bibinfo {pages} {3} (\bibinfo {year} {2025})}\BibitemShut {NoStop}%
\bibitem [{\citenamefont {Chatziioannou}\ \emph {et~al.}(2012)\citenamefont {Chatziioannou}, \citenamefont {Yunes},\ and\ \citenamefont {Cornish}}]{Chatziioannou:2012rf}%
  \BibitemOpen
  \bibfield  {author} {\bibinfo {author} {\bibfnamefont {K.}~\bibnamefont {Chatziioannou}}, \bibinfo {author} {\bibfnamefont {N.}~\bibnamefont {Yunes}},\ and\ \bibinfo {author} {\bibfnamefont {N.}~\bibnamefont {Cornish}},\ }\bibfield  {title} {\bibinfo {title} {{Model-Independent Test of General Relativity: An Extended post-Einsteinian Framework with Complete Polarization Content}},\ }\href {https://doi.org/10.1103/PhysRevD.86.022004} {\bibfield  {journal} {\bibinfo  {journal} {Phys. Rev. D}\ }\textbf {\bibinfo {volume} {86}},\ \bibinfo {pages} {022004} (\bibinfo {year} {2012})},\ \bibinfo {note} {[Erratum: Phys.Rev.D 95, 129901 (2017)]},\ \Eprint {https://arxiv.org/abs/1204.2585} {arXiv:1204.2585 [gr-qc]} \BibitemShut {NoStop}%
\bibitem [{\citenamefont {Loutrel}\ \emph {et~al.}(2023)\citenamefont {Loutrel}, \citenamefont {Pani},\ and\ \citenamefont {Yunes}}]{Loutrel:2022xok}%
  \BibitemOpen
  \bibfield  {author} {\bibinfo {author} {\bibfnamefont {N.}~\bibnamefont {Loutrel}}, \bibinfo {author} {\bibfnamefont {P.}~\bibnamefont {Pani}},\ and\ \bibinfo {author} {\bibfnamefont {N.}~\bibnamefont {Yunes}},\ }\bibfield  {title} {\bibinfo {title} {{Parametrized post-Einsteinian framework for precessing binaries}},\ }\href {https://doi.org/10.1103/PhysRevD.107.044046} {\bibfield  {journal} {\bibinfo  {journal} {Phys. Rev. D}\ }\textbf {\bibinfo {volume} {107}},\ \bibinfo {pages} {044046} (\bibinfo {year} {2023})},\ \Eprint {https://arxiv.org/abs/2210.10571} {arXiv:2210.10571 [gr-qc]} \BibitemShut {NoStop}%
\bibitem [{\citenamefont {Mezzasoma}\ and\ \citenamefont {Yunes}(2022)}]{Mezzasoma:2022pjb}%
  \BibitemOpen
  \bibfield  {author} {\bibinfo {author} {\bibfnamefont {S.}~\bibnamefont {Mezzasoma}}\ and\ \bibinfo {author} {\bibfnamefont {N.}~\bibnamefont {Yunes}},\ }\bibfield  {title} {\bibinfo {title} {{Theory-agnostic framework for inspiral tests of general relativity with higher-harmonic gravitational waves}},\ }\href {https://doi.org/10.1103/PhysRevD.106.024026} {\bibfield  {journal} {\bibinfo  {journal} {Phys. Rev. D}\ }\textbf {\bibinfo {volume} {106}},\ \bibinfo {pages} {024026} (\bibinfo {year} {2022})},\ \Eprint {https://arxiv.org/abs/2203.15934} {arXiv:2203.15934 [gr-qc]} \BibitemShut {NoStop}%
\bibitem [{\citenamefont {Mehta}\ \emph {et~al.}(2023)\citenamefont {Mehta}, \citenamefont {Buonanno}, \citenamefont {Cotesta}, \citenamefont {Ghosh}, \citenamefont {Sennett},\ and\ \citenamefont {Steinhoff}}]{Mehta:2022pcn}%
  \BibitemOpen
  \bibfield  {author} {\bibinfo {author} {\bibfnamefont {A.~K.}\ \bibnamefont {Mehta}}, \bibinfo {author} {\bibfnamefont {A.}~\bibnamefont {Buonanno}}, \bibinfo {author} {\bibfnamefont {R.}~\bibnamefont {Cotesta}}, \bibinfo {author} {\bibfnamefont {A.}~\bibnamefont {Ghosh}}, \bibinfo {author} {\bibfnamefont {N.}~\bibnamefont {Sennett}},\ and\ \bibinfo {author} {\bibfnamefont {J.}~\bibnamefont {Steinhoff}},\ }\bibfield  {title} {\bibinfo {title} {{Tests of general relativity with gravitational-wave observations using a flexible theory-independent method}},\ }\href {https://doi.org/10.1103/PhysRevD.107.044020} {\bibfield  {journal} {\bibinfo  {journal} {Phys. Rev. D}\ }\textbf {\bibinfo {volume} {107}},\ \bibinfo {pages} {044020} (\bibinfo {year} {2023})},\ \Eprint {https://arxiv.org/abs/2203.13937} {arXiv:2203.13937 [gr-qc]} \BibitemShut {NoStop}%
\bibitem [{\citenamefont {Bhat}\ \emph {et~al.}(2024)\citenamefont {Bhat}, \citenamefont {Saini}, \citenamefont {Favata}, \citenamefont {Gandevikar}, \citenamefont {Mishra},\ and\ \citenamefont {Arun}}]{Bhat:2024hyb}%
  \BibitemOpen
  \bibfield  {author} {\bibinfo {author} {\bibfnamefont {S.~A.}\ \bibnamefont {Bhat}}, \bibinfo {author} {\bibfnamefont {P.}~\bibnamefont {Saini}}, \bibinfo {author} {\bibfnamefont {M.}~\bibnamefont {Favata}}, \bibinfo {author} {\bibfnamefont {C.}~\bibnamefont {Gandevikar}}, \bibinfo {author} {\bibfnamefont {C.~K.}\ \bibnamefont {Mishra}},\ and\ \bibinfo {author} {\bibfnamefont {K.~G.}\ \bibnamefont {Arun}},\ }\bibfield  {title} {\bibinfo {title} {{Parametrized tests of general relativity using eccentric compact binaries}},\ }\href {https://doi.org/10.1103/PhysRevD.110.124062} {\bibfield  {journal} {\bibinfo  {journal} {Phys. Rev. D}\ }\textbf {\bibinfo {volume} {110}},\ \bibinfo {pages} {124062} (\bibinfo {year} {2024})},\ \Eprint {https://arxiv.org/abs/2408.14132} {arXiv:2408.14132 [gr-qc]} \BibitemShut {NoStop}%
\bibitem [{\citenamefont {Bonilla}\ \emph {et~al.}(2023)\citenamefont {Bonilla}, \citenamefont {Kumar},\ and\ \citenamefont {Teukolsky}}]{Bonilla:2022dyt}%
  \BibitemOpen
  \bibfield  {author} {\bibinfo {author} {\bibfnamefont {G.~S.}\ \bibnamefont {Bonilla}}, \bibinfo {author} {\bibfnamefont {P.}~\bibnamefont {Kumar}},\ and\ \bibinfo {author} {\bibfnamefont {S.~A.}\ \bibnamefont {Teukolsky}},\ }\bibfield  {title} {\bibinfo {title} {{Modeling compact binary merger waveforms beyond general relativity}},\ }\href {https://doi.org/10.1103/PhysRevD.107.024015} {\bibfield  {journal} {\bibinfo  {journal} {Phys. Rev. D}\ }\textbf {\bibinfo {volume} {107}},\ \bibinfo {pages} {024015} (\bibinfo {year} {2023})},\ \Eprint {https://arxiv.org/abs/2203.14026} {arXiv:2203.14026 [gr-qc]} \BibitemShut {NoStop}%
\bibitem [{\citenamefont {Maggio}\ \emph {et~al.}(2023)\citenamefont {Maggio}, \citenamefont {Silva}, \citenamefont {Buonanno},\ and\ \citenamefont {Ghosh}}]{Maggio:2022hre}%
  \BibitemOpen
  \bibfield  {author} {\bibinfo {author} {\bibfnamefont {E.}~\bibnamefont {Maggio}}, \bibinfo {author} {\bibfnamefont {H.~O.}\ \bibnamefont {Silva}}, \bibinfo {author} {\bibfnamefont {A.}~\bibnamefont {Buonanno}},\ and\ \bibinfo {author} {\bibfnamefont {A.}~\bibnamefont {Ghosh}},\ }\bibfield  {title} {\bibinfo {title} {{Tests of general relativity in the nonlinear regime: A parametrized plunge-merger-ringdown gravitational waveform model}},\ }\href {https://doi.org/10.1103/PhysRevD.108.024043} {\bibfield  {journal} {\bibinfo  {journal} {Phys. Rev. D}\ }\textbf {\bibinfo {volume} {108}},\ \bibinfo {pages} {024043} (\bibinfo {year} {2023})},\ \Eprint {https://arxiv.org/abs/2212.09655} {arXiv:2212.09655 [gr-qc]} \BibitemShut {NoStop}%
\bibitem [{\citenamefont {Watarai}\ \emph {et~al.}(2024)\citenamefont {Watarai}, \citenamefont {Nishizawa},\ and\ \citenamefont {Cannon}}]{Watarai:2023yky}%
  \BibitemOpen
  \bibfield  {author} {\bibinfo {author} {\bibfnamefont {D.}~\bibnamefont {Watarai}}, \bibinfo {author} {\bibfnamefont {A.}~\bibnamefont {Nishizawa}},\ and\ \bibinfo {author} {\bibfnamefont {K.}~\bibnamefont {Cannon}},\ }\bibfield  {title} {\bibinfo {title} {{Physically consistent gravitational waveform for capturing beyond general relativity effects in the compact object merger phase}},\ }\href {https://doi.org/10.1103/PhysRevD.109.084058} {\bibfield  {journal} {\bibinfo  {journal} {Phys. Rev. D}\ }\textbf {\bibinfo {volume} {109}},\ \bibinfo {pages} {084058} (\bibinfo {year} {2024})},\ \Eprint {https://arxiv.org/abs/2309.14061} {arXiv:2309.14061 [gr-qc]} \BibitemShut {NoStop}%
\bibitem [{\citenamefont {Watarai}\ \emph {et~al.}(2025)\citenamefont {Watarai}, \citenamefont {Nishizawa}, \citenamefont {Takeda}, \citenamefont {Imafuku},\ and\ \citenamefont {Cannon}}]{Watarai:2025hsb}%
  \BibitemOpen
  \bibfield  {author} {\bibinfo {author} {\bibfnamefont {D.}~\bibnamefont {Watarai}}, \bibinfo {author} {\bibfnamefont {A.}~\bibnamefont {Nishizawa}}, \bibinfo {author} {\bibfnamefont {H.}~\bibnamefont {Takeda}}, \bibinfo {author} {\bibfnamefont {H.}~\bibnamefont {Imafuku}},\ and\ \bibinfo {author} {\bibfnamefont {K.}~\bibnamefont {Cannon}},\ }\href@noop {} {\bibinfo {title} {{Observational constraints on the nonlinear regime of gravity with a parametrized beyond-GR gravitational waveform model}}} (\bibinfo {year} {2025}),\ \Eprint {https://arxiv.org/abs/2509.17592} {arXiv:2509.17592 [gr-qc]} \BibitemShut {NoStop}%
\bibitem [{\citenamefont {Xie}\ \emph {et~al.}(2024)\citenamefont {Xie}, \citenamefont {Chatterjee}, \citenamefont {Narayan},\ and\ \citenamefont {Yunes}}]{Xie:2024ubm}%
  \BibitemOpen
  \bibfield  {author} {\bibinfo {author} {\bibfnamefont {Y.}~\bibnamefont {Xie}}, \bibinfo {author} {\bibfnamefont {D.}~\bibnamefont {Chatterjee}}, \bibinfo {author} {\bibfnamefont {G.}~\bibnamefont {Narayan}},\ and\ \bibinfo {author} {\bibfnamefont {N.}~\bibnamefont {Yunes}},\ }\bibfield  {title} {\bibinfo {title} {{Neural post-Einsteinian framework for efficient theory-agnostic tests of general relativity with gravitational waves}},\ }\href {https://doi.org/10.1103/PhysRevD.110.024036} {\bibfield  {journal} {\bibinfo  {journal} {Phys. Rev. D}\ }\textbf {\bibinfo {volume} {110}},\ \bibinfo {pages} {024036} (\bibinfo {year} {2024})},\ \Eprint {https://arxiv.org/abs/2403.18936} {arXiv:2403.18936 [gr-qc]} \BibitemShut {NoStop}%
\bibitem [{\citenamefont {Sampson}\ \emph {et~al.}(2014)\citenamefont {Sampson}, \citenamefont {Cornish},\ and\ \citenamefont {Yunes}}]{Sampson:2013jpa}%
  \BibitemOpen
  \bibfield  {author} {\bibinfo {author} {\bibfnamefont {L.}~\bibnamefont {Sampson}}, \bibinfo {author} {\bibfnamefont {N.}~\bibnamefont {Cornish}},\ and\ \bibinfo {author} {\bibfnamefont {N.}~\bibnamefont {Yunes}},\ }\bibfield  {title} {\bibinfo {title} {{Mismodeling in gravitational-wave astronomy: The trouble with templates}},\ }\href {https://doi.org/10.1103/PhysRevD.89.064037} {\bibfield  {journal} {\bibinfo  {journal} {Phys. Rev. D}\ }\textbf {\bibinfo {volume} {89}},\ \bibinfo {pages} {064037} (\bibinfo {year} {2014})},\ \Eprint {https://arxiv.org/abs/1311.4898} {arXiv:1311.4898 [gr-qc]} \BibitemShut {NoStop}%
\bibitem [{\citenamefont {Seymour}\ and\ \citenamefont {Chen}(2024)}]{Seymour:2024kcd}%
  \BibitemOpen
  \bibfield  {author} {\bibinfo {author} {\bibfnamefont {B.~C.}\ \bibnamefont {Seymour}}\ and\ \bibinfo {author} {\bibfnamefont {Y.}~\bibnamefont {Chen}},\ }\href@noop {} {\bibinfo {title} {{Gravitational-wave signatures of non-violent non-locality}}} (\bibinfo {year} {2024}),\ \Eprint {https://arxiv.org/abs/2411.13714} {arXiv:2411.13714 [gr-qc]} \BibitemShut {NoStop}%
\bibitem [{\citenamefont {Flanagan}\ and\ \citenamefont {Hughes}(1998)}]{Flanagan:1997kp}%
  \BibitemOpen
  \bibfield  {author} {\bibinfo {author} {\bibfnamefont {E.~E.}\ \bibnamefont {Flanagan}}\ and\ \bibinfo {author} {\bibfnamefont {S.~A.}\ \bibnamefont {Hughes}},\ }\bibfield  {title} {\bibinfo {title} {{Measuring gravitational waves from binary black hole coalescences: 2. The Waves' information and its extraction, with and without templates}},\ }\href {https://doi.org/10.1103/PhysRevD.57.4566} {\bibfield  {journal} {\bibinfo  {journal} {Phys. Rev. D}\ }\textbf {\bibinfo {volume} {57}},\ \bibinfo {pages} {4566} (\bibinfo {year} {1998})},\ \Eprint {https://arxiv.org/abs/gr-qc/9710129} {arXiv:gr-qc/9710129} \BibitemShut {NoStop}%
\bibitem [{\citenamefont {Lindblom}\ \emph {et~al.}(2008)\citenamefont {Lindblom}, \citenamefont {Owen},\ and\ \citenamefont {Brown}}]{Lindblom:2008cm}%
  \BibitemOpen
  \bibfield  {author} {\bibinfo {author} {\bibfnamefont {L.}~\bibnamefont {Lindblom}}, \bibinfo {author} {\bibfnamefont {B.~J.}\ \bibnamefont {Owen}},\ and\ \bibinfo {author} {\bibfnamefont {D.~A.}\ \bibnamefont {Brown}},\ }\bibfield  {title} {\bibinfo {title} {{Model Waveform Accuracy Standards for Gravitational Wave Data Analysis}},\ }\href {https://doi.org/10.1103/PhysRevD.78.124020} {\bibfield  {journal} {\bibinfo  {journal} {Phys. Rev. D}\ }\textbf {\bibinfo {volume} {78}},\ \bibinfo {pages} {124020} (\bibinfo {year} {2008})},\ \Eprint {https://arxiv.org/abs/0809.3844} {arXiv:0809.3844 [gr-qc]} \BibitemShut {NoStop}%
\bibitem [{\citenamefont {Kumar}\ \emph {et~al.}(2015)\citenamefont {Kumar}, \citenamefont {Barkett}, \citenamefont {Bhagwat}, \citenamefont {Afshari}, \citenamefont {Brown}, \citenamefont {Lovelace}, \citenamefont {Scheel},\ and\ \citenamefont {Szil{\'a}gyi}}]{Kumar:2015tha}%
  \BibitemOpen
  \bibfield  {author} {\bibinfo {author} {\bibfnamefont {P.}~\bibnamefont {Kumar}}, \bibinfo {author} {\bibfnamefont {K.}~\bibnamefont {Barkett}}, \bibinfo {author} {\bibfnamefont {S.}~\bibnamefont {Bhagwat}}, \bibinfo {author} {\bibfnamefont {N.}~\bibnamefont {Afshari}}, \bibinfo {author} {\bibfnamefont {D.~A.}\ \bibnamefont {Brown}}, \bibinfo {author} {\bibfnamefont {G.}~\bibnamefont {Lovelace}}, \bibinfo {author} {\bibfnamefont {M.~A.}\ \bibnamefont {Scheel}},\ and\ \bibinfo {author} {\bibfnamefont {B.}~\bibnamefont {Szil{\'a}gyi}},\ }\bibfield  {title} {\bibinfo {title} {{Accuracy and precision of gravitational-wave models of inspiraling neutron star-black hole binaries with spin: Comparison with matter-free numerical relativity in the low-frequency regime}},\ }\href {https://doi.org/10.1103/PhysRevD.92.102001} {\bibfield  {journal} {\bibinfo  {journal} {Phys. Rev. D}\ }\textbf {\bibinfo {volume} {92}},\ \bibinfo {pages} {102001} (\bibinfo {year} {2015})},\ \Eprint {https://arxiv.org/abs/1507.00103} {arXiv:1507.00103 [gr-qc]} \BibitemShut {NoStop}%
\bibitem [{\citenamefont {Chatziioannou}\ \emph {et~al.}(2017)\citenamefont {Chatziioannou}, \citenamefont {Klein}, \citenamefont {Yunes},\ and\ \citenamefont {Cornish}}]{Chatziioannou:2017tdw}%
  \BibitemOpen
  \bibfield  {author} {\bibinfo {author} {\bibfnamefont {K.}~\bibnamefont {Chatziioannou}}, \bibinfo {author} {\bibfnamefont {A.}~\bibnamefont {Klein}}, \bibinfo {author} {\bibfnamefont {N.}~\bibnamefont {Yunes}},\ and\ \bibinfo {author} {\bibfnamefont {N.}~\bibnamefont {Cornish}},\ }\bibfield  {title} {\bibinfo {title} {{Constructing Gravitational Waves from Generic Spin-Precessing Compact Binary Inspirals}},\ }\href {https://doi.org/10.1103/PhysRevD.95.104004} {\bibfield  {journal} {\bibinfo  {journal} {Phys. Rev. D}\ }\textbf {\bibinfo {volume} {95}},\ \bibinfo {pages} {104004} (\bibinfo {year} {2017})},\ \Eprint {https://arxiv.org/abs/1703.03967} {arXiv:1703.03967 [gr-qc]} \BibitemShut {NoStop}%
\bibitem [{\citenamefont {P{\"u}rrer}\ and\ \citenamefont {Haster}(2020)}]{Purrer:2019jcp}%
  \BibitemOpen
  \bibfield  {author} {\bibinfo {author} {\bibfnamefont {M.}~\bibnamefont {P{\"u}rrer}}\ and\ \bibinfo {author} {\bibfnamefont {C.-J.}\ \bibnamefont {Haster}},\ }\bibfield  {title} {\bibinfo {title} {{Gravitational waveform accuracy requirements for future ground-based detectors}},\ }\href {https://doi.org/10.1103/PhysRevResearch.2.023151} {\bibfield  {journal} {\bibinfo  {journal} {Phys. Rev. Res.}\ }\textbf {\bibinfo {volume} {2}},\ \bibinfo {pages} {023151} (\bibinfo {year} {2020})},\ \Eprint {https://arxiv.org/abs/1912.10055} {arXiv:1912.10055 [gr-qc]} \BibitemShut {NoStop}%
\bibitem [{\citenamefont {Hu}\ and\ \citenamefont {Veitch}(2022)}]{Hu:2022rjq}%
  \BibitemOpen
  \bibfield  {author} {\bibinfo {author} {\bibfnamefont {Q.}~\bibnamefont {Hu}}\ and\ \bibinfo {author} {\bibfnamefont {J.}~\bibnamefont {Veitch}},\ }\bibfield  {title} {\bibinfo {title} {{Assessing the model waveform accuracy of gravitational waves}},\ }\href {https://doi.org/10.1103/PhysRevD.106.044042} {\bibfield  {journal} {\bibinfo  {journal} {Phys. Rev. D}\ }\textbf {\bibinfo {volume} {106}},\ \bibinfo {pages} {044042} (\bibinfo {year} {2022})},\ \Eprint {https://arxiv.org/abs/2205.08448} {arXiv:2205.08448 [gr-qc]} \BibitemShut {NoStop}%
\bibitem [{\citenamefont {McWilliams}\ \emph {et~al.}(2010)\citenamefont {McWilliams}, \citenamefont {Kelly},\ and\ \citenamefont {Baker}}]{McWilliams:2010eq}%
  \BibitemOpen
  \bibfield  {author} {\bibinfo {author} {\bibfnamefont {S.~T.}\ \bibnamefont {McWilliams}}, \bibinfo {author} {\bibfnamefont {B.~J.}\ \bibnamefont {Kelly}},\ and\ \bibinfo {author} {\bibfnamefont {J.~G.}\ \bibnamefont {Baker}},\ }\bibfield  {title} {\bibinfo {title} {{Observing mergers of non-spinning black-hole binaries}},\ }\href {https://doi.org/10.1103/PhysRevD.82.024014} {\bibfield  {journal} {\bibinfo  {journal} {Phys. Rev. D}\ }\textbf {\bibinfo {volume} {82}},\ \bibinfo {pages} {024014} (\bibinfo {year} {2010})},\ \Eprint {https://arxiv.org/abs/1004.0961} {arXiv:1004.0961 [gr-qc]} \BibitemShut {NoStop}%
\bibitem [{\citenamefont {Toubiana}\ and\ \citenamefont {Gair}(2024)}]{Toubiana:2024car}%
  \BibitemOpen
  \bibfield  {author} {\bibinfo {author} {\bibfnamefont {A.}~\bibnamefont {Toubiana}}\ and\ \bibinfo {author} {\bibfnamefont {J.~R.}\ \bibnamefont {Gair}},\ }\href@noop {} {\bibinfo {title} {{Indistinguishability criterion and estimating the presence of biases}}} (\bibinfo {year} {2024}),\ \Eprint {https://arxiv.org/abs/2401.06845} {arXiv:2401.06845 [gr-qc]} \BibitemShut {NoStop}%
\bibitem [{\citenamefont {Knapp}\ \emph {et~al.}(2026)\citenamefont {Knapp}, \citenamefont {Chatziioannou}, \citenamefont {Mitman}, \citenamefont {Scheel}, \citenamefont {Boyle}, \citenamefont {Kidder},\ and\ \citenamefont {Pfeiffer}}]{Knapp:2025ecw}%
  \BibitemOpen
  \bibfield  {author} {\bibinfo {author} {\bibfnamefont {T.}~\bibnamefont {Knapp}}, \bibinfo {author} {\bibfnamefont {K.}~\bibnamefont {Chatziioannou}}, \bibinfo {author} {\bibfnamefont {K.}~\bibnamefont {Mitman}}, \bibinfo {author} {\bibfnamefont {M.~A.}\ \bibnamefont {Scheel}}, \bibinfo {author} {\bibfnamefont {M.}~\bibnamefont {Boyle}}, \bibinfo {author} {\bibfnamefont {L.~E.}\ \bibnamefont {Kidder}},\ and\ \bibinfo {author} {\bibfnamefont {H.}~\bibnamefont {Pfeiffer}},\ }\bibfield  {title} {\bibinfo {title} {{Comprehensive look into the accuracy of spectral Einstein code binary black hole waveforms}},\ }\href {https://doi.org/10.1103/dmpn-q3fk} {\bibfield  {journal} {\bibinfo  {journal} {Phys. Rev. D}\ }\textbf {\bibinfo {volume} {113}},\ \bibinfo {pages} {124001} (\bibinfo {year} {2026})},\ \Eprint {https://arxiv.org/abs/2510.06393} {arXiv:2510.06393 [gr-qc]} \BibitemShut {NoStop}%
\bibitem [{\citenamefont {Thompson}\ \emph {et~al.}(2025)\citenamefont {Thompson}, \citenamefont {Hoy}, \citenamefont {Fauchon-Jones},\ and\ \citenamefont {Hannam}}]{Thompson:2025hhc}%
  \BibitemOpen
  \bibfield  {author} {\bibinfo {author} {\bibfnamefont {J.~E.}\ \bibnamefont {Thompson}}, \bibinfo {author} {\bibfnamefont {C.}~\bibnamefont {Hoy}}, \bibinfo {author} {\bibfnamefont {E.}~\bibnamefont {Fauchon-Jones}},\ and\ \bibinfo {author} {\bibfnamefont {M.}~\bibnamefont {Hannam}},\ }\bibfield  {title} {\bibinfo {title} {{Use and interpretation of signal-model indistinguishability measures for gravitational-wave astronomy}},\ }\href {https://doi.org/10.1103/ddz7-x9zz} {\bibfield  {journal} {\bibinfo  {journal} {Phys. Rev. D}\ }\textbf {\bibinfo {volume} {112}},\ \bibinfo {pages} {064011} (\bibinfo {year} {2025})},\ \Eprint {https://arxiv.org/abs/2506.10530} {arXiv:2506.10530 [gr-qc]} \BibitemShut {NoStop}%
\bibitem [{\citenamefont {Cutler}\ and\ \citenamefont {Vallisneri}(2007)}]{Cutler:2007mi}%
  \BibitemOpen
  \bibfield  {author} {\bibinfo {author} {\bibfnamefont {C.}~\bibnamefont {Cutler}}\ and\ \bibinfo {author} {\bibfnamefont {M.}~\bibnamefont {Vallisneri}},\ }\bibfield  {title} {\bibinfo {title} {{LISA detections of massive black hole inspirals: Parameter extraction errors due to inaccurate template waveforms}},\ }\href {https://doi.org/10.1103/PhysRevD.76.104018} {\bibfield  {journal} {\bibinfo  {journal} {Phys. Rev. D}\ }\textbf {\bibinfo {volume} {76}},\ \bibinfo {pages} {104018} (\bibinfo {year} {2007})},\ \Eprint {https://arxiv.org/abs/0707.2982} {arXiv:0707.2982 [gr-qc]} \BibitemShut {NoStop}%
\bibitem [{\citenamefont {Hu}\ and\ \citenamefont {Veitch}(2023)}]{Hu:2022bji}%
  \BibitemOpen
  \bibfield  {author} {\bibinfo {author} {\bibfnamefont {Q.}~\bibnamefont {Hu}}\ and\ \bibinfo {author} {\bibfnamefont {J.}~\bibnamefont {Veitch}},\ }\bibfield  {title} {\bibinfo {title} {{Accumulating Errors in Tests of General Relativity with Gravitational Waves: Overlapping Signals and Inaccurate Waveforms}},\ }\href {https://doi.org/10.3847/1538-4357/acbc18} {\bibfield  {journal} {\bibinfo  {journal} {Astrophys. J.}\ }\textbf {\bibinfo {volume} {945}},\ \bibinfo {pages} {103} (\bibinfo {year} {2023})},\ \Eprint {https://arxiv.org/abs/2210.04769} {arXiv:2210.04769 [gr-qc]} \BibitemShut {NoStop}%
\bibitem [{\citenamefont {Owen}\ \emph {et~al.}(2023)\citenamefont {Owen}, \citenamefont {Haster}, \citenamefont {Perkins}, \citenamefont {Cornish},\ and\ \citenamefont {Yunes}}]{Owen:2023mid}%
  \BibitemOpen
  \bibfield  {author} {\bibinfo {author} {\bibfnamefont {C.~B.}\ \bibnamefont {Owen}}, \bibinfo {author} {\bibfnamefont {C.-J.}\ \bibnamefont {Haster}}, \bibinfo {author} {\bibfnamefont {S.}~\bibnamefont {Perkins}}, \bibinfo {author} {\bibfnamefont {N.~J.}\ \bibnamefont {Cornish}},\ and\ \bibinfo {author} {\bibfnamefont {N.}~\bibnamefont {Yunes}},\ }\bibfield  {title} {\bibinfo {title} {{Waveform accuracy and systematic uncertainties in current gravitational wave observations}},\ }\href {https://doi.org/10.1103/PhysRevD.108.044018} {\bibfield  {journal} {\bibinfo  {journal} {Phys. Rev. D}\ }\textbf {\bibinfo {volume} {108}},\ \bibinfo {pages} {044018} (\bibinfo {year} {2023})},\ \Eprint {https://arxiv.org/abs/2301.11941} {arXiv:2301.11941 [gr-qc]} \BibitemShut {NoStop}%
\bibitem [{\citenamefont {Kapil}\ \emph {et~al.}(2024)\citenamefont {Kapil}, \citenamefont {Reali}, \citenamefont {Cotesta},\ and\ \citenamefont {Berti}}]{Kapil:2024zdn}%
  \BibitemOpen
  \bibfield  {author} {\bibinfo {author} {\bibfnamefont {V.}~\bibnamefont {Kapil}}, \bibinfo {author} {\bibfnamefont {L.}~\bibnamefont {Reali}}, \bibinfo {author} {\bibfnamefont {R.}~\bibnamefont {Cotesta}},\ and\ \bibinfo {author} {\bibfnamefont {E.}~\bibnamefont {Berti}},\ }\bibfield  {title} {\bibinfo {title} {{Systematic bias from waveform modeling for binary black hole populations in next-generation gravitational wave detectors}},\ }\href {https://doi.org/10.1103/PhysRevD.109.104043} {\bibfield  {journal} {\bibinfo  {journal} {Phys. Rev. D}\ }\textbf {\bibinfo {volume} {109}},\ \bibinfo {pages} {104043} (\bibinfo {year} {2024})},\ \Eprint {https://arxiv.org/abs/2404.00090} {arXiv:2404.00090 [gr-qc]} \BibitemShut {NoStop}%
\bibitem [{\citenamefont {Dhani}\ \emph {et~al.}(2025)\citenamefont {Dhani}, \citenamefont {V{\"o}lkel}, \citenamefont {Buonanno}, \citenamefont {Estelles}, \citenamefont {Gair}, \citenamefont {Pfeiffer}, \citenamefont {Pompili},\ and\ \citenamefont {Toubiana}}]{Dhani:2024jja}%
  \BibitemOpen
  \bibfield  {author} {\bibinfo {author} {\bibfnamefont {A.}~\bibnamefont {Dhani}}, \bibinfo {author} {\bibfnamefont {S.~H.}\ \bibnamefont {V{\"o}lkel}}, \bibinfo {author} {\bibfnamefont {A.}~\bibnamefont {Buonanno}}, \bibinfo {author} {\bibfnamefont {H.}~\bibnamefont {Estelles}}, \bibinfo {author} {\bibfnamefont {J.}~\bibnamefont {Gair}}, \bibinfo {author} {\bibfnamefont {H.~P.}\ \bibnamefont {Pfeiffer}}, \bibinfo {author} {\bibfnamefont {L.}~\bibnamefont {Pompili}},\ and\ \bibinfo {author} {\bibfnamefont {A.}~\bibnamefont {Toubiana}},\ }\bibfield  {title} {\bibinfo {title} {{Systematic Biases in Estimating the Properties of Black Holes Due to Inaccurate Gravitational-Wave Models}},\ }\href {https://doi.org/10.1103/5pks-qz6b} {\bibfield  {journal} {\bibinfo  {journal} {Phys. Rev. X}\ }\textbf {\bibinfo {volume} {15}},\ \bibinfo {pages} {031036} (\bibinfo {year} {2025})},\ \Eprint {https://arxiv.org/abs/2404.05811} {arXiv:2404.05811 [gr-qc]} \BibitemShut {NoStop}%
\bibitem [{\citenamefont {Capuano}\ \emph {et~al.}(2025)\citenamefont {Capuano}, \citenamefont {Vaglio}, \citenamefont {Chandramouli}, \citenamefont {Pitte}, \citenamefont {Kuntz},\ and\ \citenamefont {Barausse}}]{Capuano:2025kkl}%
  \BibitemOpen
  \bibfield  {author} {\bibinfo {author} {\bibfnamefont {L.}~\bibnamefont {Capuano}}, \bibinfo {author} {\bibfnamefont {M.}~\bibnamefont {Vaglio}}, \bibinfo {author} {\bibfnamefont {R.~S.}\ \bibnamefont {Chandramouli}}, \bibinfo {author} {\bibfnamefont {C.~L.}\ \bibnamefont {Pitte}}, \bibinfo {author} {\bibfnamefont {A.}~\bibnamefont {Kuntz}},\ and\ \bibinfo {author} {\bibfnamefont {E.}~\bibnamefont {Barausse}},\ }\bibfield  {title} {\bibinfo {title} {{Systematic bias in LISA ringdown analysis due to waveform inaccuracy}},\ }\href {https://doi.org/10.1103/86yd-x1sl} {\bibfield  {journal} {\bibinfo  {journal} {Phys. Rev. D}\ }\textbf {\bibinfo {volume} {112}},\ \bibinfo {pages} {104031} (\bibinfo {year} {2025})},\ \Eprint {https://arxiv.org/abs/2506.21181} {arXiv:2506.21181 [gr-qc]} \BibitemShut {NoStop}%
\bibitem [{\citenamefont {V{\"o}lkel}\ and\ \citenamefont {Dhani}(2025)}]{Volkel:2025jdx}%
  \BibitemOpen
  \bibfield  {author} {\bibinfo {author} {\bibfnamefont {S.~H.}\ \bibnamefont {V{\"o}lkel}}\ and\ \bibinfo {author} {\bibfnamefont {A.}~\bibnamefont {Dhani}},\ }\bibfield  {title} {\bibinfo {title} {{Quantifying systematic biases in black hole spectroscopy}},\ }\href {https://doi.org/10.1103/g6sz-dw28} {\bibfield  {journal} {\bibinfo  {journal} {Phys. Rev. D}\ }\textbf {\bibinfo {volume} {112}},\ \bibinfo {pages} {084076} (\bibinfo {year} {2025})},\ \Eprint {https://arxiv.org/abs/2507.22122} {arXiv:2507.22122 [gr-qc]} \BibitemShut {NoStop}%
\bibitem [{\citenamefont {Vallisneri}(2012)}]{Vallisneri:2012qq}%
  \BibitemOpen
  \bibfield  {author} {\bibinfo {author} {\bibfnamefont {M.}~\bibnamefont {Vallisneri}},\ }\bibfield  {title} {\bibinfo {title} {{Testing general relativity with gravitational waves: a reality check}},\ }\href {https://doi.org/10.1103/PhysRevD.86.082001} {\bibfield  {journal} {\bibinfo  {journal} {Phys. Rev. D}\ }\textbf {\bibinfo {volume} {86}},\ \bibinfo {pages} {082001} (\bibinfo {year} {2012})},\ \Eprint {https://arxiv.org/abs/1207.4759} {arXiv:1207.4759 [gr-qc]} \BibitemShut {NoStop}%
\bibitem [{\citenamefont {Vallisneri}\ and\ \citenamefont {Yunes}(2013)}]{Vallisneri:2013rc}%
  \BibitemOpen
  \bibfield  {author} {\bibinfo {author} {\bibfnamefont {M.}~\bibnamefont {Vallisneri}}\ and\ \bibinfo {author} {\bibfnamefont {N.}~\bibnamefont {Yunes}},\ }\bibfield  {title} {\bibinfo {title} {{Stealth Bias in Gravitational-Wave Parameter Estimation}},\ }\href {https://doi.org/10.1103/PhysRevD.87.102002} {\bibfield  {journal} {\bibinfo  {journal} {Phys. Rev. D}\ }\textbf {\bibinfo {volume} {87}},\ \bibinfo {pages} {102002} (\bibinfo {year} {2013})},\ \Eprint {https://arxiv.org/abs/1301.2627} {arXiv:1301.2627 [gr-qc]} \BibitemShut {NoStop}%
\bibitem [{\citenamefont {Vitale}\ and\ \citenamefont {Del~Pozzo}(2014)}]{Vitale:2013bma}%
  \BibitemOpen
  \bibfield  {author} {\bibinfo {author} {\bibfnamefont {S.}~\bibnamefont {Vitale}}\ and\ \bibinfo {author} {\bibfnamefont {W.}~\bibnamefont {Del~Pozzo}},\ }\bibfield  {title} {\bibinfo {title} {{How serious can the stealth bias be in gravitational wave parameter estimation?}},\ }\href {https://doi.org/10.1103/PhysRevD.89.022002} {\bibfield  {journal} {\bibinfo  {journal} {Phys. Rev. D}\ }\textbf {\bibinfo {volume} {89}},\ \bibinfo {pages} {022002} (\bibinfo {year} {2014})},\ \Eprint {https://arxiv.org/abs/1311.2057} {arXiv:1311.2057 [gr-qc]} \BibitemShut {NoStop}%
\bibitem [{\citenamefont {Moore}\ \emph {et~al.}(2021)\citenamefont {Moore}, \citenamefont {Finch}, \citenamefont {Buscicchio},\ and\ \citenamefont {Gerosa}}]{Moore:2021eok}%
  \BibitemOpen
  \bibfield  {author} {\bibinfo {author} {\bibfnamefont {C.~J.}\ \bibnamefont {Moore}}, \bibinfo {author} {\bibfnamefont {E.}~\bibnamefont {Finch}}, \bibinfo {author} {\bibfnamefont {R.}~\bibnamefont {Buscicchio}},\ and\ \bibinfo {author} {\bibfnamefont {D.}~\bibnamefont {Gerosa}},\ }\bibfield  {title} {\bibinfo {title} {{Testing general relativity with gravitational-wave catalogs: The insidious nature of waveform systematics}},\ }\href {https://doi.org/10.1016/j.isci.2021.102577} {\bibfield  {journal} {\bibinfo  {journal} {iScience}\ }\textbf {\bibinfo {volume} {24}},\ \bibinfo {pages} {102577} (\bibinfo {year} {2021})},\ \Eprint {https://arxiv.org/abs/2103.16486} {arXiv:2103.16486 [gr-qc]} \BibitemShut {NoStop}%
\bibitem [{\citenamefont {Perkins}\ and\ \citenamefont {Yunes}(2022)}]{Perkins:2022fhr}%
  \BibitemOpen
  \bibfield  {author} {\bibinfo {author} {\bibfnamefont {S.}~\bibnamefont {Perkins}}\ and\ \bibinfo {author} {\bibfnamefont {N.}~\bibnamefont {Yunes}},\ }\bibfield  {title} {\bibinfo {title} {{Are parametrized tests of general relativity with gravitational waves robust to unknown higher post-Newtonian order effects?}},\ }\href {https://doi.org/10.1103/PhysRevD.105.124047} {\bibfield  {journal} {\bibinfo  {journal} {Phys. Rev. D}\ }\textbf {\bibinfo {volume} {105}},\ \bibinfo {pages} {124047} (\bibinfo {year} {2022})},\ \Eprint {https://arxiv.org/abs/2201.02542} {arXiv:2201.02542 [gr-qc]} \BibitemShut {NoStop}%
\bibitem [{\citenamefont {Favata}(2014)}]{Favata:2013rwa}%
  \BibitemOpen
  \bibfield  {author} {\bibinfo {author} {\bibfnamefont {M.}~\bibnamefont {Favata}},\ }\bibfield  {title} {\bibinfo {title} {{Systematic parameter errors in inspiraling neutron star binaries}},\ }\href {https://doi.org/10.1103/PhysRevLett.112.101101} {\bibfield  {journal} {\bibinfo  {journal} {Phys. Rev. Lett.}\ }\textbf {\bibinfo {volume} {112}},\ \bibinfo {pages} {101101} (\bibinfo {year} {2014})},\ \Eprint {https://arxiv.org/abs/1310.8288} {arXiv:1310.8288 [gr-qc]} \BibitemShut {NoStop}%
\bibitem [{\citenamefont {Chandramouli}\ \emph {et~al.}(2024)\citenamefont {Chandramouli}, \citenamefont {Prokup}, \citenamefont {Berti},\ and\ \citenamefont {Yunes}}]{Chandramouli:2024vhw}%
  \BibitemOpen
  \bibfield  {author} {\bibinfo {author} {\bibfnamefont {R.~S.}\ \bibnamefont {Chandramouli}}, \bibinfo {author} {\bibfnamefont {K.}~\bibnamefont {Prokup}}, \bibinfo {author} {\bibfnamefont {E.}~\bibnamefont {Berti}},\ and\ \bibinfo {author} {\bibfnamefont {N.}~\bibnamefont {Yunes}},\ }\href@noop {} {\bibinfo {title} {{Systematic biases in parametrized tests of general relativity due to waveform mismodeling: the impact of neglecting spin precession and higher modes}}} (\bibinfo {year} {2024}),\ \Eprint {https://arxiv.org/abs/2410.06254} {arXiv:2410.06254 [gr-qc]} \BibitemShut {NoStop}%
\bibitem [{\citenamefont {Gupta}\ \emph {et~al.}(2024)\citenamefont {Gupta} \emph {et~al.}}]{Gupta:2024gun}%
  \BibitemOpen
  \bibfield  {author} {\bibinfo {author} {\bibfnamefont {A.}~\bibnamefont {Gupta}} \emph {et~al.},\ }\bibfield  {title} {\bibinfo {title} {{Possible causes of false general relativity violations in gravitational wave observations}},\ }\bibfield  {journal} {\bibinfo  {journal} {SciPost Phys.Comm.Rep.}\ }\href {https://doi.org/10.21468/SciPostPhysCommRep.5} {10.21468/SciPostPhysCommRep.5} (\bibinfo {year} {2024}),\ \Eprint {https://arxiv.org/abs/2405.02197} {arXiv:2405.02197 [gr-qc]} \BibitemShut {NoStop}%
\bibitem [{\citenamefont {Saini}\ \emph {et~al.}(2022)\citenamefont {Saini}, \citenamefont {Favata},\ and\ \citenamefont {Arun}}]{Saini:2022igm}%
  \BibitemOpen
  \bibfield  {author} {\bibinfo {author} {\bibfnamefont {P.}~\bibnamefont {Saini}}, \bibinfo {author} {\bibfnamefont {M.}~\bibnamefont {Favata}},\ and\ \bibinfo {author} {\bibfnamefont {K.~G.}\ \bibnamefont {Arun}},\ }\bibfield  {title} {\bibinfo {title} {{Systematic bias on parametrized tests of general relativity due to neglect of orbital eccentricity}},\ }\href {https://doi.org/10.1103/PhysRevD.106.084031} {\bibfield  {journal} {\bibinfo  {journal} {Phys. Rev. D}\ }\textbf {\bibinfo {volume} {106}},\ \bibinfo {pages} {084031} (\bibinfo {year} {2022})},\ \Eprint {https://arxiv.org/abs/2203.04634} {arXiv:2203.04634 [gr-qc]} \BibitemShut {NoStop}%
\bibitem [{\citenamefont {Bhat}\ \emph {et~al.}(2023)\citenamefont {Bhat}, \citenamefont {Saini}, \citenamefont {Favata},\ and\ \citenamefont {Arun}}]{Bhat:2022amc}%
  \BibitemOpen
  \bibfield  {author} {\bibinfo {author} {\bibfnamefont {S.~A.}\ \bibnamefont {Bhat}}, \bibinfo {author} {\bibfnamefont {P.}~\bibnamefont {Saini}}, \bibinfo {author} {\bibfnamefont {M.}~\bibnamefont {Favata}},\ and\ \bibinfo {author} {\bibfnamefont {K.~G.}\ \bibnamefont {Arun}},\ }\bibfield  {title} {\bibinfo {title} {{Systematic bias on the inspiral-merger-ringdown consistency test due to neglect of orbital eccentricity}},\ }\href {https://doi.org/10.1103/PhysRevD.107.024009} {\bibfield  {journal} {\bibinfo  {journal} {Phys. Rev. D}\ }\textbf {\bibinfo {volume} {107}},\ \bibinfo {pages} {024009} (\bibinfo {year} {2023})},\ \Eprint {https://arxiv.org/abs/2207.13761} {arXiv:2207.13761 [gr-qc]} \BibitemShut {NoStop}%
\bibitem [{\citenamefont {Saini}\ \emph {et~al.}(2024)\citenamefont {Saini}, \citenamefont {Bhat}, \citenamefont {Favata},\ and\ \citenamefont {Arun}}]{Saini:2023rto}%
  \BibitemOpen
  \bibfield  {author} {\bibinfo {author} {\bibfnamefont {P.}~\bibnamefont {Saini}}, \bibinfo {author} {\bibfnamefont {S.~A.}\ \bibnamefont {Bhat}}, \bibinfo {author} {\bibfnamefont {M.}~\bibnamefont {Favata}},\ and\ \bibinfo {author} {\bibfnamefont {K.~G.}\ \bibnamefont {Arun}},\ }\bibfield  {title} {\bibinfo {title} {{Eccentricity-induced systematic error on parametrized tests of general relativity: Hierarchical Bayesian inference applied to a binary black hole population}},\ }\href {https://doi.org/10.1103/PhysRevD.109.084056} {\bibfield  {journal} {\bibinfo  {journal} {Phys. Rev. D}\ }\textbf {\bibinfo {volume} {109}},\ \bibinfo {pages} {084056} (\bibinfo {year} {2024})},\ \Eprint {https://arxiv.org/abs/2311.08033} {arXiv:2311.08033 [gr-qc]} \BibitemShut {NoStop}%
\bibitem [{\citenamefont {Narayan}\ \emph {et~al.}(2023)\citenamefont {Narayan}, \citenamefont {Johnson-McDaniel},\ and\ \citenamefont {Gupta}}]{Narayan:2023vhm}%
  \BibitemOpen
  \bibfield  {author} {\bibinfo {author} {\bibfnamefont {P.}~\bibnamefont {Narayan}}, \bibinfo {author} {\bibfnamefont {N.~K.}\ \bibnamefont {Johnson-McDaniel}},\ and\ \bibinfo {author} {\bibfnamefont {A.}~\bibnamefont {Gupta}},\ }\bibfield  {title} {\bibinfo {title} {{Effect of ignoring eccentricity in testing general relativity with gravitational waves}},\ }\href {https://doi.org/10.1103/PhysRevD.108.064003} {\bibfield  {journal} {\bibinfo  {journal} {Phys. Rev. D}\ }\textbf {\bibinfo {volume} {108}},\ \bibinfo {pages} {064003} (\bibinfo {year} {2023})},\ \Eprint {https://arxiv.org/abs/2306.04068} {arXiv:2306.04068 [gr-qc]} \BibitemShut {NoStop}%
\bibitem [{\citenamefont {Garg}\ \emph {et~al.}(2024)\citenamefont {Garg}, \citenamefont {Sberna}, \citenamefont {Speri}, \citenamefont {Duque},\ and\ \citenamefont {Gair}}]{Garg:2024qxq}%
  \BibitemOpen
  \bibfield  {author} {\bibinfo {author} {\bibfnamefont {M.}~\bibnamefont {Garg}}, \bibinfo {author} {\bibfnamefont {L.}~\bibnamefont {Sberna}}, \bibinfo {author} {\bibfnamefont {L.}~\bibnamefont {Speri}}, \bibinfo {author} {\bibfnamefont {F.}~\bibnamefont {Duque}},\ and\ \bibinfo {author} {\bibfnamefont {J.}~\bibnamefont {Gair}},\ }\bibfield  {title} {\bibinfo {title} {{Systematics in tests of general relativity using LISA massive black hole binaries}},\ }\href {https://doi.org/10.1093/mnras/stae2605} {\bibfield  {journal} {\bibinfo  {journal} {Mon. Not. Roy. Astron. Soc.}\ }\textbf {\bibinfo {volume} {535}},\ \bibinfo {pages} {3283} (\bibinfo {year} {2024})},\ \Eprint {https://arxiv.org/abs/2410.02910} {arXiv:2410.02910 [astro-ph.GA]} \BibitemShut {NoStop}%
\bibitem [{\citenamefont {Kejriwal}\ \emph {et~al.}(2024)\citenamefont {Kejriwal}, \citenamefont {Speri},\ and\ \citenamefont {Chua}}]{Kejriwal:2023djc}%
  \BibitemOpen
  \bibfield  {author} {\bibinfo {author} {\bibfnamefont {S.}~\bibnamefont {Kejriwal}}, \bibinfo {author} {\bibfnamefont {L.}~\bibnamefont {Speri}},\ and\ \bibinfo {author} {\bibfnamefont {A.~J.~K.}\ \bibnamefont {Chua}},\ }\bibfield  {title} {\bibinfo {title} {{Impact of correlations on the modeling and inference of beyond vacuum\textendash{}general relativistic effects in extreme-mass-ratio inspirals}},\ }\href {https://doi.org/10.1103/PhysRevD.110.084060} {\bibfield  {journal} {\bibinfo  {journal} {Phys. Rev. D}\ }\textbf {\bibinfo {volume} {110}},\ \bibinfo {pages} {084060} (\bibinfo {year} {2024})},\ \Eprint {https://arxiv.org/abs/2312.13028} {arXiv:2312.13028 [gr-qc]} \BibitemShut {NoStop}%
\bibitem [{\citenamefont {Yunes}\ \emph {et~al.}(2011)\citenamefont {Yunes}, \citenamefont {Coleman~Miller},\ and\ \citenamefont {Thornburg}}]{Yunes:2010sm}%
  \BibitemOpen
  \bibfield  {author} {\bibinfo {author} {\bibfnamefont {N.}~\bibnamefont {Yunes}}, \bibinfo {author} {\bibfnamefont {M.}~\bibnamefont {Coleman~Miller}},\ and\ \bibinfo {author} {\bibfnamefont {J.}~\bibnamefont {Thornburg}},\ }\bibfield  {title} {\bibinfo {title} {{The Effect of Massive Perturbers on Extreme Mass-Ratio Inspiral Waveforms}},\ }\href {https://doi.org/10.1103/PhysRevD.83.044030} {\bibfield  {journal} {\bibinfo  {journal} {Phys. Rev. D}\ }\textbf {\bibinfo {volume} {83}},\ \bibinfo {pages} {044030} (\bibinfo {year} {2011})},\ \Eprint {https://arxiv.org/abs/1010.1721} {arXiv:1010.1721 [astro-ph.GA]} \BibitemShut {NoStop}%
\bibitem [{\citenamefont {Chamberlain}\ \emph {et~al.}(2019)\citenamefont {Chamberlain}, \citenamefont {Moore}, \citenamefont {Gerosa},\ and\ \citenamefont {Yunes}}]{Chamberlain:2018snj}%
  \BibitemOpen
  \bibfield  {author} {\bibinfo {author} {\bibfnamefont {K.}~\bibnamefont {Chamberlain}}, \bibinfo {author} {\bibfnamefont {C.~J.}\ \bibnamefont {Moore}}, \bibinfo {author} {\bibfnamefont {D.}~\bibnamefont {Gerosa}},\ and\ \bibinfo {author} {\bibfnamefont {N.}~\bibnamefont {Yunes}},\ }\bibfield  {title} {\bibinfo {title} {{Frequency-domain waveform approximants capturing Doppler shifts}},\ }\href {https://doi.org/10.1103/PhysRevD.99.024025} {\bibfield  {journal} {\bibinfo  {journal} {Phys. Rev. D}\ }\textbf {\bibinfo {volume} {99}},\ \bibinfo {pages} {024025} (\bibinfo {year} {2019})},\ \Eprint {https://arxiv.org/abs/1809.04799} {arXiv:1809.04799 [gr-qc]} \BibitemShut {NoStop}%
\bibitem [{\citenamefont {Robson}\ \emph {et~al.}(2018)\citenamefont {Robson}, \citenamefont {Cornish}, \citenamefont {Tamanini},\ and\ \citenamefont {Toonen}}]{Robson:2018svj}%
  \BibitemOpen
  \bibfield  {author} {\bibinfo {author} {\bibfnamefont {T.}~\bibnamefont {Robson}}, \bibinfo {author} {\bibfnamefont {N.~J.}\ \bibnamefont {Cornish}}, \bibinfo {author} {\bibfnamefont {N.}~\bibnamefont {Tamanini}},\ and\ \bibinfo {author} {\bibfnamefont {S.}~\bibnamefont {Toonen}},\ }\bibfield  {title} {\bibinfo {title} {{Detecting hierarchical stellar systems with LISA}},\ }\href {https://doi.org/10.1103/PhysRevD.98.064012} {\bibfield  {journal} {\bibinfo  {journal} {Phys. Rev. D}\ }\textbf {\bibinfo {volume} {98}},\ \bibinfo {pages} {064012} (\bibinfo {year} {2018})},\ \Eprint {https://arxiv.org/abs/1806.00500} {arXiv:1806.00500 [gr-qc]} \BibitemShut {NoStop}%
\bibitem [{\citenamefont {Yu}\ and\ \citenamefont {Chen}(2021)}]{Yu:2020dlm}%
  \BibitemOpen
  \bibfield  {author} {\bibinfo {author} {\bibfnamefont {H.}~\bibnamefont {Yu}}\ and\ \bibinfo {author} {\bibfnamefont {Y.}~\bibnamefont {Chen}},\ }\bibfield  {title} {\bibinfo {title} {{Direct determination of supermassive black hole properties with gravitational-wave radiation from surrounding stellar-mass black hole binaries}},\ }\href {https://doi.org/10.1103/PhysRevLett.126.021101} {\bibfield  {journal} {\bibinfo  {journal} {Phys. Rev. Lett.}\ }\textbf {\bibinfo {volume} {126}},\ \bibinfo {pages} {021101} (\bibinfo {year} {2021})},\ \Eprint {https://arxiv.org/abs/2009.02579} {arXiv:2009.02579 [gr-qc]} \BibitemShut {NoStop}%
\bibitem [{\citenamefont {Yu}\ \emph {et~al.}(2021)\citenamefont {Yu}, \citenamefont {Wang}, \citenamefont {Seymour},\ and\ \citenamefont {Chen}}]{Yu:2021dqx}%
  \BibitemOpen
  \bibfield  {author} {\bibinfo {author} {\bibfnamefont {H.}~\bibnamefont {Yu}}, \bibinfo {author} {\bibfnamefont {Y.}~\bibnamefont {Wang}}, \bibinfo {author} {\bibfnamefont {B.}~\bibnamefont {Seymour}},\ and\ \bibinfo {author} {\bibfnamefont {Y.}~\bibnamefont {Chen}},\ }\bibfield  {title} {\bibinfo {title} {{Detecting gravitational lensing in hierarchical triples in galactic nuclei with space-borne gravitational-wave observatories}},\ }\href {https://doi.org/10.1103/PhysRevD.104.103011} {\bibfield  {journal} {\bibinfo  {journal} {Phys. Rev. D}\ }\textbf {\bibinfo {volume} {104}},\ \bibinfo {pages} {103011} (\bibinfo {year} {2021})},\ \Eprint {https://arxiv.org/abs/2107.14318} {arXiv:2107.14318 [gr-qc]} \BibitemShut {NoStop}%
\bibitem [{\citenamefont {Laeuger}\ \emph {et~al.}(2024)\citenamefont {Laeuger}, \citenamefont {Seymour}, \citenamefont {Chen},\ and\ \citenamefont {Yu}}]{Laeuger:2023qyz}%
  \BibitemOpen
  \bibfield  {author} {\bibinfo {author} {\bibfnamefont {A.}~\bibnamefont {Laeuger}}, \bibinfo {author} {\bibfnamefont {B.}~\bibnamefont {Seymour}}, \bibinfo {author} {\bibfnamefont {Y.}~\bibnamefont {Chen}},\ and\ \bibinfo {author} {\bibfnamefont {H.}~\bibnamefont {Yu}},\ }\bibfield  {title} {\bibinfo {title} {{Measuring supermassive black hole properties via gravitational radiation from eccentrically orbiting stellar mass black hole binaries}},\ }\href {https://doi.org/10.1103/PhysRevD.109.064086} {\bibfield  {journal} {\bibinfo  {journal} {Phys. Rev. D}\ }\textbf {\bibinfo {volume} {109}},\ \bibinfo {pages} {064086} (\bibinfo {year} {2024})},\ \Eprint {https://arxiv.org/abs/2310.16799} {arXiv:2310.16799 [gr-qc]} \BibitemShut {NoStop}%
\bibitem [{\citenamefont {Deme}\ \emph {et~al.}(2020)\citenamefont {Deme}, \citenamefont {Hoang}, \citenamefont {Naoz},\ and\ \citenamefont {Kocsis}}]{Deme:2020ewx}%
  \BibitemOpen
  \bibfield  {author} {\bibinfo {author} {\bibfnamefont {B.}~\bibnamefont {Deme}}, \bibinfo {author} {\bibfnamefont {B.-M.}\ \bibnamefont {Hoang}}, \bibinfo {author} {\bibfnamefont {S.}~\bibnamefont {Naoz}},\ and\ \bibinfo {author} {\bibfnamefont {B.}~\bibnamefont {Kocsis}},\ }\bibfield  {title} {\bibinfo {title} {{Detecting Kozai{\textendash}Lidov Imprints on the Gravitational Waves of Intermediate-mass Black Holes in Galactic Nuclei}},\ }\href {https://doi.org/10.3847/1538-4357/abafa3} {\bibfield  {journal} {\bibinfo  {journal} {Astrophys. J.}\ }\textbf {\bibinfo {volume} {901}},\ \bibinfo {pages} {125} (\bibinfo {year} {2020})},\ \Eprint {https://arxiv.org/abs/2005.03677} {arXiv:2005.03677 [astro-ph.HE]} \BibitemShut {NoStop}%
\bibitem [{\citenamefont {Chandramouli}\ and\ \citenamefont {Yunes}(2022)}]{Chandramouli:2021kts}%
  \BibitemOpen
  \bibfield  {author} {\bibinfo {author} {\bibfnamefont {R.~S.}\ \bibnamefont {Chandramouli}}\ and\ \bibinfo {author} {\bibfnamefont {N.}~\bibnamefont {Yunes}},\ }\bibfield  {title} {\bibinfo {title} {{Ready-to-use analytic model for gravitational waves from a hierarchical triple with Kozai-Lidov oscillations}},\ }\href {https://doi.org/10.1103/PhysRevD.105.064009} {\bibfield  {journal} {\bibinfo  {journal} {Phys. Rev. D}\ }\textbf {\bibinfo {volume} {105}},\ \bibinfo {pages} {064009} (\bibinfo {year} {2022})},\ \Eprint {https://arxiv.org/abs/2107.00741} {arXiv:2107.00741 [gr-qc]} \BibitemShut {NoStop}%
\bibitem [{\citenamefont {Gupta}\ \emph {et~al.}(2020{\natexlab{a}})\citenamefont {Gupta}, \citenamefont {Suzuki}, \citenamefont {Okawa},\ and\ \citenamefont {Maeda}}]{Gupta:2019unn}%
  \BibitemOpen
  \bibfield  {author} {\bibinfo {author} {\bibfnamefont {P.}~\bibnamefont {Gupta}}, \bibinfo {author} {\bibfnamefont {H.}~\bibnamefont {Suzuki}}, \bibinfo {author} {\bibfnamefont {H.}~\bibnamefont {Okawa}},\ and\ \bibinfo {author} {\bibfnamefont {K.-i.}\ \bibnamefont {Maeda}},\ }\bibfield  {title} {\bibinfo {title} {{Gravitational Waves from Hierarchical Triple Systems with Kozai-Lidov Oscillation}},\ }\href {https://doi.org/10.1103/PhysRevD.101.104053} {\bibfield  {journal} {\bibinfo  {journal} {Phys. Rev. D}\ }\textbf {\bibinfo {volume} {101}},\ \bibinfo {pages} {104053} (\bibinfo {year} {2020}{\natexlab{a}})},\ \Eprint {https://arxiv.org/abs/1911.11318} {arXiv:1911.11318 [gr-qc]} \BibitemShut {NoStop}%
\bibitem [{\citenamefont {Abbott}\ \emph {et~al.}(2016{\natexlab{c}})\citenamefont {Abbott} \emph {et~al.}}]{LIGOScientific:2016gtq}%
  \BibitemOpen
  \bibfield  {author} {\bibinfo {author} {\bibfnamefont {B.~P.}\ \bibnamefont {Abbott}} \emph {et~al.} (\bibinfo {collaboration} {LIGO Scientific, Virgo}),\ }\bibfield  {title} {\bibinfo {title} {{Characterization of transient noise in Advanced LIGO relevant to gravitational wave signal GW150914}},\ }\href {https://doi.org/10.1088/0264-9381/33/13/134001} {\bibfield  {journal} {\bibinfo  {journal} {Class. Quant. Grav.}\ }\textbf {\bibinfo {volume} {33}},\ \bibinfo {pages} {134001} (\bibinfo {year} {2016}{\natexlab{c}})},\ \Eprint {https://arxiv.org/abs/1602.03844} {arXiv:1602.03844 [gr-qc]} \BibitemShut {NoStop}%
\bibitem [{\citenamefont {Ghonge}\ \emph {et~al.}(2024)\citenamefont {Ghonge}, \citenamefont {Brandt}, \citenamefont {Sullivan}, \citenamefont {Millhouse}, \citenamefont {Chatziioannou}, \citenamefont {Clark}, \citenamefont {Littenberg}, \citenamefont {Cornish}, \citenamefont {Hourihane},\ and\ \citenamefont {Cadonati}}]{Ghonge:2023ksb}%
  \BibitemOpen
  \bibfield  {author} {\bibinfo {author} {\bibfnamefont {S.}~\bibnamefont {Ghonge}}, \bibinfo {author} {\bibfnamefont {J.}~\bibnamefont {Brandt}}, \bibinfo {author} {\bibfnamefont {J.~M.}\ \bibnamefont {Sullivan}}, \bibinfo {author} {\bibfnamefont {M.}~\bibnamefont {Millhouse}}, \bibinfo {author} {\bibfnamefont {K.}~\bibnamefont {Chatziioannou}}, \bibinfo {author} {\bibfnamefont {J.~A.}\ \bibnamefont {Clark}}, \bibinfo {author} {\bibfnamefont {T.}~\bibnamefont {Littenberg}}, \bibinfo {author} {\bibfnamefont {N.}~\bibnamefont {Cornish}}, \bibinfo {author} {\bibfnamefont {S.}~\bibnamefont {Hourihane}},\ and\ \bibinfo {author} {\bibfnamefont {L.}~\bibnamefont {Cadonati}},\ }\bibfield  {title} {\bibinfo {title} {{Assessing and mitigating the impact of glitches on gravitational-wave parameter estimation: A model agnostic approach}},\ }\href {https://doi.org/10.1103/PhysRevD.110.122002} {\bibfield  {journal} {\bibinfo  {journal} {Phys. Rev. D}\ }\textbf {\bibinfo {volume} {110}},\ \bibinfo {pages} {122002} (\bibinfo {year} {2024})},\ \Eprint {https://arxiv.org/abs/2311.09159} {arXiv:2311.09159 [gr-qc]} \BibitemShut {NoStop}%
\bibitem [{\citenamefont {Kwok}\ \emph {et~al.}(2022)\citenamefont {Kwok}, \citenamefont {Lo}, \citenamefont {Weinstein},\ and\ \citenamefont {Li}}]{Kwok:2021zny}%
  \BibitemOpen
  \bibfield  {author} {\bibinfo {author} {\bibfnamefont {J.~Y.~L.}\ \bibnamefont {Kwok}}, \bibinfo {author} {\bibfnamefont {R.~K.~L.}\ \bibnamefont {Lo}}, \bibinfo {author} {\bibfnamefont {A.~J.}\ \bibnamefont {Weinstein}},\ and\ \bibinfo {author} {\bibfnamefont {T.~G.~F.}\ \bibnamefont {Li}},\ }\bibfield  {title} {\bibinfo {title} {{Investigation of the effects of non-Gaussian noise transients and their mitigation in parameterized gravitational-wave tests of general relativity}},\ }\href {https://doi.org/10.1103/PhysRevD.105.024066} {\bibfield  {journal} {\bibinfo  {journal} {Phys. Rev. D}\ }\textbf {\bibinfo {volume} {105}},\ \bibinfo {pages} {024066} (\bibinfo {year} {2022})},\ \Eprint {https://arxiv.org/abs/2109.07642} {arXiv:2109.07642 [gr-qc]} \BibitemShut {NoStop}%
\bibitem [{\citenamefont {Ashton}(2023)}]{Ashton:2022ztk}%
  \BibitemOpen
  \bibfield  {author} {\bibinfo {author} {\bibfnamefont {G.}~\bibnamefont {Ashton}},\ }\bibfield  {title} {\bibinfo {title} {{Gaussian processes for glitch-robust gravitational-wave astronomy}},\ }\href {https://doi.org/10.1093/mnras/stad341} {\bibfield  {journal} {\bibinfo  {journal} {Mon. Not. Roy. Astron. Soc.}\ }\textbf {\bibinfo {volume} {520}},\ \bibinfo {pages} {2983} (\bibinfo {year} {2023})},\ \Eprint {https://arxiv.org/abs/2209.15547} {arXiv:2209.15547 [gr-qc]} \BibitemShut {NoStop}%
\bibitem [{\citenamefont {Spadaro}\ \emph {et~al.}(2023)\citenamefont {Spadaro}, \citenamefont {Buscicchio}, \citenamefont {Vetrugno}, \citenamefont {Klein}, \citenamefont {Vitale}, \citenamefont {Dolesi}, \citenamefont {Weber},\ and\ \citenamefont {Colpi}}]{Spadaro:2023muy}%
  \BibitemOpen
  \bibfield  {author} {\bibinfo {author} {\bibfnamefont {A.}~\bibnamefont {Spadaro}}, \bibinfo {author} {\bibfnamefont {R.}~\bibnamefont {Buscicchio}}, \bibinfo {author} {\bibfnamefont {D.}~\bibnamefont {Vetrugno}}, \bibinfo {author} {\bibfnamefont {A.}~\bibnamefont {Klein}}, \bibinfo {author} {\bibfnamefont {S.}~\bibnamefont {Vitale}}, \bibinfo {author} {\bibfnamefont {R.}~\bibnamefont {Dolesi}}, \bibinfo {author} {\bibfnamefont {W.~J.}\ \bibnamefont {Weber}},\ and\ \bibinfo {author} {\bibfnamefont {M.}~\bibnamefont {Colpi}},\ }\bibfield  {title} {\bibinfo {title} {{Glitch systematics on the observation of massive black-hole binaries with LISA}},\ }\href {https://doi.org/10.1103/PhysRevD.108.123029} {\bibfield  {journal} {\bibinfo  {journal} {Phys. Rev. D}\ }\textbf {\bibinfo {volume} {108}},\ \bibinfo {pages} {123029} (\bibinfo {year} {2023})},\ \Eprint {https://arxiv.org/abs/2306.03923} {arXiv:2306.03923 [gr-qc]} \BibitemShut {NoStop}%
\bibitem [{\citenamefont {Pai}\ and\ \citenamefont {Arun}(2013)}]{Pai:2012mv}%
  \BibitemOpen
  \bibfield  {author} {\bibinfo {author} {\bibfnamefont {A.}~\bibnamefont {Pai}}\ and\ \bibinfo {author} {\bibfnamefont {K.~G.}\ \bibnamefont {Arun}},\ }\bibfield  {title} {\bibinfo {title} {{Singular value decomposition in parametrised tests of post-Newtonian theory}},\ }\href {https://doi.org/10.1088/0264-9381/30/2/025011} {\bibfield  {journal} {\bibinfo  {journal} {Class. Quant. Grav.}\ }\textbf {\bibinfo {volume} {30}},\ \bibinfo {pages} {025011} (\bibinfo {year} {2013})},\ \Eprint {https://arxiv.org/abs/1207.1943} {arXiv:1207.1943 [gr-qc]} \BibitemShut {NoStop}%
\bibitem [{\citenamefont {Arun}\ and\ \citenamefont {Pai}(2013)}]{Arun:2013bp}%
  \BibitemOpen
  \bibfield  {author} {\bibinfo {author} {\bibfnamefont {K.~G.}\ \bibnamefont {Arun}}\ and\ \bibinfo {author} {\bibfnamefont {A.}~\bibnamefont {Pai}},\ }\bibfield  {title} {\bibinfo {title} {{Tests of General Relativity and Alternative theories of gravity using Gravitational Wave observations}},\ }\href {https://doi.org/10.1142/S0218271813410125} {\bibfield  {journal} {\bibinfo  {journal} {Int. J. Mod. Phys. D}\ }\textbf {\bibinfo {volume} {22}},\ \bibinfo {pages} {1341012} (\bibinfo {year} {2013})},\ \Eprint {https://arxiv.org/abs/1302.2198} {arXiv:1302.2198 [gr-qc]} \BibitemShut {NoStop}%
\bibitem [{\citenamefont {Cannon}\ \emph {et~al.}(2010)\citenamefont {Cannon}, \citenamefont {Chapman}, \citenamefont {Hanna}, \citenamefont {Keppel}, \citenamefont {Searle},\ and\ \citenamefont {Weinstein}}]{Cannon:2010qh}%
  \BibitemOpen
  \bibfield  {author} {\bibinfo {author} {\bibfnamefont {K.}~\bibnamefont {Cannon}}, \bibinfo {author} {\bibfnamefont {A.}~\bibnamefont {Chapman}}, \bibinfo {author} {\bibfnamefont {C.}~\bibnamefont {Hanna}}, \bibinfo {author} {\bibfnamefont {D.}~\bibnamefont {Keppel}}, \bibinfo {author} {\bibfnamefont {A.~C.}\ \bibnamefont {Searle}},\ and\ \bibinfo {author} {\bibfnamefont {A.~J.}\ \bibnamefont {Weinstein}},\ }\bibfield  {title} {\bibinfo {title} {{Singular value decomposition applied to compact binary coalescence gravitational-wave signals}},\ }\href {https://doi.org/10.1103/PhysRevD.82.044025} {\bibfield  {journal} {\bibinfo  {journal} {Phys. Rev. D}\ }\textbf {\bibinfo {volume} {82}},\ \bibinfo {pages} {044025} (\bibinfo {year} {2010})},\ \Eprint {https://arxiv.org/abs/1005.0012} {arXiv:1005.0012 [gr-qc]} \BibitemShut {NoStop}%
\bibitem [{\citenamefont {Cannon}\ \emph {et~al.}(2011)\citenamefont {Cannon}, \citenamefont {Hanna},\ and\ \citenamefont {Keppel}}]{Cannon:2011xk}%
  \BibitemOpen
  \bibfield  {author} {\bibinfo {author} {\bibfnamefont {K.}~\bibnamefont {Cannon}}, \bibinfo {author} {\bibfnamefont {C.}~\bibnamefont {Hanna}},\ and\ \bibinfo {author} {\bibfnamefont {D.}~\bibnamefont {Keppel}},\ }\bibfield  {title} {\bibinfo {title} {{Efficiently enclosing the compact binary parameter space by singular-value decomposition}},\ }\href {https://doi.org/10.1103/PhysRevD.84.084003} {\bibfield  {journal} {\bibinfo  {journal} {Phys. Rev. D}\ }\textbf {\bibinfo {volume} {84}},\ \bibinfo {pages} {084003} (\bibinfo {year} {2011})},\ \Eprint {https://arxiv.org/abs/1101.4939} {arXiv:1101.4939 [gr-qc]} \BibitemShut {NoStop}%
\bibitem [{\citenamefont {Cannon}\ \emph {et~al.}(2012)\citenamefont {Cannon}, \citenamefont {Hanna},\ and\ \citenamefont {Keppel}}]{Cannon:2011rj}%
  \BibitemOpen
  \bibfield  {author} {\bibinfo {author} {\bibfnamefont {K.}~\bibnamefont {Cannon}}, \bibinfo {author} {\bibfnamefont {C.}~\bibnamefont {Hanna}},\ and\ \bibinfo {author} {\bibfnamefont {D.}~\bibnamefont {Keppel}},\ }\bibfield  {title} {\bibinfo {title} {{Interpolating compact binary waveforms using the singular value decomposition}},\ }\href {https://doi.org/10.1103/PhysRevD.85.081504} {\bibfield  {journal} {\bibinfo  {journal} {Phys. Rev. D}\ }\textbf {\bibinfo {volume} {85}},\ \bibinfo {pages} {081504} (\bibinfo {year} {2012})},\ \Eprint {https://arxiv.org/abs/1108.5618} {arXiv:1108.5618 [gr-qc]} \BibitemShut {NoStop}%
\bibitem [{\citenamefont {Keppel}(2013)}]{Keppel:2012nb}%
  \BibitemOpen
  \bibfield  {author} {\bibinfo {author} {\bibfnamefont {D.}~\bibnamefont {Keppel}},\ }\bibfield  {title} {\bibinfo {title} {{Use of Singular-Value Decomposition in Gravitational-Wave Data Analysis}},\ }\href {https://doi.org/10.1142/S2010194513011136} {\bibfield  {journal} {\bibinfo  {journal} {Int. J. Mod. Phys. Conf. Ser.}\ }\textbf {\bibinfo {volume} {23}},\ \bibinfo {pages} {01113} (\bibinfo {year} {2013})},\ \Eprint {https://arxiv.org/abs/1201.1739} {arXiv:1201.1739 [physics.data-an]} \BibitemShut {NoStop}%
\bibitem [{\citenamefont {Tiglio}\ and\ \citenamefont {Villanueva}(2022)}]{Tiglio:2021ysj}%
  \BibitemOpen
  \bibfield  {author} {\bibinfo {author} {\bibfnamefont {M.}~\bibnamefont {Tiglio}}\ and\ \bibinfo {author} {\bibfnamefont {A.}~\bibnamefont {Villanueva}},\ }\bibfield  {title} {\bibinfo {title} {{Reduced order and surrogate models for gravitational waves}},\ }\href {https://doi.org/10.1007/s41114-022-00035-w} {\bibfield  {journal} {\bibinfo  {journal} {Living Rev. Rel.}\ }\textbf {\bibinfo {volume} {25}},\ \bibinfo {pages} {2} (\bibinfo {year} {2022})},\ \Eprint {https://arxiv.org/abs/2101.11608} {arXiv:2101.11608 [gr-qc]} \BibitemShut {NoStop}%
\bibitem [{\citenamefont {Datta}\ \emph {et~al.}(2024)\citenamefont {Datta}, \citenamefont {Saleem}, \citenamefont {Arun},\ and\ \citenamefont {Sathyaprakash}}]{Datta:2022izc}%
  \BibitemOpen
  \bibfield  {author} {\bibinfo {author} {\bibfnamefont {S.}~\bibnamefont {Datta}}, \bibinfo {author} {\bibfnamefont {M.}~\bibnamefont {Saleem}}, \bibinfo {author} {\bibfnamefont {K.~G.}\ \bibnamefont {Arun}},\ and\ \bibinfo {author} {\bibfnamefont {B.~S.}\ \bibnamefont {Sathyaprakash}},\ }\bibfield  {title} {\bibinfo {title} {{Multiparameter tests of general relativity using a principle component analysis with next-generation gravitational-wave detectors}},\ }\href {https://doi.org/10.1103/PhysRevD.109.044036} {\bibfield  {journal} {\bibinfo  {journal} {Phys. Rev. D}\ }\textbf {\bibinfo {volume} {109}},\ \bibinfo {pages} {044036} (\bibinfo {year} {2024})},\ \Eprint {https://arxiv.org/abs/2208.07757} {arXiv:2208.07757 [gr-qc]} \BibitemShut {NoStop}%
\bibitem [{\citenamefont {Saleem}\ \emph {et~al.}(2022{\natexlab{b}})\citenamefont {Saleem}, \citenamefont {Datta}, \citenamefont {Arun},\ and\ \citenamefont {Sathyaprakash}}]{Saleem:2021nsb}%
  \BibitemOpen
  \bibfield  {author} {\bibinfo {author} {\bibfnamefont {M.}~\bibnamefont {Saleem}}, \bibinfo {author} {\bibfnamefont {S.}~\bibnamefont {Datta}}, \bibinfo {author} {\bibfnamefont {K.~G.}\ \bibnamefont {Arun}},\ and\ \bibinfo {author} {\bibfnamefont {B.~S.}\ \bibnamefont {Sathyaprakash}},\ }\bibfield  {title} {\bibinfo {title} {{Parametrized tests of post-Newtonian theory using principal component analysis}},\ }\href {https://doi.org/10.1103/PhysRevD.105.084062} {\bibfield  {journal} {\bibinfo  {journal} {Phys. Rev. D}\ }\textbf {\bibinfo {volume} {105}},\ \bibinfo {pages} {084062} (\bibinfo {year} {2022}{\natexlab{b}})},\ \Eprint {https://arxiv.org/abs/2110.10147} {arXiv:2110.10147 [gr-qc]} \BibitemShut {NoStop}%
\bibitem [{\citenamefont {Datta}(2023)}]{Datta:2023muk}%
  \BibitemOpen
  \bibfield  {author} {\bibinfo {author} {\bibfnamefont {S.}~\bibnamefont {Datta}},\ }\href@noop {} {\bibinfo {title} {{Enhancing the performance of multiparameter tests of general relativity with LISA using Principal Component Analysis}}} (\bibinfo {year} {2023}),\ \Eprint {https://arxiv.org/abs/2303.04399} {arXiv:2303.04399 [gr-qc]} \BibitemShut {NoStop}%
\bibitem [{\citenamefont {Ma}\ \emph {et~al.}(2024)\citenamefont {Ma}, \citenamefont {Niu},\ and\ \citenamefont {Zhao}}]{Ma:2024kkz}%
  \BibitemOpen
  \bibfield  {author} {\bibinfo {author} {\bibfnamefont {Z.-C.}\ \bibnamefont {Ma}}, \bibinfo {author} {\bibfnamefont {R.}~\bibnamefont {Niu}},\ and\ \bibinfo {author} {\bibfnamefont {W.}~\bibnamefont {Zhao}},\ }\bibfield  {title} {\bibinfo {title} {{The Multi-parameter Test of Gravitational Wave Dispersion with Principal Component Analysis}},\ }\href {https://doi.org/10.1088/1674-4527/ad3c70} {\bibfield  {journal} {\bibinfo  {journal} {Res. Astron. Astrophys.}\ }\textbf {\bibinfo {volume} {24}},\ \bibinfo {pages} {055012} (\bibinfo {year} {2024})},\ \Eprint {https://arxiv.org/abs/2401.17666} {arXiv:2401.17666 [gr-qc]} \BibitemShut {NoStop}%
\bibitem [{\citenamefont {Mahapatra}\ \emph {et~al.}(2025)\citenamefont {Mahapatra} \emph {et~al.}}]{Mahapatra:2025cwk}%
  \BibitemOpen
  \bibfield  {author} {\bibinfo {author} {\bibfnamefont {P.}~\bibnamefont {Mahapatra}} \emph {et~al.},\ }\bibfield  {title} {\bibinfo {title} {{Confronting general relativity with principal component analysis: Simulations and results from GWTC-3 events}},\ }\href {https://doi.org/10.1103/c1sj-jc4v} {\bibfield  {journal} {\bibinfo  {journal} {Phys. Rev. D}\ }\textbf {\bibinfo {volume} {112}},\ \bibinfo {pages} {104007} (\bibinfo {year} {2025})},\ \Eprint {https://arxiv.org/abs/2508.06862} {arXiv:2508.06862 [gr-qc]} \BibitemShut {NoStop}%
\bibitem [{\citenamefont {Shoom}\ \emph {et~al.}(2023)\citenamefont {Shoom}, \citenamefont {Gupta}, \citenamefont {Krishnan}, \citenamefont {Nielsen},\ and\ \citenamefont {Capano}}]{Shoom:2021mdj}%
  \BibitemOpen
  \bibfield  {author} {\bibinfo {author} {\bibfnamefont {A.~A.}\ \bibnamefont {Shoom}}, \bibinfo {author} {\bibfnamefont {P.~K.}\ \bibnamefont {Gupta}}, \bibinfo {author} {\bibfnamefont {B.}~\bibnamefont {Krishnan}}, \bibinfo {author} {\bibfnamefont {A.~B.}\ \bibnamefont {Nielsen}},\ and\ \bibinfo {author} {\bibfnamefont {C.~D.}\ \bibnamefont {Capano}},\ }\bibfield  {title} {\bibinfo {title} {{Testing the post-Newtonian expansion with GW170817}},\ }\href {https://doi.org/10.1007/s10714-023-03100-z} {\bibfield  {journal} {\bibinfo  {journal} {Gen. Rel. Grav.}\ }\textbf {\bibinfo {volume} {55}},\ \bibinfo {pages} {55} (\bibinfo {year} {2023})},\ \Eprint {https://arxiv.org/abs/2105.02191} {arXiv:2105.02191 [gr-qc]} \BibitemShut {NoStop}%
\bibitem [{\citenamefont {Arun}\ \emph {et~al.}(2006{\natexlab{a}})\citenamefont {Arun}, \citenamefont {Iyer}, \citenamefont {Qusailah},\ and\ \citenamefont {Sathyaprakash}}]{Arun:2006yw}%
  \BibitemOpen
  \bibfield  {author} {\bibinfo {author} {\bibfnamefont {K.~G.}\ \bibnamefont {Arun}}, \bibinfo {author} {\bibfnamefont {B.~R.}\ \bibnamefont {Iyer}}, \bibinfo {author} {\bibfnamefont {M.~S.~S.}\ \bibnamefont {Qusailah}},\ and\ \bibinfo {author} {\bibfnamefont {B.~S.}\ \bibnamefont {Sathyaprakash}},\ }\bibfield  {title} {\bibinfo {title} {{Testing post-Newtonian theory with gravitational wave observations}},\ }\href {https://doi.org/10.1088/0264-9381/23/9/L01} {\bibfield  {journal} {\bibinfo  {journal} {Class. Quant. Grav.}\ }\textbf {\bibinfo {volume} {23}},\ \bibinfo {pages} {L37} (\bibinfo {year} {2006}{\natexlab{a}})},\ \Eprint {https://arxiv.org/abs/gr-qc/0604018} {arXiv:gr-qc/0604018} \BibitemShut {NoStop}%
\bibitem [{\citenamefont {Arun}\ \emph {et~al.}(2006{\natexlab{b}})\citenamefont {Arun}, \citenamefont {Iyer}, \citenamefont {Qusailah},\ and\ \citenamefont {Sathyaprakash}}]{Arun:2006hn}%
  \BibitemOpen
  \bibfield  {author} {\bibinfo {author} {\bibfnamefont {K.~G.}\ \bibnamefont {Arun}}, \bibinfo {author} {\bibfnamefont {B.~R.}\ \bibnamefont {Iyer}}, \bibinfo {author} {\bibfnamefont {M.~S.~S.}\ \bibnamefont {Qusailah}},\ and\ \bibinfo {author} {\bibfnamefont {B.~S.}\ \bibnamefont {Sathyaprakash}},\ }\bibfield  {title} {\bibinfo {title} {{Probing the non-linear structure of general relativity with black hole binaries}},\ }\href {https://doi.org/10.1103/PhysRevD.74.024006} {\bibfield  {journal} {\bibinfo  {journal} {Phys. Rev. D}\ }\textbf {\bibinfo {volume} {74}},\ \bibinfo {pages} {024006} (\bibinfo {year} {2006}{\natexlab{b}})},\ \Eprint {https://arxiv.org/abs/gr-qc/0604067} {arXiv:gr-qc/0604067} \BibitemShut {NoStop}%
\bibitem [{\citenamefont {V{\"o}lkel}\ \emph {et~al.}(2022)\citenamefont {V{\"o}lkel}, \citenamefont {Franchini},\ and\ \citenamefont {Barausse}}]{Volkel:2022aca}%
  \BibitemOpen
  \bibfield  {author} {\bibinfo {author} {\bibfnamefont {S.~H.}\ \bibnamefont {V{\"o}lkel}}, \bibinfo {author} {\bibfnamefont {N.}~\bibnamefont {Franchini}},\ and\ \bibinfo {author} {\bibfnamefont {E.}~\bibnamefont {Barausse}},\ }\bibfield  {title} {\bibinfo {title} {{Theory-agnostic reconstruction of potential and couplings from quasinormal modes}},\ }\href {https://doi.org/10.1103/PhysRevD.105.084046} {\bibfield  {journal} {\bibinfo  {journal} {Phys. Rev. D}\ }\textbf {\bibinfo {volume} {105}},\ \bibinfo {pages} {084046} (\bibinfo {year} {2022})},\ \Eprint {https://arxiv.org/abs/2202.08655} {arXiv:2202.08655 [gr-qc]} \BibitemShut {NoStop}%
\bibitem [{\citenamefont {Cutler}\ and\ \citenamefont {Flanagan}(1994)}]{Cutler:1994ys}%
  \BibitemOpen
  \bibfield  {author} {\bibinfo {author} {\bibfnamefont {C.}~\bibnamefont {Cutler}}\ and\ \bibinfo {author} {\bibfnamefont {E.~E.}\ \bibnamefont {Flanagan}},\ }\bibfield  {title} {\bibinfo {title} {{Gravitational waves from merging compact binaries: How accurately can one extract the binary's parameters from the inspiral wave form?}},\ }\href {https://doi.org/10.1103/PhysRevD.49.2658} {\bibfield  {journal} {\bibinfo  {journal} {Phys. Rev. D}\ }\textbf {\bibinfo {volume} {49}},\ \bibinfo {pages} {2658} (\bibinfo {year} {1994})},\ \Eprint {https://arxiv.org/abs/gr-qc/9402014} {arXiv:gr-qc/9402014} \BibitemShut {NoStop}%
\bibitem [{\citenamefont {Vallisneri}(2008)}]{Vallisneri:2007ev}%
  \BibitemOpen
  \bibfield  {author} {\bibinfo {author} {\bibfnamefont {M.}~\bibnamefont {Vallisneri}},\ }\bibfield  {title} {\bibinfo {title} {{Use and abuse of the Fisher information matrix in the assessment of gravitational-wave parameter-estimation prospects}},\ }\href {https://doi.org/10.1103/PhysRevD.77.042001} {\bibfield  {journal} {\bibinfo  {journal} {Phys. Rev. D}\ }\textbf {\bibinfo {volume} {77}},\ \bibinfo {pages} {042001} (\bibinfo {year} {2008})},\ \Eprint {https://arxiv.org/abs/gr-qc/0703086} {arXiv:gr-qc/0703086} \BibitemShut {NoStop}%
\bibitem [{\citenamefont {Rodriguez}\ \emph {et~al.}(2013)\citenamefont {Rodriguez}, \citenamefont {Farr}, \citenamefont {Farr},\ and\ \citenamefont {Mandel}}]{Rodriguez:2013mla}%
  \BibitemOpen
  \bibfield  {author} {\bibinfo {author} {\bibfnamefont {C.~L.}\ \bibnamefont {Rodriguez}}, \bibinfo {author} {\bibfnamefont {B.}~\bibnamefont {Farr}}, \bibinfo {author} {\bibfnamefont {W.~M.}\ \bibnamefont {Farr}},\ and\ \bibinfo {author} {\bibfnamefont {I.}~\bibnamefont {Mandel}},\ }\bibfield  {title} {\bibinfo {title} {{Inadequacies of the Fisher Information Matrix in gravitational-wave parameter estimation}},\ }\href {https://doi.org/10.1103/PhysRevD.88.084013} {\bibfield  {journal} {\bibinfo  {journal} {Phys. Rev. D}\ }\textbf {\bibinfo {volume} {88}},\ \bibinfo {pages} {084013} (\bibinfo {year} {2013})},\ \Eprint {https://arxiv.org/abs/1308.1397} {arXiv:1308.1397 [astro-ph.IM]} \BibitemShut {NoStop}%
\bibitem [{\citenamefont {Finn}(1992)}]{Finn:1992wt}%
  \BibitemOpen
  \bibfield  {author} {\bibinfo {author} {\bibfnamefont {L.~S.}\ \bibnamefont {Finn}},\ }\bibfield  {title} {\bibinfo {title} {{Detection, measurement and gravitational radiation}},\ }\href {https://doi.org/10.1103/PhysRevD.46.5236} {\bibfield  {journal} {\bibinfo  {journal} {Phys. Rev. D}\ }\textbf {\bibinfo {volume} {46}},\ \bibinfo {pages} {5236} (\bibinfo {year} {1992})},\ \Eprint {https://arxiv.org/abs/gr-qc/9209010} {arXiv:gr-qc/9209010} \BibitemShut {NoStop}%
\bibitem [{\citenamefont {Mishra}\ \emph {et~al.}(2010)\citenamefont {Mishra}, \citenamefont {Arun}, \citenamefont {Iyer},\ and\ \citenamefont {Sathyaprakash}}]{Mishra:2010tp}%
  \BibitemOpen
  \bibfield  {author} {\bibinfo {author} {\bibfnamefont {C.~K.}\ \bibnamefont {Mishra}}, \bibinfo {author} {\bibfnamefont {K.~G.}\ \bibnamefont {Arun}}, \bibinfo {author} {\bibfnamefont {B.~R.}\ \bibnamefont {Iyer}},\ and\ \bibinfo {author} {\bibfnamefont {B.~S.}\ \bibnamefont {Sathyaprakash}},\ }\bibfield  {title} {\bibinfo {title} {{Parametrized tests of post-Newtonian theory using Advanced LIGO and Einstein Telescope}},\ }\href {https://doi.org/10.1103/PhysRevD.82.064010} {\bibfield  {journal} {\bibinfo  {journal} {Phys. Rev. D}\ }\textbf {\bibinfo {volume} {82}},\ \bibinfo {pages} {064010} (\bibinfo {year} {2010})},\ \Eprint {https://arxiv.org/abs/1005.0304} {arXiv:1005.0304 [gr-qc]} \BibitemShut {NoStop}%
\bibitem [{\citenamefont {Ajith}\ \emph {et~al.}(2011)\citenamefont {Ajith} \emph {et~al.}}]{Ajith:2009bn}%
  \BibitemOpen
  \bibfield  {author} {\bibinfo {author} {\bibfnamefont {P.}~\bibnamefont {Ajith}} \emph {et~al.},\ }\bibfield  {title} {\bibinfo {title} {{Inspiral-merger-ringdown waveforms for black-hole binaries with non-precessing spins}},\ }\href {https://doi.org/10.1103/PhysRevLett.106.241101} {\bibfield  {journal} {\bibinfo  {journal} {Phys. Rev. Lett.}\ }\textbf {\bibinfo {volume} {106}},\ \bibinfo {pages} {241101} (\bibinfo {year} {2011})},\ \Eprint {https://arxiv.org/abs/0909.2867} {arXiv:0909.2867 [gr-qc]} \BibitemShut {NoStop}%
\bibitem [{\citenamefont {Schmidt}\ \emph {et~al.}(2015)\citenamefont {Schmidt}, \citenamefont {Ohme},\ and\ \citenamefont {Hannam}}]{Schmidt:2014iyl}%
  \BibitemOpen
  \bibfield  {author} {\bibinfo {author} {\bibfnamefont {P.}~\bibnamefont {Schmidt}}, \bibinfo {author} {\bibfnamefont {F.}~\bibnamefont {Ohme}},\ and\ \bibinfo {author} {\bibfnamefont {M.}~\bibnamefont {Hannam}},\ }\bibfield  {title} {\bibinfo {title} {{Towards models of gravitational waveforms from generic binaries II: Modelling precession effects with a single effective precession parameter}},\ }\href {https://doi.org/10.1103/PhysRevD.91.024043} {\bibfield  {journal} {\bibinfo  {journal} {Phys. Rev. D}\ }\textbf {\bibinfo {volume} {91}},\ \bibinfo {pages} {024043} (\bibinfo {year} {2015})},\ \Eprint {https://arxiv.org/abs/1408.1810} {arXiv:1408.1810 [gr-qc]} \BibitemShut {NoStop}%
\bibitem [{\citenamefont {Ohme}\ \emph {et~al.}(2013)\citenamefont {Ohme}, \citenamefont {Nielsen}, \citenamefont {Keppel},\ and\ \citenamefont {Lundgren}}]{Ohme:2013nsa}%
  \BibitemOpen
  \bibfield  {author} {\bibinfo {author} {\bibfnamefont {F.}~\bibnamefont {Ohme}}, \bibinfo {author} {\bibfnamefont {A.~B.}\ \bibnamefont {Nielsen}}, \bibinfo {author} {\bibfnamefont {D.}~\bibnamefont {Keppel}},\ and\ \bibinfo {author} {\bibfnamefont {A.}~\bibnamefont {Lundgren}},\ }\bibfield  {title} {\bibinfo {title} {{Statistical and systematic errors for gravitational-wave inspiral signals: A principal component analysis}},\ }\href {https://doi.org/10.1103/PhysRevD.88.042002} {\bibfield  {journal} {\bibinfo  {journal} {Phys. Rev. D}\ }\textbf {\bibinfo {volume} {88}},\ \bibinfo {pages} {042002} (\bibinfo {year} {2013})},\ \Eprint {https://arxiv.org/abs/1304.7017} {arXiv:1304.7017 [gr-qc]} \BibitemShut {NoStop}%
\bibitem [{\citenamefont {Gupta}\ \emph {et~al.}(2020{\natexlab{b}})\citenamefont {Gupta}, \citenamefont {Datta}, \citenamefont {Kastha}, \citenamefont {Borhanian}, \citenamefont {Arun},\ and\ \citenamefont {Sathyaprakash}}]{Gupta:2020lxa}%
  \BibitemOpen
  \bibfield  {author} {\bibinfo {author} {\bibfnamefont {A.}~\bibnamefont {Gupta}}, \bibinfo {author} {\bibfnamefont {S.}~\bibnamefont {Datta}}, \bibinfo {author} {\bibfnamefont {S.}~\bibnamefont {Kastha}}, \bibinfo {author} {\bibfnamefont {S.}~\bibnamefont {Borhanian}}, \bibinfo {author} {\bibfnamefont {K.~G.}\ \bibnamefont {Arun}},\ and\ \bibinfo {author} {\bibfnamefont {B.~S.}\ \bibnamefont {Sathyaprakash}},\ }\bibfield  {title} {\bibinfo {title} {{Multiparameter tests of general relativity using multiband gravitational-wave observations}},\ }\href {https://doi.org/10.1103/PhysRevLett.125.201101} {\bibfield  {journal} {\bibinfo  {journal} {Phys. Rev. Lett.}\ }\textbf {\bibinfo {volume} {125}},\ \bibinfo {pages} {201101} (\bibinfo {year} {2020}{\natexlab{b}})},\ \Eprint {https://arxiv.org/abs/2005.09607} {arXiv:2005.09607 [gr-qc]} \BibitemShut {NoStop}%
\bibitem [{\citenamefont {Seymour}\ \emph {et~al.}(2026)\citenamefont {Seymour}, \citenamefont {Golomb},\ and\ \citenamefont {Chen}}]{seymour_2026_waveform_geometry}%
  \BibitemOpen
  \bibfield  {author} {\bibinfo {author} {\bibfnamefont {B.~C.}\ \bibnamefont {Seymour}}, \bibinfo {author} {\bibfnamefont {J.}~\bibnamefont {Golomb}},\ and\ \bibinfo {author} {\bibfnamefont {Y.}~\bibnamefont {Chen}},\ }\href {https://doi.org/10.5281/zenodo.20931755} {\bibinfo {title} {waveform-geometry-testing-gr}} (\bibinfo {year} {2026}),\ \bibinfo {note} {software}\BibitemShut {NoStop}%
\bibitem [{\citenamefont {Payne}\ \emph {et~al.}(2023)\citenamefont {Payne}, \citenamefont {Isi}, \citenamefont {Chatziioannou},\ and\ \citenamefont {Farr}}]{Payne:2023kwj}%
  \BibitemOpen
  \bibfield  {author} {\bibinfo {author} {\bibfnamefont {E.}~\bibnamefont {Payne}}, \bibinfo {author} {\bibfnamefont {M.}~\bibnamefont {Isi}}, \bibinfo {author} {\bibfnamefont {K.}~\bibnamefont {Chatziioannou}},\ and\ \bibinfo {author} {\bibfnamefont {W.~M.}\ \bibnamefont {Farr}},\ }\bibfield  {title} {\bibinfo {title} {{Fortifying gravitational-wave tests of general relativity against astrophysical assumptions}},\ }\href {https://doi.org/10.1103/PhysRevD.108.124060} {\bibfield  {journal} {\bibinfo  {journal} {Phys. Rev. D}\ }\textbf {\bibinfo {volume} {108}},\ \bibinfo {pages} {124060} (\bibinfo {year} {2023})},\ \Eprint {https://arxiv.org/abs/2309.04528} {arXiv:2309.04528 [gr-qc]} \BibitemShut {NoStop}%
\end{thebibliography}%

\end{document}